\documentclass[12pt]{article}
\usepackage{import}

\usepackage[margin=1in]{geometry}
\usepackage{setspace}
\usepackage[T1]{fontenc}
\usepackage[leftmargin = 1in, rightmargin = 0in, vskip = 0in]{quoting}
\usepackage{microtype}
\usepackage[utf8]{inputenc}
\usepackage[english]{babel}
\usepackage{amsmath, dirtytalk, xcolor, enumerate, graphicx, latexsym, amsfonts, amssymb, amsthm, euscript, mathrsfs, csquotes, verbatim, dsfont, hyperref, import, bbm}
\usepackage[tiny]{titlesec}
\usepackage[title]{appendix}
\usepackage{tikz, tikz-feynman}
\usetikzlibrary{shapes.geometric}

\usepackage[affil-it]{authblk}

\usepackage[style=numeric]{biblatex}
\renewbibmacro{in:}{}
\usepackage{hyperref}
    \hypersetup{
    colorlinks=true,
    linkcolor=purple,
    citecolor=blue,     
    urlcolor=blue,
    }

\allowdisplaybreaks

\newtheorem{thm}{Theorem}[section]
\newtheorem{prop}[thm]{Proposition}
\newtheorem{cor}[thm]{Corollary}

\theoremstyle{definition}
\newtheorem{defn}[thm]{Definition}

\newtheorem{lem}[thm]{Lemma}

\theoremstyle{remark}

\newcommand{\supp}{\text{supp}}
\newcommand{\R}{\mathbb R}
\newcommand{\C}{\mathbb C}

\graphicspath{ {./images/} }

\usepackage{orcidlink}

\usepackage{makeidx}
\makeindex

\title{Scaling Limits as Extensions of \\ Continuous Bundles of C*-algebras}
\author[1,2,*]{Kade Cicchella \orcidlink{0009-0004-0632-9189}}
\author[1]{Benjamin H. Feintzeig\orcidlink{0000-0001-7692-5155}}
\date{}

\affil[1]{Department of Philosophy, University of Washington}

\affil[2]{Department of Physics, University of Washington}

\affil[*]{Corresponding author: \normalfont \texttt{kadec@uw.edu}}

\begin{document}

\setstretch{1}
\maketitle
\setlength{\parindent}{.5in}

\maketitle

\section{Introduction}

This paper investigates the notion of a \emph{scaling limit} inherent in renormalization group methods in quantum field theory.  We aim to continue existing investigations that employ the C*-algebraic approach to quantum field theory: \textcite{buchholzScalingAlgebrasRenormalization1995,buchholzScalingAlgebrasRenormalization1998} initiated the study of what they call \emph{scaling algebras}---C*-algebras whose elements are certain maps
\begin{align}
    \lambda\in (0,1]\mapsto \underline{A}_\lambda\in \mathfrak{A}(\lambda\mathcal{O})
\end{align}
associating to each value $\lambda$ of a scaling parameter an observable $\underline{A}_\lambda$ contained in the C*-algebra $\mathfrak{A}(\lambda\mathcal{O})$ of local quantities in the region $\lambda\mathcal{O}$ of Minkowski spacetime.  The thought is that as $\lambda\to 0$, the map $\underline{A}$ captures the scaling behavior of a quantity on smaller and smaller distances, which suffices to encode the ultraviolet, or small-distance scaling limit of a quantum field theory.  We will draw a connection between scaling algebras and bundles of C*-algebras, which have been studied for some time in operator theory generally \parencite{fellStructureAlgebrasOperator1961,fellRepresentationsAlgebrasLocally1988,dixmierAlgebras1977,kirchberg_operations_1995}, as well as in application to other physically interesting limits like the classical $\hbar\to 0$ limit involved in strict deformation quantization \parencite{Rieffel1993,landsmanMathematicalTopicsClassical1998}.  It is a short step to see the maps $\underline{A}$ in a scaling algebra as \emph{sections} of a bundle whose fibers consist in distinct algebraic quantum field theories, each carrying their own nets of local algebras and dynamical automorphisms.  We hope to show that this reframing of the central concept of a scaling algebra leads to interesting further questions and useful tools for the analysis of scaling limits.

Seeing elements of the scaling algebra as sections of a bundle raises the question of whether one can form \emph{continuous} bundles from any algebras of observables appearing in quantum field theory.  Indeed, while in a scaling algebra the map
\begin{align}
\label{eqn:section norm}
    \lambda\mapsto ||\underline{A}_\lambda||
\end{align}
is merely required to be bounded, the literature on bundles of C*-algebras \parencite{rieffelContinuousFieldsAlgebras1989,kirchbergOperationsContinuousBundles1995,steegerExtensionsBundlesAlgebras2021} distinguishes among sections for which Eq.~\eqref{eqn:section norm} is lower semicontinuous, upper semicontinuous, continuous, or uniformly continuous.  Different continuity properties of this sort (in a neighborhood of $\lambda = 0$) generally correspond to different strengths of the approximation one makes by taking the $\lambda\to 0$ limit.  In particular, \textcite{steegerExtensionsBundlesAlgebras2021} show that uniform continuity of the map in Eq.~\eqref{eqn:section norm} plays a crucial role in guaranteeing the existence of a unique C*-algebra that can be associated as the observables of the $\lambda = 0$ limit theory, which also carries canonical automorphisms determined through the $\lambda\to 0$ limit of the dynamics.  This motivates our investigation in the current paper of the continuity properties of the sections that arise in scaling limits of algebraic quantum field theories.

We shall find that the continuity properties of sections of rescaled observables in quantum field theories vary significantly among different classes of observables.  We provide examples of local observables for the real scalar field for which the map in Eq.~\eqref{eqn:section norm} is lower semicontinuous, but \emph{not} upper semicontinuous.  On the other hand, we construct an algebra of nonlocal observables for the real scalar field for which the map in Eq.~\eqref{eqn:section norm} is uniformly continuous.  By translating the existing scaling algebra framework into the language of bundles of C*-algebras, we hope to provide tools for further investigation of the approximations that arise in scaling limits.  We demonstrate how the conditions of lower semicontinuity and uniform continuity have distinct implications for the existence and uniqueness of the scaling limit.  And we point out what we believe are some conceptual advantages of the framework of bundles of C*-algebras developed here, in particular for analyzing scaling limits in a way that relies only on the abstract C*-algebraic structure of the observables, unlike the scaling algebra approach which proceeds through concrete representations on vacuum Hilbert spaces.

The paper proceeds as follows.  We provide background on the existing scaling algebra approach and develop a general theory of scaling limits via bundles of C*-algebras in \S\ref{sec:general scaling limits}.  This discussion proceeds abstractly before focusing on a concrete quantum field theory for the remainder of the paper.  In \S\ref{sec:scalar field background} we introduce an example of a simple algebraic quantum field theory: the real scalar field.  We construct several distinct nets of C*-algebras of local observables for the real scalar field, consisting of Weyl unitaries and certain weak operator integrals thereof.  In \S\ref{sec:lower semicontinuous scalar field}, we establish that in each of these nets, rescaling local observables leads to a construction of sections of a lower semicontinuous bundle of C*-algebras, which thus forms what \S\ref{sec:general scaling limits} calls a \emph{scaling bundle}.  In \S\ref{sec:uniformly continuous scalar field}, we show that the continuity properties of the bundle can be improved at the cost of sacrificing the structure of a local net.  We define an algebra of nonlocal observables obtained through quantization of phase space functions with anti-Wick ordering, and we show that rescaling these nonlocal observables leads to a construction of sections of a uniformly continuous bundle of C*-algebras representing the scaling limit.  By establishing the different continuity properties of the different algebras of rescaled sections, we hope to motivate moving beyond the framework of general scaling algebras and toward a focus on particular section algebras of rescaled observables.

\section{General Theory of Scaling Limits}
\label{sec:general scaling limits}

We begin by developing a general theory of scaling limits via bundles of C*-algebras and comparing our methods with existing work.

\subsection{Buchholz-Verch Scaling Algebras}
\label{sec:buchholz and verch scaling algebras}

In this section, we present the existing framework developed by \textcite{buchholzScalingAlgebrasRenormalization1995,buchholzScalingAlgebrasRenormalization1998}.  We begin with background on algebraic quantum field theory; see also \parencite{haagAlgebraicApproachQuantum1964,haagLocalQuantumPhysics1996,arakiMathematicalTheoryQuantum1999,fewsterAlgebraicQuantumField2019,halvorsonAlgebraicQuantumField2007}.  The objects of algebraic quantum field theory are Poincar\'{e} nets satisfying a collection of axioms.  We will be primarily concerned with the Poincar\'{e} net structure and leave discussion of further axioms to future work.  In what follows, we take local regions of Minkowski spacetime to be double cones.

\begin{defn}\label{defn:Poincare net}
    Let $M$ be a Minkowski spacetime and let $\mathfrak{A}: \mathcal{O}\mapsto\mathfrak{A}(\mathcal{O})$ be an assignment of a C*-algebra to each local region.  We call $\mathfrak{A}(\mathcal{O})$ the algebra of local observables in $\mathcal{O}$.  The assignment satisfies \textbf{isotony} if whenever $\mathcal{O}_1\subseteq\mathcal{O}_2$,
    \begin{align}
        \label{req:isotony}
        \mathfrak{A}(\mathcal{O}_1)\subseteq\mathfrak{A}(\mathcal{O}_2).
    \end{align}
    The inductive limit of this net is also labeled $\mathfrak{A}$ and called the quasiolocal algebra.  Further, let $\alpha: (\Lambda,x)\mapsto \alpha_{\Lambda,x}$ be a representation of the identity component of the Poincar\'{e} group by *-automorphisms $\alpha_{\Lambda,x}$ of the quasilocal algebra $\mathfrak{A}$.  The pair $(\mathfrak{A},\alpha)$ is called a \textbf{Poincar\'{e} (covariant) net} if it satisfies isotony and for every local region $\mathcal{O}$,
    \begin{align}
        \alpha_{\Lambda,x}\left(\mathfrak{A}(\mathcal{O})\right) = \mathfrak{A}\left(\Lambda\mathcal{O} + x\right).\label{req:compatibility of alpha and net}
    \end{align}
\end{defn}

\begin{defn}\label{defn:net isomorphism}
        A \textbf{net homomorphism} between Poincar\'e nets $(\mathfrak A, \alpha)$ and $(\mathfrak B, \beta)$ over the same Minkowski spacetime is a *-homomorphism $\psi: \mathfrak A \to \mathfrak B$ between the quasilocal algebras such that for every local region $\mathcal O$,
    	\begin{align} 
        \psi \big( \mathfrak A(\mathcal O) \big) \subseteq \mathfrak B(\mathcal O)
        \label{req:net homo preserves net}
        \end{align}
and for every Poincar\'e transformation $(\Lambda,x)$,
	   \begin{align} 
       \alpha_{\Lambda,x} \circ \psi = \psi \circ \beta_{\Lambda, x}.
       \label{req:net homo preserves poincare transformations}
       \end{align}
A net homomorphism $\psi$ is a \textbf{net epimorphism} if $\psi(\mathfrak{A}(\mathcal{O})) = \mathfrak{B}(\mathcal{O})$ for every double cone. A \textbf{net isomorphism} is a net epimorphism that is also a *-isomorphism. 
\end{defn}

With this background in place, we turn to the framework developed by \textcite{buchholzScalingAlgebrasRenormalization1995,buchholzScalingAlgebrasRenormalization1998} for the analysis of scaling limits of Poincar\'{e} nets.  To begin, assume we are given a Poincar\'{e} net $(\mathfrak{A},\alpha)$.  \textcite{buchholzScalingAlgebrasRenormalization1995} represent the families of rescaled observables by maps $\underline{A}: (0,1]\to\mathfrak{A}$ denoted $\lambda\mapsto \underline{A}_\lambda$.  The idea is that each $\lambda\in (0,1]$ determines a scaled net $(\mathfrak{A}_\lambda,\alpha^\lambda)$ defined by
\begin{align}
    \mathfrak{A}_\lambda(\mathcal{O}) = \mathfrak{A}(\lambda \mathcal{O}) && \alpha^\lambda_{\Lambda,x} = \alpha_{\Lambda,\lambda x}.
\end{align}
On this conception, one can lift the automorphisms $\alpha^\lambda_{\Lambda,x}$ to maps on the families of rescaled observables $\underline{\alpha}_{\Lambda,x}: \prod_{\lambda\in (0,1]}\mathfrak{A}_\lambda\to \prod_{\lambda\in (0,1]}\mathfrak{A}_\lambda$ defined by
\begin{align}
        \big(\underline{\alpha}_{\Lambda,x} (\underline A) \big)_\lambda = \alpha^\lambda_{\Lambda, x}(\underline A_\lambda)
\end{align}
for all $\underline{A}: (0,1]\to \mathfrak{A}$ with $\underline{A}_\lambda\in\mathfrak{A}_\lambda$.  For \textcite{buchholzScalingAlgebrasRenormalization1995}, the \textbf{scaling algebra} $\underline{\mathfrak{A}}$ is the collection of all maps $\underline{A}: (0,1]\to \mathfrak{A}$ satisfying the conditions:
\begin{enumerate}
    \item for all $\lambda\in (0,1]$, if $\underline{A}_1\in \mathfrak{A}(\mathcal{O})$, then $\underline{A}_\lambda\in \mathfrak{A}(\lambda\mathcal{O})$;
    \item the function $\lambda\mapsto ||\underline{A}_\lambda||$ is bounded; and
    \item $||\underline{\alpha}_{\Lambda,x}(\underline{A}) - \underline{A}||\to 0$ as $(\Lambda,x)\to (1,0)$.
\end{enumerate}
The scaling algebra $\underline{\mathfrak{A}}$ is a C*-algebra with pointwise algebraic operations and norm
\begin{align}
    ||\underline{A}|| = \sup_{\lambda\in (0,1]}||\underline{A}_\lambda||.
\end{align}
The scaling algebra also has the structure of a Poincar\'{e} net $(\underline{\mathfrak{A}},\underline{\alpha})$ with $\underline{\alpha}$ defined as above and
\begin{align}
    \underline{\mathfrak{A}}(\mathcal{O}) = \{\underline{A}\in \underline{\mathfrak{A}}\ |\ \underline{A}_1\in\mathfrak{A}_1(\mathcal{O})\}.
\end{align}
Notice that the definition of the scaled net $\mathfrak{A}_\lambda$ implies that for any $\underline{A}\in\underline{\mathfrak{A}}$, $\underline{A}\in\underline{\mathfrak{A}}(\mathcal{O})$ if and only if $\underline{A}_\lambda\in \mathfrak{A}_\lambda(\mathcal{O})$ for all $\lambda\in (0,1]$.  In this sense, given an observable $A\in\mathfrak{A}(\mathcal{O})$, the algebra $\underline{\mathfrak{A}}(\mathcal{O})$ describes all the possible scaling behaviors (for different scaling maps $\underline{A}\in\underline{\mathfrak{A}}$ with $\underline{A}_1 = A$) of the observable $A$.

\textcite{buchholzScalingAlgebrasRenormalization1995} then use the scaling algebra $\underline{\mathfrak{A}}$ to construct the \emph{scaling limit} as $\lambda\to 0$.  The construction of the scaling limit begins with the choice of a state $\omega$ on $\mathfrak{A}$ that is locally normal to a given underlying vacuum representation.  For each $\lambda\in (0,1]$, the state $\omega$ can be lifted to a state $\underline{\omega}_\lambda$ on the scaling algebra $\underline{\mathfrak{A}}$ defined by
\begin{align}
    \underline{\omega}_\lambda(\underline{A}) = \omega(\underline{A}_\lambda)
\end{align}
for all $\underline{A}\in\underline{\mathfrak{A}}$.  By the Banach-Alaoglu theorem \parencite[][p. 115]{reedMethodsModernMathematical1972}, the family of states $\{\underline{\omega}_\lambda\}_{\lambda\in (0,1]}$ has subnets which converge in the weak* topology as $\lambda\to 0$.  Let $\underline{\omega}_{0,\iota}$ be one such limit and denote its GNS representation by $\underline{\pi}_{0,\iota}$.  The \textbf{scaling limit} of the net $(\mathfrak{A},\alpha)$ with respect to the choice $\underline{\omega}_{0,\iota}$ of limit state is defined as the Poincar\'{e} net
\begin{align}
    \mathfrak{A}_{0,\iota}(\mathcal{O}) :=\underline{\pi}_{0,\iota}(\underline{\mathfrak{A}}(\mathcal{O})) && \alpha_{\Lambda,x}^{0,\iota}(\underline{\pi}_{0,\iota}(\underline{A}):= \underline{\pi}_{0,\iota}(\underline{\alpha}_{\Lambda,x}(\underline{A}))
\end{align}
for all double cones $\mathcal{O}$ and $\underline{A}\in\underline{\mathfrak{A}}$.  While this scaling limit does depend on the choice of limit state $\underline{\omega}_{0,\iota}$, \textcite[][p. 24, Prop. 4.4]{buchholzScalingAlgebrasRenormalization1995} show that the collection of all scaling limits is independent of the original choice of state $\omega$ as long as $\omega$ is constrained to be locally normal to the chosen vacuum representation.  They also prove a classification theorem of the collection of scaling limits of a given net.

In this framework, \textcite{buchholzScalingAlgebrasRenormalization1998} show that every scaling limit of the Poincar\'{e} net of the free scalar field of mass $m$ in 3 or 4 spacetime dimensions is net-isomorphic to the net of the massless free scalar field, as expected.  One way to see this is to interpret Buchholz and Verch's scaling maps as encoding spacetime transformations such that $\hbar$ and $c$ are held fixed, so that change of mass is reciprocal to a change in spatiotemporal quantities.  In other words, if we have a transformation $x\mapsto \lambda^{-1}x$, then it is equivalent to $m\mapsto \lambda m$ and in the limit $\lambda\to 0$ the value of the effective mass $\lambda m$ also approaches $0$.  The framework of scaling algebras makes this heuristic argument precise, and we shall return to a modified form of this argument below.

The scaling algebra approach is powerful and has much to recommend.  It is general and intrinsic in the sense that once one is given a Poincar\'{e} net $(\mathfrak{A},\alpha)$, its collection of scaling limits is completely determined and can be constructed without the need to add further data.  Moreover, scaling algebras employ a very broad notion of a scaling map $\underline{A}: (0,1]\to \mathfrak{A}$ making seemingly very few assumptions in order to construct the collection of scaling limits as well as a classification of the collection of scaling limits.  And the notion of a scaling limit that comes from the scaling algebra framework agrees with our expectations in simple models like the free scalar field.  Finally, we note that the use of scaling algebras has given rise to a productive research program for analyzing small-distance behavior, described in further work \parencite{buchholzQuarksGluonsColour1996,morsellaStructureChargesUltraviolet2002,dantoniScalingAlgebrasCharged2004,dantoniScalingAlgebrasSuperselection2006,bostelmannScalingAlgebrasPointlike2009,bostelmannDilationSymmetriesArising2010,contiScalingLimitSubsystems2009,contiAsymptoticMorphismsSuperselection2014,contiAsymptoticMorphismsSuperselection2020}.

On the other hand, we believe that the scaling algebra approach of \textcite{buchholzScalingAlgebrasRenormalization1995,buchholzScalingAlgebrasRenormalization1998} has a few conceptual drawbacks, which will motivate our exploration of a modification to the framework in the remainder of this paper.  First, the definition of the scaling limit does not guarantee the limit is unique.  Of course, \textcite{buchholzScalingAlgebrasRenormalization1995} see non-uniqueness of the scaling limit as a feature of their framework, which they argue allows different scaling limits to capture different scaling behavior inherent in the original net.  Ultimately, we believe a useful way to understand the distinct scaling limits in the scaling algebra framework is by thinking of them as distinct representations of a \emph{single, unique} underlying net that is more properly thought of as the scaling limit theory.  This is only a slight change of perspective, and one that we think agrees with the original discussion \textcite{buchholzScalingAlgebrasRenormalization1995} give of non-unique scaling limits.  Nevertheless, we believe that searching for conditions that guarantees uniqueness of the scaling limit net itself, and leaves non-uniqueness only at the level of representations of the net, can be illuminating.

Second, the approach of \textcite{buchholzScalingAlgebrasRenormalization1995} to defining scaling limits relies on choosing GNS representations of only those states that are locally normal to a given vacuum representation.  Their work suggests the open question of whether one could construct scaling limits in a representation-independent way.  Of course, the Hilbert space representations of a scaling limit are still expected to encode physically interesting content, but seeing these as representations of a single object (per our remarks on uniqueness) may shed some light on the structure of the scaling limit theory.

Finally, the definition of a scaling limit from \textcite{buchholzScalingAlgebrasRenormalization1995} leaves opaque how local operators in the scaling limit net at $\lambda = 0$ approximate local observables of the rescaled nets for $\lambda>0$.  A definition of scaling limits that clarifies the notion of convergence may aid the physical understanding of the role of a scaling limit net in providing a structural approximation to the underlying net.  This gives motivation to explore whether additional continuity conditions might be added---beyond the condition that $\lambda\mapsto ||\underline{A}_\lambda||$ is bounded---in order clarify the approximation involved in the scaling limit.  In the next section, we introduce a mathematical structure that allows us to consider such further conditions on scaling limits.

\subsection{Bundles of C*-algebras} \label{sec:bundles of c* algebras}

In this section, we introduce the notion of a bundle of C*-algebras, which we will later use to supplement the conditions of a scaling algebra.  Bundles of C*-algebras arise in the study of the classical $\hbar\to 0$ limit of quantum theories through deformation quantization \parencite{rieffelQuestionsQuantization1998}, which is known to give rise to continuous bundles of C*-algebras \parencite{dixmierAlgebras1977, rieffelContinuousFieldsAlgebras1989, kirchberg_operations_1995}.  In this context, the full quantum theory is represented partially by a C*-algebra indexed by the value $\hbar =1$ and the classical theory by a C*-algebra indexed by $\hbar = 0$. ‘Quantization maps’ are defined so that they trace out sections of the bundle, i.e., for each $\hbar \in (0,1]$ there is a quantization map $Q_\hbar$ that takes an element $A$ of the classical algebra and outputs the element $Q_h(A)$ in the quantum algebra for $\hbar$ \parencite{landsmanMathematicalTopicsClassical1998, landsmanFoundationsQuantumTheory2017}. This approach to quantization and the classical limit is often called \textit{strict deformation quantization} in contrast to the more common \textit{formal deformation quantization} which treats $\hbar$ as a formal parameter and quantum mechanics with formal power series in $\hbar$ \parencite{bayenDeformationTheoryQuantization1978,bayenDeformationTheoryQuantization1978b,fedosovSimpleGeometricalConstruction1994,kontsevichDeformationQuantizationPoisson2003}.  The maps $\hbar\mapsto Q_\hbar(A)$ generate a C*-algebra of functions over the space $(0,1]$, and this algebra is the centerpiece of a bundle of C*-algebras \parencite{steegerExtensionsBundlesAlgebras2021}.

We employ the following definition of a bundle of C*-algebras.  In what follows, we use $UC_b(I)$ to denote the uniformly continuous and bounded complex-valued functions on a locally compact metric space $I$.

\begin{defn} \label{defn:N-bundle of algebras} Let $I$ be a locally compact metric space, and let $\mathcal{N}$ be a set of bounded functions on $I$. An \textbf{$\mathcal{N}$-bundle of C*-algebras} over $I$ consists of a C*-algebra $\underline{\mathfrak A}$, a collection of C*-algebras $\{\mathfrak A_\lambda \}_{\lambda \in I},$ and surjective *-homomorphisms $\phi_\lambda: \underline{\mathfrak A} \rightarrow \mathfrak A_\lambda$ for each $\lambda \in I,$ such that:
\begin{enumerate}
    \item  \label{req:norm is sup norm}
    writing $\|\cdot\|_{\lambda}$ for the norm in $\mathfrak A_{\lambda},$ the norm of any $\underline A \in \underline{\mathfrak A}$ is given by
        \begin{align}
        \|\underline A\|=\sup _{\lambda \in I}\left\|\phi_\lambda(\underline A)\right\|_{\lambda},
        \end{align}
    \item \label{req:norm is ucb}
    the function $\lambda \mapsto\left\|\phi_\lambda(\underline A)\right\|_{\lambda}$ is in $\mathcal{N}$  for each $\underline A \in \underline{\mathfrak A}$, and
    \item \label{req:closed under multiplication by ucb functions}
    for any $f \in UC_b(I)$ and $\underline A \in \underline{\mathfrak A}$, there is an element $f \underline A \in \underline{\mathfrak A}$ such that for each $\lambda \in I$, 
        \begin{align}
        \phi_\lambda(f \underline A)=f(\lambda) \phi_\lambda(\underline A) .
        \end{align}
\end{enumerate}
An element of the algebra $\underline{\mathfrak A}$ is known as a \textbf{($\mathcal{N}$-)section} of the bundle, and the algebras $\{\mathfrak{A}_\lambda\}_{\lambda\in I}$ are known as the \textbf{fibers}, while the maps $\phi_\lambda$ are known as \textbf{evaluation maps}.
\end{defn}

\noindent For simplicity of notation, we will denote an $\mathcal{N}$-bundle simply by $(\underline{\mathfrak A}, \phi, I)$ where it is understood that the base space is $I$, the corresponding fiber algebras are $\{ \mathfrak A_\lambda \}$ for every $\lambda \in I$, and the evaluation maps are surjective *-homomorphisms are $\phi_\lambda : \underline{\mathfrak A} \to \mathfrak A_\lambda$.

We establish some special terminology for particular choices of $\mathcal{N}$, which we think of as conditions on the strength of continuity of sections of the bundle.  If $\mathcal{N} = UC_b(I)$, then we call a $UC_b(I)$-bundle a \textbf{uniformly continuous bundle of C*-algebras}.  If $\mathcal{N} = LSC_b(I)$ is the set of lower semicontinuous and bounded functions on $I$, then we call a $LSC_b(I)$-bundle a \textbf{lower semicontinuous bundle of C*-algebras}.  We fix our sections to possess an action of $UC_b(I)$ by left multiplication specifically in condition \ref{req:closed under multiplication by ucb functions}, regardless of the continuity conditions on sections.  As we will see, this is appropriate for the problem of extending a bundle of C*-algebras to a larger base space since every uniformly continuous and bounded function on $I$ has a continuous extension to boundary points of $I$.

Whenever the base space $I$ is compact, the definition of a uniformly continuous bundle agrees with the more well known definition of a continuous bundle of C*-algebras.  In the context of strict quantization, one often takes values $\hbar\in [0,1]$ for a known classical theory at $\hbar = 0$, and so one may set $I = [0,1]$ to be compact.  However, for our purposes of analyzing scaling limits, we begin only with data for values $\lambda\in (0,1]$ so $I=(0,1]$ will be understood as a non-compact base space.  This is analogous to the setting of the classical $\hbar\to 0$ limit if one begins only with data for values $\hbar\in (0,1]$ in order to investigate, for example, the existence and uniqueness of the classical limit at $\hbar=0$.  \textcite{steegerExtensionsBundlesAlgebras2021, steegerClassicalLimitSingular2021} undertake such an investigation of the classical $\hbar\to 0$ limit through the use of uniformly continuous bundles of C*-algebras over $(0,1]$.  We will follow suit here for our investigation of the scaling limit of quantum field theories, but we must grapple with the fact that for rescalings of local observables in the infinite-dimensional setting of quantum field theory, uniform continuity of the maps $\lambda\mapsto ||\underline{A}_\lambda||$ sometimes fails.  Later on, we will encounter both uniformly continuous bundles of C*-algebras of nonlocal observables and merely lower semicontinuous bundles of C*-algebras for local observables.  Thus, we will require not only existence and uniqueness results for uniformly continuous bundles as provided by \textcite{steegerExtensionsBundlesAlgebras2021}, but also a generalization of their analysis to lower semicontinuous bundles of C*-algebras.

In order to consider extensions of bundles and their uniqueness, we will use the following notion of a structure-preserving map between bundles.  In what follows, when $I$ and $J$ are metric spaces, a metric map is a function $s: I \to J$ satisfying $d_I(s(x),s(y)) \leq d_J(x,y)$ for all $x,y \in I$.

\begin{defn} \label{defn:bundle homomorphism}
    A \textbf{bundle homomorphism} between $\mathcal{N}$-bundles of C*-algebras $(\underline{\mathfrak A}, \phi, I)$ and $(\underline{\mathfrak B}, \chi, J)$ is a pair of maps $(s, \underline \sigma)$ such that $s : I \to J$ is a metric map and $\underline \sigma : \underline{\mathfrak A} \to \underline{\mathfrak B}$ is a *-homomorphism such that for every $\underline A, \underline A' \in \underline{\mathfrak A}$ and $\lambda \in I$, $\underline{\sigma}(\ker \phi_\lambda)\subseteq \ker \chi_{s(\lambda)}$. A bundle homomorphism $(s,\underline \sigma)$ is an \textbf{isomorphism} if $s$ is an isometry and $\underline \sigma$ is a *-isomorphism. 
\end{defn}

\begin{defn}
    A bundle homomorphism $(s, \underline \sigma)$ is said to be an \textbf{extension} if both $s$ and $\underline \sigma$ are injective and a \textbf{minimal extension} if $s$ and $\underline \sigma$ are both dense embeddings.
\end{defn}

\textcite{steegerExtensionsBundlesAlgebras2021} show that the details of a uniformly continuous bundle of C*-algebras on a dense subset of the base space are sufficient to determine the entire bundle of C*-algebras, in the following sense.

\begin{thm} \label{thm:steeger and feintzeig extension theorem}
Let $I$ be a locally compact metric space and $J$ be an open dense subset of $I$. Every uniformly continuous bundle of C*-algebras over $J$ has a unique (up to isomorphism) minimal extension to a uniformly continuous bundle over $I$. \parencite[Thm. 4.1]{steegerExtensionsBundlesAlgebras2021}
\end{thm}

\noindent In the context of quantization, \textcite{steegerClassicalLimitSingular2021} argue that this shows that information about the scaling behavior of observables in a quantum theory with varying values of $\hbar$ determines a unique classical theory. The scaling behavior of observables with respect to $\hbar$ is encoded in a uniformly continuous bundle of C*-algebras over $(0,1]$ and the resulting extension to a bundle over $[0,1]$ gives a theory associated with $\hbar = 0$. It can then be shown that this algebra is abelian and can be interpreted as a classical limit of the original quantum theory \parencite{steegerClassicalLimitSingular2021}.

In fact, not only is the existence and uniqueness of such an extension guaranteed, but the assignment which sends a bundle $(\underline{\mathfrak{A}},\phi,I)$ to the fiber algebra $\mathfrak{A}_\lambda$ in the extension at any point $\lambda$ in the closure of $I$ is functorial in the following sense.

\begin{prop}\parencite[Prop. 5.2]{steegerExtensionsBundlesAlgebras2021} \label{prop:steeger and feintzeig functoriality of extension}
Given a uniformly continuous bundle $(\underline{\mathfrak A}, \phi)$ over $(0,1]$, we can assign to every bundle automorphism $(\operatorname{id}, \underline{\alpha})$ from $\underline{\mathfrak A}$ to itself a *-automorphism $\alpha_0$ from the C*-algebra $\mathfrak A_0$ to itself which extends $\underline{\alpha}$ in the sense that
\begin{align}
    \phi_0\circ\underline{\alpha} = \alpha_0\circ\phi_0.
\end{align}
Moreover, the extension $\underline{\alpha}\mapsto \alpha_0$ is functorial in the sense that for any bundle automorphisms $\underline (\operatorname{id},\alpha)$ and $(\operatorname{id},\underline \beta)$ of $(\underline{\mathfrak{A}},\phi)$,
    \begin{align}
    \underline \alpha \circ \underline \beta \mapsto \alpha_0 \circ \beta_0.
    \end{align}
\end{prop}

\noindent In the context of quantization, this implies that the dynamical automorphisms of the quantum theories determine corresponding dynamical automorphisms of the classical theory \parencite[\S 5.2]{steegerExtensionsBundlesAlgebras2021}. 

We will see later on that when analyzing scaling limits of local algebras in quantum field theories, we often lack a uniformly continuous bundle of C*-algebras and instead have a merely lower semicontinuous bundle of C*-algebras.  Still, one can establish the following existence and uniqueness results following the strategy outlined in \ref{thm:steeger and feintzeig extension theorem}.  To prepare, note that our notion of a minimal extension guarantees that the C*-algebra of sections of the extended bundle on $I$ is *-isomorphic to the C*-algebra of sections of the original bundle on $J$.  The extension of a uniformly continuous bundle $(\underline{\mathfrak{A}},\phi,(0,1])$ in Theorem \ref{thm:steeger and feintzeig extension theorem} is accomplished then by setting $\mathfrak{A}_0 = \mathfrak{A}/\mathcal{I}_0$ for the closed two-sided ideal $\mathcal{I}_0 = \{\underline{A}\in\underline{\mathfrak{A}}\ |\ \lim_{\lambda\to 0}||\underline{A}_\lambda||_\lambda = 0\}$, with $\phi_0: \mathfrak{A}\to \mathfrak{A}_0$ defined as the quotient map.  In fact, upper semicontinuity of the extension already forces $\mathcal{I}_0 = \{f\underline{A}\in\underline{\mathfrak{A}}\ |\ \lim_{\lambda\to 0} f(\lambda) = 0\}$ and so uniquely picks out the ideal for the quotient construction \textcite[][Prop. 1.2-1.3]{rieffelContinuousFieldsAlgebras1989}.  However, when the extension is not required to be upper semicontinuous but merely lower semicontinuous, one generally has many possible choices for the ideal $\mathcal{I}_0$.  We show next that even when one begins with a merely lower semicontinuous bundle of C*-algebras, it is always possible to take the smallest possible ideal to construct an extension.

\begin{thm}
\label{thm:lower semicontinuous extension}
    Suppose $(\underline{\mathfrak{A}},\tilde{\phi},(0,1])$ is a lower semicontinuous bundle of C*-algebras.  Then there is a minimal extension $(\operatorname{id},\operatorname{id})$ to a lower semicontinuous bundle $(\underline{\mathfrak{A}},\phi,[0,1])$.  Moreover, for any other minimal extension $(\operatorname{id},\operatorname{id})$ to a lower semicontinuous bundle $(\underline{\mathfrak{A}},\psi,[0,1])$, there is a unique surjective *-homomorphism $\alpha: \phi_0(\underline{\mathfrak{A}})\to \psi_0(\underline{\mathfrak{A}})$ satisfying $\alpha\circ \phi_0 = \psi_0$.   
\end{thm}

Note again that considering minimal extensions to have the same C*-algebra of sections $\underline{\mathfrak{A}}$ as the original bundle---so that the bundle homomorphism has the form $(\operatorname{id},\operatorname{id})$---is no restriction, since the definition of a minimal extension guarantees the C*-algebras of sections will be *-isomorphic.  The only object that may differ across different extensions is the fiber at the boundary point $0\in [0,1]$ and the corresponding evaluation map at $\lambda = 0$.

\begin{proof}
Let $\mathcal{P}_0$ be the collection of all C*-seminorms on $\underline{\mathfrak{A}}$ such that
\begin{align}
    p(\underline{A})\leq \liminf_{\lambda\to 0}||\underline{A}_\lambda||, && p(f\underline{A}) = \lim_{\lambda\to 0} \left|f(x)\right| p(\underline{A})
\end{align}
for all $f\in UC_b\left((0,1]\right)$.  Then define a closed two-sided ideal and corresponding quotient:
\begin{align}
  \mathcal{J}_0 &:= \bigcap_{p\in\mathcal{P}_0} \ker p = \{ \underline{A}\in \underline{\mathfrak{A}}\ |\ p(\underline{A}) = 0\text{ for all } p\in\mathcal{P}_0\}\\
  \mathfrak{A}_0 &:= \underline{\mathfrak{A}}/\mathcal{J}_0.
\end{align}
Define $\phi_\lambda = \tilde{\phi}_\lambda$ for all $\lambda\in (0,1]$ and let $\phi_0:\underline{\mathfrak{A}}\to \underline{\mathfrak{A}}/\mathcal{J}_0$ be the quotient map.  We have $\mathcal{J}_0 = \ker p_0$ for the C*-seminorm $p_0$ on $\underline{\mathfrak{A}}$ defined by
\begin{align}
    p_0(\underline{A}) = \sup_{p\in\mathcal{P}_0}p(\underline{A}).
\end{align}
Clearly, $p_0(\underline{A}) = ||\phi_0(\underline{A})||_0$, where the right hand side is the norm in $\mathfrak{A}_0$, because $p_0$ defines another C*-norm on $\mathfrak{A}_0$ and any injective homomorphism between C*-algebras is isometric; see \textcite[][\S1.8]{dixmierAlgebras1977} for relevant discussion of quotients of C*-algebras.

Hence, it follows that for all $\underline{A}\in \underline{\mathfrak{A}}$,
\begin{align}
    ||\phi_0(\underline{A})||_0 = p_0(\underline{A}) \leq \liminf_{\lambda\to 0}||\underline{A}_\lambda||_\lambda.
\end{align}
Therefore, the bundle $(\underline{\mathfrak{A}},\phi,[0,1])$ is lower semicontinuous.  

Now, suppose $(\underline{\mathfrak{A}},\psi,[0,1])$ is another lower semicontinuous bundle extending $(\underline{\mathfrak{A}},\tilde{\phi},(0,1])$.  Then for any $\underline{A}\in\underline{\mathfrak{A}}$, $||\psi_0(\underline{A})||_0\leq p_0(\underline{A})$ so that $\mathcal{J}_0 = \ker \phi_0 \subseteq \ker \psi_0$.  So define $\alpha: \phi_0(\underline{\mathfrak{A}})\to \psi_0(\underline{\mathfrak{A}})$ by
\begin{align}
\alpha(\phi_0(\underline{A})) = \psi_0(\underline{A}).
\end{align}
This is well-defined because $\phi_0(\underline{A}) = \phi_0(\underline{B})$ implies $\phi_0(\underline{A}-\underline{B})= 0$, and then since $\ker \phi_0 = \mathcal{J}_0$, it follows that $\psi_0(\underline{A}-\underline{B}) = 0$, thus establishing that $\psi_0(\underline{A}) = \psi_0(\underline{B})$.  Moreover, $\alpha$ is clearly a surjective *-homomorphism since $\psi_0$ is surjective.

Suppose $\alpha': \phi_0(\underline{\mathfrak{A}})\to \psi(\underline{\mathfrak{A}})$ is another surjective *-homomorphism satisfying $\alpha'\circ \phi_0 = \psi_0$.  Then since $\phi_0$ is surjective, it follows from $\alpha'(\phi_0(\underline{A})) = \psi_0(\underline{A}) = \alpha(\phi_0(\underline{A}))$ that $\alpha' = \alpha$.
\end{proof}

Theorem \ref{thm:lower semicontinuous extension} will be important for our purposes, since it will guarantee the existence of a C*-algebra for a certain kind of scaling limit of a quantum field theory.  Unlike in the case of a uniformly continuous bundle of C*-algebras, which has a unique limiting fiber at a boundary point in the base space, a lower semicontinuous bundle of C*-algebras generally can be extended by many different limiting fibers at a boundary point.  Thus, we have only shown that there is a \emph{largest} limiting fiber $\mathfrak{A}_0$ with surjective *-homomorphism $\phi_0: \underline{\mathfrak{A}}\to\mathfrak{A}_0$, where our notion of ``largest fiber" is captured by the universal property appearing in Thm \ref{thm:lower semicontinuous extension}:
\begin{enumerate}
    \item[(UP)] for any other minimal extension to a lower semicontinuous bundle $(\underline{\mathfrak{A}},\psi,[0,1])$, there is a unique surjective *-homomorphism $\alpha: \phi_0(\underline{\mathfrak{A}})\to \psi_0(\underline{\mathfrak{A}})$ satisfying $\alpha\circ \phi_0 = \psi_0$.
\end{enumerate}
Fortunately, for lower semicontinuous bundles of C*-algebras the largest limiting fiber is still unique in the following sense.
\begin{cor}
\label{cor:uniqueness of extension of lower semicontinuous bundles}
    Suppose $(\underline{\mathfrak{A}},\tilde{\phi},(0,1])$ is a lower semicontinuous bundle of C*-algebras.  Suppose $(\underline{\mathfrak{A}},\phi,[0,1])$ and $(\underline{\mathfrak{A}},\psi,[0,1])$ are lower semicontinuous bundles that are  minimal extensions of $(\underline{\mathfrak{A}},\tilde{\phi},(0,1])$ satisfying (UP).  Then there is a unique *-isomorphism $\beta: \phi_0(\underline{\mathfrak{A}})\to \pi_0(\underline{\mathfrak{A}})$ such that $\beta\circ \phi_0 = \pi_0$.
\end{cor}
\noindent The corollary follows immediately from the universal property.  It will be essential to our understanding of the uniqueness of scaling limits of quantum field theories.

Moreover, one also has a restricted sense of functoriality for extensions of lower semicontinuous bundles, as follows.
\begin{thm}
\label{thm:functoriality of extensions of lower semicontinuous bundles}
    Let $(\operatorname{id},\underline{\sigma})$ be a bundle homomorphism from $(\underline{\mathfrak{A}},\tilde{\phi},(0,1])$ to itself.  Suppose $(\underline{\mathfrak{A}},\phi,[0,1])$ is a minimal extension to a lower semicontinuous bundle satisfying (UP).  If $\underline{\sigma}(\ker \phi_0) \subseteq \ker \phi_0$, then there is a unique *-homomorphism $\sigma_0: \phi_0(\underline{\mathfrak{A}})\to\phi_0(\underline{\mathfrak{A}})$ satisfying $\sigma_0\circ \phi_0 = \phi_0\circ\underline{\sigma}$.
\end{thm}

\begin{proof}
    Define $\sigma_0: \phi_0(\underline{\mathfrak{A}})\to \phi_0(\underline{\mathfrak{A}})$ by
    \begin{align}
        \sigma_0(\phi_0(\underline{A})) = \phi_0(\underline{\sigma}(\underline{A})).
    \end{align}
    This is well-defined because if $\phi_0(\underline{A}) = \phi_0(\underline{B})$, then $(\underline{A}-\underline{B})\in \ker \phi_0$, which implies $\underline{\sigma}(\underline{A}-\underline{B})\in \ker\phi_0$ and hence $\phi_0(\underline{\sigma}(\underline{A})) = \phi_0(\underline{\sigma}(\underline{B}))$.  Uniqueness again follows because $\phi_0$ is surjective.
\end{proof}
\noindent Theorem \ref{thm:functoriality of extensions of lower semicontinuous bundles} will guarantee that the algebra of the scaling limit also has Poincar\'e automorphisms, turning it into a full-fledged Poincar\'e net. In the next section, we apply the formalism and results of the theory of lower semicontinuous bundles of C*-algebras to the context of scaling in the sense of Buchholz and Verch's scaling algebras.

\subsection{Scaling Bundles} \label{sec:scaling bundles}

In this section, we introduce the notion of an $\mathcal{N}$-scaling bundle. Essentially, an $\mathcal{N}-$scaling bundle is a scaling algebra in the sense of Buchholz and Verch that satisfies the additional requirements of an $\mathcal{N}$-bundle of C*-algebras.

\textcite[15]{buchholzScalingAlgebrasRenormalization1995} acknowledge that “it may sometimes be convenient to impose further constraints on the orbits (of observables under the renormalization group transformations) which amounts to proceeding to subnets.” We follow this approach by adding the further constraints of an $\mathcal{N}$-bundle. The resulting structure is slightly more restrictive than a scaling algebra in the scaling behavior permitted for possible sections, but we believe the loss of generality is compensated by some conceptual advantages. The procedure for taking the scaling limit of a scaling bundle reflects that of taking the classical limit of a uniformly continuous bundle describing quantum kinematics. For our scaling limits, although uniform continuity is replaced by lower semicontinuity, once an algebra of sections is specified, the largest resulting limiting net is unique and does not rely on a choice of physical state. Moreover, the resulting scaling limit also has the added benefit of furnishing state-independent approximations between the scaling limit net and nets for small values of $\lambda$. 

In order to present this scaling bundle framework we first introduce the structure of an \textit{$\mathcal{N}$-bundle of Poincar\'e nets}, or \textit{net $\mathcal{N}$-bundle} for short. Put simply, a net $\mathcal{N}$-bundle is an $\mathcal{N}$-bundle of C*-algebras where each fiber is assumed to have the additional structure of a Poincar\'e net. Moreover these Poincar\'e nets are assumed to be compatible with the bundle structure. Roughly, this means that the Poincar\'e net structure varies appropriately over the base space.

\begin{defn} \label{defn:uniformly continuous bundle of P-nets}
    Let $I$ be a locally compact metric space and $M$ be a Minkowski space. An \textbf{$\mathcal{N}$-bundle of Poincar\'e nets} (or \textbf{net $\mathcal{N}$-bundle} for short) over $I$ is a triple of a Poincar\'e net $\big( \underline{\mathfrak A}, \underline \alpha  \big)$ over $M$, a collection of Poincar\'e nets $\big \{ (\mathfrak A_\lambda, \alpha^\lambda) \big \}_{\lambda \in I}$ over $M$, and surjective net homomorphisms $\phi_\lambda : (\underline{\mathfrak A}, \underline \alpha) \to (\mathfrak A_\lambda, \alpha^\lambda)$ between the inductive limits of the corresponding nets such that $( \underline{\mathfrak A}, \phi )$ is a $\mathcal{N}$-bundle of C*-algebras. 
    For simplicity we will label such a uniformly continuous bundle as $(\underline{\mathfrak A}, \underline{\alpha}, \phi, I)$.
\end{defn}

As before, when $\mathcal{N} = LSC_b(I)$, we call a net $LSC_b(I)$-bundle a lower semicontinuous bundle of Poincar\'e nets, or a lower semicontinuous net bundle for short.

\begin{defn} \label{defn:net bundle homomorphism}
    A \textbf{net bundle homomorphism} between net $\mathcal{N}$-bundles $(\underline{\mathfrak A}, \underline{\alpha}, \phi, I)$ and $(\underline{\mathfrak B}, \underline{\beta}, \chi, J)$ is a bundle homomorphism $(s, \underline \sigma) : \underline{\mathfrak A} \to \underline{\mathfrak B}$ such that $\underline \sigma$ is also a net homomorphism. A net bundle homomorphism is a \textbf{(minimal) net bundle extension} if it is an (minimal) extension as a bundle homomorphism.
\end{defn}


Now that we have all the necessary definitions we can state the existence result for extensions of lower semicontinuous bundles of Poincar\'e nets as follows from Theorem \ref{thm:lower semicontinuous extension}. If one has a net bundle over $(0,1]$ then there is a unique largest net bundle over $[0,1]$ that extends the original net bundle. In particular, there is a unique (up to isomorphism) largest net above the base point $\lambda = 0$. 



\begin{thm} \label{thm:extension of net bundle}
Let $(\underline{\mathfrak A}, \underline \alpha, \phi, (0,1])$ be a lower semicontinuous bundle of Poincar\'e nets.  Then the following hold:
\begin{itemize}
    \item There is a C*-algebra $\mathfrak{A}_0$ and surjective *-homomorphism $\phi_0: \underline{\mathfrak{A}}\to \mathfrak{A}_0$ providing a minimal extension of $\underline{\mathfrak{A}}$ to a lower semicontinuous bundle over $[0,1]$ satisfying (UP).
    \item If the Poincar\'e transformations $\underline{\alpha}_{\Lambda,x}$ satisfy $\underline{\alpha}_{\Lambda,x}(\ker \phi_0)\subseteq \ker \phi_0$, then there are unique extensions of the Poincar\'e transformations to *-automorphisms $\alpha^0_{\Lambda,x}$ of $\mathfrak{A}_0$ such that $\alpha^0_{\Lambda,x}\circ \phi_0 = \phi_0\circ \underline{\alpha}_{\Lambda,x}$.  With the Poincar\'e transformations, $(\mathfrak{A}_0,\alpha^0)$ is a Poincar\'e net and the minimal extension of $\underline{\mathfrak{A}}$ so described is a lower semicontinuous net bundle.
    \item Under the conditions just stated, any other minimal extension of $\underline{\mathfrak{A}}$ to a lower semicontinuous bundle satsifying (UP) can be given a net structure from the extension of Poincar\'e transformations, which makes it net bundle isomorphic to $(\mathfrak{A}_0,\alpha^0)$.
\end{itemize}
\end{thm}

\begin{proof}
One easily finds from Theorem \ref{thm:lower semicontinuous extension} that the local algebras have lower semicontinuous bundles with sections $\underline{\mathfrak{A}}(\mathcal{O})$, whose minimal extensions to fiber algebras $\mathfrak{A}_0(\mathcal{O})$ satisfy (UP).  Theorem \ref{thm:functoriality of extensions of lower semicontinuous bundles} guarantees the existence and uniqueness of the Poincar\'e transformations between local algebras $\mathfrak{A}_0(\mathcal{O})$.  Moreover, Corollary \ref{cor:uniqueness of extension of lower semicontinuous bundles} implies that the resulting net bundle $(\mathfrak{A}_0,\alpha^0)$ is unique up to net bundle isomorphism. 
\end{proof}

We next define lower semicontinuous scaling bundles as particular kinds of lower semicontinuous net bundles capturing a small-distance scaling limit.  We present lower semicontinuous scaling bundles as instances of more general objects, called merely scaling bundles.  Our definition of a scaling bundle is a small generalization of the definition of a scaling algebra from \textcite{buchholzScalingAlgebrasRenormalization1995}.
A lower semicontinuous scaling bundle is then a scaling algebra that also has the structure of a lower semicontinuous bundle of C*-algebras in a compatible way.

\begin{defn} \label{defn:scaling bundle}
    Let $I \subseteq \mathbb R_{>0}$. A \textbf{scaling bundle} for the Poincar\'e net $(\mathfrak A, \alpha)$ over $I$, is a *-algebra $\underline{\mathfrak A}$ with a net structure and Poincar\'e automorphisms $\underline{\alpha}$ along with a family of Poincar\'e nets $(\mathfrak A_\lambda, \alpha^\lambda)$ for every $\lambda \in I$ and surjective *-homomorphisms $\phi_\lambda: \underline{\mathfrak A} \to \mathfrak A_\lambda$ for every $\lambda \in I$ that satisfies
    \begin{enumerate}
        \label{req:1 fiber net is underlying net} 
        \item $\mathfrak A_\lambda(\mathcal O) = \mathfrak A(\lambda \mathcal O)$ for every local region $\mathcal O$, and \label{req:fiber nets are scaled nets}
        \item $\alpha^{\lambda}_{\Lambda,x} = \alpha_{\Lambda, \lambda x}$ for every Poincar\'e transformation $(\Lambda ,x)$. \label{req:fiber automorphisms are scaled automorphisms} 
    \end{enumerate}
    A \textbf{lower semicontinuous scaling bundle} is a scaling bundle that is also a lower semicontinuous bundle of Poincar\'e nets. 
\end{defn}

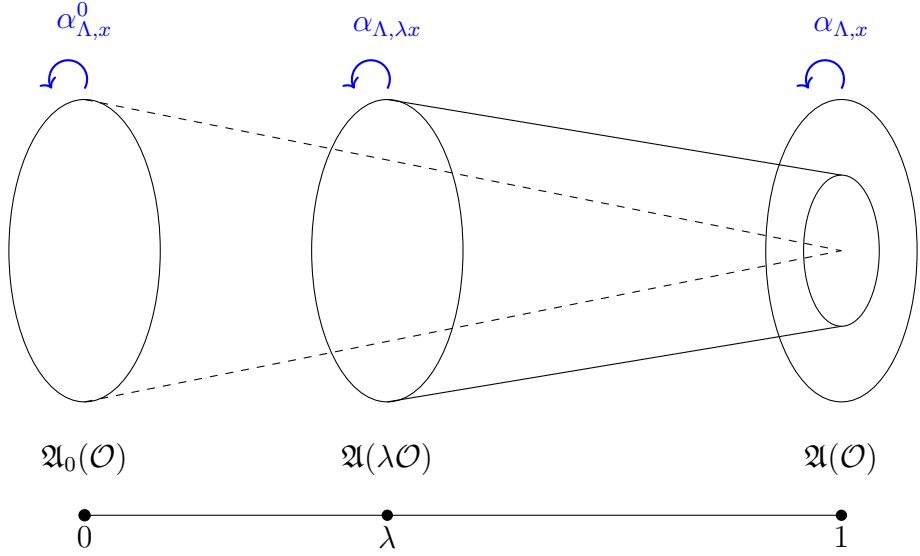
\begin{figure}
    \centering
    \begin{tikzpicture}
\draw (0,-0.5) node[anchor=north] {$0$} -- (4,-0.5) node[anchor=north] {$\lambda$};
\draw (4,-0.5) node[anchor=north] {} -- (10,-0.5) node[anchor=north] {$1$};
\node[circle,draw, fill,inner sep=1.5pt] (p0) at (0,-0.5) {};
\node[circle,fill,inner sep=1.5pt] (plamdba) at (4,-0.5) {};
\node[circle,fill,inner sep=1.5pt] (p1) at (10,-0.5) {};
\node[ellipse, draw, minimum width = 2cm, minimum height = 4cm] (a0) at (0,3) {};
\node[ellipse, draw, minimum width = 2cm, minimum height = 4cm] (alambda) at (4,3) {};
\node[ellipse, draw, fill=white, minimum width = 2cm, minimum height = 4cm] (a1) at (10,3) {};
\node[ellipse, draw, minimum width = 1cm, minimum height = 2cm] (a1) at (10,3) {};
\node[] at (0,0.25) {$\mathfrak A_0(\mathcal O)$};
\node[] at (4,0.25) {$\mathfrak A(\lambda \mathcal O)$};
\node[] at (10,0.25) {$\mathfrak A(\mathcal O)$};
\draw (4,5) -- (10,4);
\draw (4,1) -- (10,2);
\draw[dashed] (0,5) -- (10,3);
\draw[dashed] (0,1) -- (10,3);
\node[above = 2 cm] (alpha1) at (a1) {};
\draw[blue,thick,->] (alpha1) arc (-30:210:.25cm);
\node[text=blue, above = .5cm] at (alpha1) {$\alpha_{\Lambda, x}$};
\node[above = 2 cm] (alphalambda) at (alambda) {};
\draw[blue,thick,->] (alphalambda) arc (-30:210:.25cm);
\node[text=blue, above = .5cm] at (alphalambda) {$\alpha_{\Lambda, \lambda x}$};
\node[above = 2 cm] (alpha0) at (a0) {};
\draw[blue,thick,->] (alpha0) arc (-30:210:.25cm);
\node[text=blue, above = .5cm] at (alpha0) {$\alpha^0_{\Lambda, x}$};
\end{tikzpicture}
    \caption{A scaling bundle for the Poincar\'e net $(\mathfrak A, \alpha)$ over the base space $(0,1]$ and its extension to a net bundle over $[0,1]$}
    \label{fig:scaling bundle}
\end{figure}

Our notion of a general scaling bundle relaxes the requirements of a scaling algebra from \textcite{buchholzScalingAlgebrasRenormalization1995}.  They assume that the algebra of sections is a C*-algebra so that rescaled operators have bounded norm across all values of $\lambda\in (0,1]$ in the sense that the maps $\lambda\mapsto ||\underline{A}_\lambda||$ are required to be bounded.  We leave open the possibility that the algebra of sections $\underline{\mathfrak A}$ is only a *-algebra so that the norms of rescaled operators (sections) may become unbounded as $\lambda\to 0$.  Our motivation is that some quantum field theories are thought to have less well-behaved scaling behavior giving rise to divergences at small distances, for example if a theory lacks UV completeness.  We leave open the possibility that in those cases, one might want a sense in which the quantum field theory lacks a small-distance scaling limit altogether, which is prohibited by the boundedness requirement of a scaling algebra, but allowed in our more general notion of a scaling bundle.  This brings to the fore a point implicit in the work of \textcite{buchholzScalingAlgebrasRenormalization1995}, that the conditions on sections of a bundle are what allow for the analysis of the scaling limit.  The fact that norms of sections must be bounded in a scaling algebra leads to the existence of the scaling limit.  Our point here is that the further continuity  requirement of a lower semicontinuous scaling bundle leads additionally to uniqueness for the largest scaling limit algebras.

So we see different specifications of sections of a scaling bundle as data that may encode different scaling behaviors of a quantum field theory.  Compare this conception to \textcite{buchholzScalingAlgebrasRenormalization1995}, who see distinct subnets of a family of rescaled states as encoding different scaling behaviors.

In some sense, we retain the non-uniqueness of scaling limits in the framework of uniformly continuous scaling bundles because it may be possible to choose different algebras of sections, and so different families of rescaled operators, through the same fiber algebras.  In other words, it is possible to construct distinct bundles of C*-algebras that represent the scaling limit of the same quantum field theory at $\lambda = 1$.  Our point is only that the existence and uniqueness theorems for extensions of bundles guarantee that \emph{once the scaling behavior of observables is specified} through a choice of sections, then the largest scaling limit theory is uniquely determined.  We believe this perspective can be fruitful as it highlights the substantive physical considerations that may go into the choice of an algebra of rescaled operators that form the sections of a bundle.  We will illustrate in \S\ref{sec:lower semicontinuous scalar field}-\ref{sec:uniformly continuous scalar field} how consideration of physical rescaling of particular operators can inform this choice, which we hope will lead to better understanding of scaling limits.

Thus, in our framework of scaling bundles, we initiate the study of \emph{subalgebras} of the scaling algebra which satisfy the conditions of an algebra of sections of a suitably continuous bundle of C*-algebras.  We hope to have given motivation for how such subalgebras of sections can be significant for the analysis of scaling limits.  We intend our investigation of lower semicontinuous and uniformly continuous bundles of C*-algebras to extend and supplement work on scaling algebras rather than compete with the existing framework.

To see this, let us compare our construction of the limiting net with that of \textcite{buchholzScalingAlgebrasRenormalization1995}. Suppose we are given a lower semicontinuous scaling bundle $(\underline{\mathfrak{A}}, \underline{\alpha},\phi,(0,1])$.  Recall that \textcite{buchholzScalingAlgebrasRenormalization1995} recommend considering a state $\omega$ on $\mathfrak{A}_1$ that is locally normal to a vacuum representation, and defining a lifted state $\underline{\omega}_\lambda$ on $\underline{\mathfrak{A}}$ by
\begin{align}
\underline{\omega}_\lambda(\underline{A})= \omega(\underline{A}_\lambda)
\end{align}
for $\underline{A}\in\underline{\mathfrak{A}}$.  Then moving from the net $\{\underline{\omega}_{\lambda}\}_{\lambda\in (0,1]}$ to a weak* convergent subnet $\{\underline{\omega}_{\lambda,\iota}\}_{\lambda\in (0,1]}$, they pick out the scaling limit as the Poincar\'e net $\mathfrak{A}_{0,\iota}(\mathcal{O}) =\underline{\pi}_{0,\iota}(\underline{\mathfrak{A}}(\mathcal{O}))$,
where $\underline{\pi}_{0,\iota}$ denotes the GNS representation of the limit state $\underline{\omega}_{0,\iota}$.  On the other hand, Theorem \ref{thm:extension of net bundle} constructs the limit net as a quotient $\mathfrak{A}_0(\mathcal{O})= \underline{\mathfrak{A}}/\mathcal{J}_0$, where $\mathcal{J}_0$ is the minimal ideal required to make the extension lower semicontinuous.  It follows immediately from the definition of $\mathcal{J}_0$ that $\mathcal{J}_0\subseteq \ker \underline{\pi}_{0,\iota}$ because the map $\underline{A}\mapsto ||\underline{\pi}_{0,\iota}(\underline{A})||$ is a C*-seminorm in $\mathcal{P}_0$.  This implies that adding the fiber $\mathfrak{A}_{0,\iota}(\mathcal{O})$ at the boundary with surjective *-homomorphism $\underline{\pi}_{0,\iota}$ yields a lower semicontinuous bundle of C*-algebras and hence there is unique surjective *-homomorphism $\beta: \mathfrak{A}_0(\mathcal{O})\to\mathfrak{A}_{0,\iota}(\mathcal{O})$ such that $\beta\circ \phi_0 = \underline{\pi}_{0,\iota}$.  In other words, every scaling limit in the sense of \textcite{buchholzScalingAlgebrasRenormalization1995} extends to a representation of the largest fiber algebra at $\lambda = 0$ obtained in the framework of lower semicontinuous bundles.

One may wish to add to the definition of a scaling bundle the further requirement that the representation $\underline{\alpha}$ of the Poincar\'{e} group is strongly continuous, as in the notion of a scaling algebra.  For \textcite{buchholzScalingAlgebrasRenormalization1995}, this condition captures the requirement that the rescaled families of operators do not alter the value $\hbar$ of Planck's constant.  We omit this requirement from our definition only because in our framework the choice of sections of a bundle can ensure that the value of $\hbar$ is held constant as $\lambda$ varies.  We will see this in our construction of a scaling bundles for the massive scalar field in \S\ref{sec:lower semicontinuous scalar field}-\ref{sec:uniformly continuous scalar field}.  In this sense, our analysis of the scaling limit establishes that strong continuity of the translation group is \emph{not necessary} for  making sense of the scaling limit; rather strong continuity of the translation group is merely a special property that holds of some choices of local algebras, but not others.

In the remainder of the paper, we focus on this new notion of a lower semicontinuous scaling bundle.  Unless otherwise stated, for brevity all net bundles and scaling bundles will be assumed to be lower semicontinuous in what follows.  When we encounter uniformly continuous bundles below, they will not carry the structure of a scaling bundle, or even a net bundle, because they will consist in nonlocal observables.

\newcommand{\onemhs}{\mathfrak h^{(m)}}
\newcommand{\ronemhs}{\onemhs_\R}

\newcommand{\mhs}{\mathcal H^{(m)}}
\newcommand{\lmhs}{\mathcal H^{(\lambda m)}}

\newcommand{\mr}{\pi^{(m)}}
\newcommand{\lmr}{\pi^{(\lambda m)}}

\newcommand{\miso}{\gamma^{(m)}}
\newcommand{\ziso}{\gamma^{(0)}}
\newcommand{\liso}{\gamma^{(\lambda m)}}

\section{Background on the Massive Free Scalar Field}
\label{sec:scalar field background}

In the remainder of the paper, we aim to provide a proof of concept for the use of bundles of C*-algebras to represent scaling limits.  We take up the small-distance scaling limit of the real scalar field.  The first task is to define scaling transformations to construct families of observables $\lambda\mapsto \underline{A}_\lambda$.  The second task is to assess the continuity conditions satisfied by the maps $\lambda\mapsto ||\underline{A}_\lambda||_\lambda$.  For the first task of defining relevant rescaled observables, in this section we will define three distinct local C*-algebras of operators for the scalar field.  In \S\ref{sec:lower semicontinuous scalar field}, we will find that all of these algebras lead to the construction of a lower semicontinuous scaling bundle capturing the small-distance limit, but at least two of the local algebras considered do not define a uniformly continuous bundle.

We begin by motivating our consideration of the distinct local algebras we define in this section.  After all, one might wonder: why not just choose a single net of local algebras to work with for the given model?  Indeed, we will first in \S\ref{subsec:weyl algebras} construct the net for the free scalar field through the \emph{Weyl algebra}, which is often treated as a canonical C*-algebra of operators.  On the other hand, some recent work has called into question the use of the Weyl algebra, instead proposing alternatives including the resolvent algebra \parencite{BuGr08,BuGr15,Bu17,Bu18,vN19,vN22,vNSt24} and the so-called dynamical C*-algebra \parencite{BuFr20,BuFr20a,BuFr20b,BrDuFrRe22}.  Moreover, the analysis by \textcite{BuVe98} of the scaling algebra for the real scalar field makes special use of certain weak operator integrals of Weyl unitaries, smeared out across spacetime.  In principle, the scaling behavior of each alternative family of operators in the small-distance scaling limit might satisfy distinct continuity conditions.  We do not have space to treat all the possible alternatives here, so we set aside the resolvent algebra and the dynamical C*-algebra for future work and focus instead on weak operator integrals of Weyl unitaries.  In \S\ref{subsec: spacetime smeared observables}, we define local C*-algebras consisting of spacetime smeared observables.  In \S\ref{subsec:phase space smeared observables}, we define local C*-algebras of phase space smeared observables motivated by Weyl quantization of certain phase space functions.  These last local algebras serve not only as a useful example, but also as a stepping stone to the nonlocal operators that we consider later in \S\ref{sec:uniformly continuous scalar field}.

\subsection{Weyl Algebras for the Real Scalar Field}
\label{subsec:weyl algebras}

We now construct a first Poincar\'{e} net for the mass $m$ free scalar field on $s+1$-dimensional spacetime \parencite{buchholzScalingAlgebrasRenormalization1998}.  Choose some inertial coordinates $(t,\mathbf{y})$ so that $M\cong \R^{s+1}$, while setting $c=1$.  Let $\Sigma\cong \R^s$ be the spatial slice of $M$ at $t=0$ and denote by $C_c^\infty(\Sigma)$ the space of smooth, compactly supported, real-valued functions on $\Sigma$. Then let $E = C_c^\infty(\Sigma)\oplus C_c^\infty(\Sigma)$ with the symplectic form
\begin{align}
    \sigma\big( (f_1, g_1), (f_2, g_2) \big) = \int_{\Sigma} \big( f_1(\mathbf{y}) g_2(\mathbf{y}) - f_2(\mathbf{y}) g_1(\mathbf{y})  \big) d^s \mathbf{y}.
    \end{align}
for all $f_1,f_2,g_1,g_2\in C_c^\infty(\Sigma)$.  For brevity, we will label elements $u = (f,g)\in E$.  

Next, define spatial translations by associating each $\mathbf x \in \mathbb R^s$ with the map $\tau_{\mathbf{x}}: E\to E$ defined by
\begin{align}
    (\tau_{\mathbf{x}}u)(\mathbf{y}) = u(\mathbf{y}-\mathbf{x})
\end{align}
for all $\mathbf{y}\in\Sigma$.

To construct time translations for the mass $m\geq 0$ scalar field, first define the one-particle energy operator
\begin{align}
    \mu_m = (m^2-\nabla_{\mathbf{x}}^2)^{1/2}
\end{align}
acting on $C_c^\infty(\Sigma)$.  Then define the map $\tau_t^{(m)}: E\to E$ by
    \begin{align}
    \tau_t^{(m)} (f,g)  = \big( \cos(t \mu_m) f - \mu_m \sin(t \mu_m) g ,  \mu_m^{-1} \sin(t \mu_m)  f  +  \cos(t \mu_m) g  \big) 
    \end{align}
for all $f,g\in C_c^\infty(\Sigma)$.  It follows that $\tau_t^{(m)}$ implements time evolution according to the mass $m$ Klein-Gordon equation
    \begin{align}
    (\partial_t^2 - \nabla_{\mathbf x}^2 + m^2) \varphi = 0 ,
    \end{align}
with initial data at $t= 0$ given by $\varphi(0,\cdot) = g$ and $(\partial_t \varphi)(0, \cdot) = f$.

We can define the action $\tau^{(m)}_{\Lambda}$ of Lorentz transformations similarly, and then construct the combined transformation $\tau^{(m)}_{\Lambda,x} = \tau_t^{(m)}\circ \tau_{\mathbf{x}}\circ\tau^{(m)}_\Lambda$ for $x=(t,\mathbf{x})$.  A straightforward calculation shows these transformations are all symplectomorphisms.

We can now build the Poincar\'{e} net of the free scalar field.  To every $u\in E$, associate a symbol $W(u)$ and define the algebraic relations
\begin{align}
    & W(u_1)W(u_2) = e^{-\frac{i}{2} \sigma(u_1, u_2)} W(u_1 + u_2) \text{ for every $u_1,u_2 \in E$, and} \label{eqn:multiplicative Weyl relation} \\
    & W(u)^* = W(-u) \text{ for every $u \in E$.} \label{eqn:involutive Weyl relation}
\end{align}
for all $u_1,u_2,u\in E$.  These are known as the \emph{Weyl relations} and are an exponentiated form of the canonical commutation relations for bosonic quantum field theories \parencite{manuceauSmallestAlgebraCanonical1973,binzConstructionUniquenessWeyl2004}.  The minimal *-algebra containing $\{W(u)\}_{u\in E}$ satisfying the Weyl relations will be denoted $\Delta$.  For every $(\Lambda,x)$ in the identity component of the Poincar\'{e} group, we define $\alpha_{\Lambda,x}^{(m)}: \Delta\to \Delta$ by $\alpha_{\Lambda,x}^{(m)}\left(W(u)\right) = W(\tau_{\Lambda,x}^{(m)}u).$  Then for every double cone $\mathcal O$ in $M$ with base $U$ in $\Sigma$ we define $E_U := C_c^\infty(U)\oplus C_c^\infty(U)\subseteq E$ and construct the local *-algebra
    \begin{align}
    \Delta^{(m)}\big( \Lambda \mathcal O + x \big) = \operatorname{span} \{ \alpha^{(m)}_{\Lambda, x} (W(u)) \, | \, \text{supp} \, u \in E_U \}.
    \end{align}
In this way, we have assigned a *-algebra to every double cone $\mathcal O$ in the spacetime $M$.  The *-algebra $\Delta$ has a unique maximal C*-norm, and we call the completion $\overline{\Delta}$ in this norm the \textbf{Weyl algebra} $\mathcal{W}$.  We also define a net structure according to the assignment
\begin{align}
    \label{eqn:definition of weyl net}
    \mathcal O \mapsto \mathcal W^{(m)}(\mathcal O) := \overline{ \Delta^{(m)}(\mathcal O) } \subseteq \mathcal W
\end{align}
for every double cone $\mathcal{O}\subseteq M$, where the completion is with respect to the C*-norm.  The automorphisms $\alpha_{\Lambda,x}^{(m)}$ extend continuously to these algebras.  The pair $(\mathcal{W}^{(m)},\alpha^{(m)})$ is a Poincar\'{e} net representing the free scalar field of mass $m$.

We mention one particular representation of $\mathcal{W}^{(m)}$, the mass $m$ vacuum representation defined as the GNS representation of the mass $m$ vacuum state $\omega^{(m)}: \mathcal{W}\to\C$ given by
\begin{align}
    \omega^{(m)}\left(W(u)\right) = e^{-\langle u,u\rangle_m/4}
\end{align}
for all $u\in E$.  Here, we define the complex inner product $\langle\cdot,\cdot\rangle_m$
on $E$ by
\begin{align}
\label{eqn:inner product}
    J_m(f,g) = (-\mu_mg,\mu_m^{-1} f) && \left\langle u,v\right\rangle_m = \sigma(u,J_m v) + i\sigma(u,v)
\end{align}
for all $f,g\in C_c^\infty(\Sigma)$ and $u,v\in E$.

We must make some caveats concerning the inner product $\langle\cdot,\cdot\rangle_m$ and the corresponding dynamics $\tau^{(m)}_t$ on $E$.  Both structures are well-defined when $m>0$, but when $m=0$ those definitions only make sense for spatial dimension $s\geq 2$.  
The case of $m=0$ and $s=1$ is complicated by the presence of distinct charge sectors. 
For the remainder of this paper we restrict attention to $s=2,3$ where the inner product and dynamics are well-defined, and so is the vacuum state $\omega^{(m)}$ that gives rise to the vacuum representation.  We denote the mass $m$ vacuum representation by $(\pi^{(m)},\mathcal{H}^{(m)})$.

We will also use a one particle structure \parencite{Kay1979} $\kappa_m: E\to L^2(\R^s ;\C)$ defined by
\begin{align}
\label{eqn: one particle structure}
    \kappa_m(f,g) = \mu_m^{-1/2}f + i\mu_m^{1/2}g
\end{align}
for $(f,g)\in E$.  It is well-known that $\kappa_m$ is unitary.  The one-particle structure gives us a natural way to represent time evolution in terms of the Fourier transform $f\mapsto \mathcal{F}[f]$ on $L^2(\R^s)$:
\begin{align}
    \mathcal{F}\big[\kappa_m(\tau^{(m)}_t u)\big](p) = e^{i\hat{\mu}_m(p) t}\mathcal{F}\left[\kappa_m(u)\right](p)
\end{align}
for $u\in E$ with the multiplication operator $\hat{\mu}_m(p) = (m^2 + p^2)^{1/2}$.

\subsection{Local Algebras of Spacetime Smeared Observables}
\label{subsec: spacetime smeared observables}

The Weyl net $(\mathcal{W}^{(m)},\alpha^{(m)})$ is not the only net of local algebras for the real scalar field.  We will construct two distinct, but closely related, nets of local algebras.  For motivation, note that some authors require as a further condition on an algebraic quantum field theory that the translation group is given a \emph{strongly continuous} representation on a Poincar\'{e} net $(\mathfrak{A},\alpha)$ in the sense that 
  \begin{align} 
    \lim_{ x \to 0} || \alpha^{(m)}_{1, x} (A) - A || = 0 
    \end{align}
for all $A \in \mathfrak A$ \parencite[III.3]{haagLocalQuantumPhysics1996}.  The representation of the translation group on the Poincar\'{e} net $(\mathcal{W}^{(m)},\alpha^{(m)})$, however, is not strongly continuous due to the fact that $||W(u) - W(u')|| = 2$ whenever $u\neq u'$.

Still, one can construct a distinct net with quasilocal algebra of observables $\mathcal{A}^{(m)}$ generated by elements of the form
\begin{align}
S^{(m)}(q,u) = \int q(x) \pi^{(m)}(\alpha^{(m)}_{1,x}( W(u) )) d^{s+1} x ,
    \label{eqn:smeared elements}
    \end{align}
where $q \in C^\infty_c(\mathbb R^{s+1})$, $u \in E$, and the integral is a weak operator integral.  Then the Poincar\'e group has a representation on $\mathcal{A}^{(m)}$ by the extension of the action
\begin{align}
    \alpha^{(m)}_{\Lambda,x}(S^{(m)}(q,u)) = \int q(x') \pi^{(m)}\big(\alpha^{(m)}_{\Lambda,x+x'}( W(u) )\big) d^{s+1} x' 
\end{align}
for any Poincar\'e transformation $(\Lambda,x)$.  It follows that the action of the translation group on $\mathcal{A}^{(m)}$ is strongly continuous.  Moreover, one can define a net by associating local regions to subalgebras as follows.  If $\mathcal{O}$ is a double cone with base $U\subseteq \Sigma$, define
\begin{align}
    \mathcal{A}^{(m)}(\Lambda\mathcal{O}+x) = C^*\left(\{ \alpha_{\Lambda,x}(S(q,u))\ |\ \tau_{1,y}^{(m)}(u)\in C_c^\infty(U) \text{ for all } y\in \supp (q)\}\right).
\end{align}
Since \textcite{buchholzScalingAlgebrasRenormalization1998} use the operators of the form $S(q,u)$ in their analysis of the scaling algebra for the free scalar field, we will treat the scaling behavior of the net $\mathcal{A}^{(m)}$.

\subsection{Local Algebras of Phase Space Smeared Observables}
\label{subsec:phase space smeared observables}

We present one further net of local algebras that follows a similar construction---with the key difference that we will smear over phase space rather than over spacetime translations.  

Our first task is to specify a class of smearing functions that will enter into weak operator integrals over phase space.  We say that a function $h: E\to\mathbb{C}$ is \textbf{tame} if there are bounded linear functionals $\ell_1,...,\ell_n\in E^*$ and a Borel measurable function $\tilde{h}_{\ell_1,...,\ell_n}: \R^n\to\C$ such that
\begin{align}
    h(u) = \tilde{h}_{\ell_1,...,\ell_n}(\ell_1(u),...,\ell_n(u))
\end{align}
for all $u\in E$.  In our setting where $E$ carries the real inner product $\Re\left(\langle\cdot,\cdot\rangle_m\right)$, it is often more convenient to use a slightly different characterization that follows from the Riesz representation theorem.  Let $\mathfrak{h}^{(m)}= \overline{E}^{\Re(\langle\cdot,\cdot\rangle_m)}$ denote the completion of $E$ with respect to the real inner product.  A function $h$ is tame
if there is a $\Re\left(\langle\cdot,\cdot\rangle_m\right)$-orthogonal projection $P^{(m)}_{F^{(m)}}: \mathfrak{h}^{(m)}\to F^{(m)}$ onto a \emph{finite-dimensional} subspace $F^{(m)}\subseteq \mathfrak{h}^{(m)}$ and a measurable function, now denoted $\tilde{h}_{F^{(m)}}: F^{(m)}\to\C$ such that
\begin{align}
    h(u) = \tilde{h}_{F^{(m)}}\big(P^{(m)}_{F^{(m)}}u\big)
\end{align}
for all $u\in E$.  We denote by $A^0(E)$ the collection of bounded, tame functions for which $\tilde{h}_{F^{(m)}}$ is continuous. While generally the set $F^{(m)}$ is not unique even for fixed $m$, for any $h\in A^0(E)$ and given $m\geq 0$ there is a unique largest finite-dimensional subspace $E^{(m)}_h$ contained in all such subspaces $F^{(m)}$.  In other words, $E^{(m)}_h$ is the smallest finite-dimensional subspace such that there is a function $\tilde{h}: E^{(m)}_h\to \mathbb{C}$ satisfying
\begin{align}\label{eqn:canonical representation of tame function}
    h(u) = \tilde{h}\big(P^{(m)}_{E^{(m)}_h}u\big)
\end{align}
for all $u\in E$.  We call Eq.~\eqref{eqn:canonical representation of tame function} the \emph{canonical presentation} of the tame function $h\in A^0(E)$, as the data $E^{(m)}_h$ and $\tilde{h}$ are now uniquely specified.  We refer to $E^{(m)}_h$ as the \textbf{base} of $h$ and $\tilde{h}$ as the \textbf{representative} of $h$.

We can now define a smeared operator for each tame function $h\in A^0(E)$ given by
\begin{align}  \label{eqn:definition of extra smeared}
    K^{(m)}(h) = \int_{E^{(m)}_h} h(u) \pi^{(m)}\left(W(u)\right)d\mu_{E^{(m)}_h}(u),
\end{align}
where the integral is a weak operator integral in the vacuum representation on $\mathcal{H}^{(m)}$.  The measure $\mu_{E^{(m)}_h}$ is the Lebesgue measure after identifying $E^{(m)}_h\cong \R^n$ with the measure normalized by employing the volume form determined by $\sigma$.  While it is well-known that there is no Lebesgue measure on the whole infinite-dimensional space $E$, by restricting to tame functions with finite-dimensional base we allow ourselves to construct ordinary finite-dimensional integrals in Eq.~\eqref{eqn:definition of extra smeared}.

\begin{prop}
    For $h\in A^0(E)$, if the representative $\tilde{h}$ is in $L^1(E^{(m)}_h,\mu_{E^{(m)}_h})$, then the weak operator integral in Eq.~\eqref{eqn:definition of extra smeared} exists.
\end{prop}
\begin{proof}
    Suppose $h\in A^0(E)$ and $\tilde{h}\in L^1(E^{(m)}_h,\mu_{E^{(m)}_h})$.  The map $u \mapsto \langle \varphi, \pi^{(m)}\left(W(u)\right)\psi\rangle$ is measurable (in fact, continuous) in the \emph{regular} vacuum representation $\pi^{(m)}$ for any $\varphi,\psi\in\mathcal{H}^{(m)}$, and so
\begin{align}
    \langle \varphi,K^{(m)}(h)\psi\rangle &= \int_{E^{(m)}_h} h(u) \langle \varphi, \pi^{(m)}\left(W(u)\right)\psi\rangle d\mu_{E^{(m)}_h}(u)\\
    &\leq ||\varphi||\cdot||\psi||\int_{E^{(m)}_h} h(u) ||\pi^{(m)}\left(W(u)\right)||d\mu_{E^{(m)}_h}(u)\\
    &\leq ||\varphi||\cdot||\psi||\cdot||\tilde{h}||_{L^1(E^{(m)}_h,\mu_{E^{(m)}_h})}
\end{align}
since $||W(u)|| = 1$.  Thus $K^{(m)}(h)$ determines a bounded sesquilinear form, and hence a bounded operator on $\mathcal{H}^{(m)}$.
\end{proof}

Such operators have been studied extensively in quantum harmonic analysis in finite dimensions \parencite{wernerQuantumHarmonicAnalysis1984}.  Moreover, the so-called Weyl calculus for such operators has been extended to infinite dimensions \parencite{KrRa78,AmNi08,AmJaNo15}.  We will restrict attention even further to tame functions $h$ such that $\tilde{h}\in C_c^\infty(E^{(m)}_h)$.  We denote the set of tame functions whose representative is smooth and compactly supported by $A^\infty_c(E)$.  Clearly, we have $A^\infty_c(E)\subseteq A^0(E)$.

Then we can construct a C*-algebra generated by operators of the form $K^{(m)}(h)$ for tame functions $h\in A^\infty_c(E)$.  Denote by $\mathcal{L}_0(E)$ the set of finite-dimensional symplectically non-degenerate subspaces $F\subseteq E$.  For each $F\in \mathcal{L}_0(E)$, define
\begin{align}
    \mathcal{K}_F^{(m)} = C^*\left(\left\{ K^{(m)}(h)\in\mathcal{B}(\mathcal{H}^{(m)})\ \middle|\ h\in A^\infty_c(E) \text{ and } E^{(m)}_h = F\right\}\right).
\end{align}
In other words, $\mathcal{K}_F^{(m)}$ is the smallest C*-algebra containing all the operators $K^{(m)}(h)$, where the base of $h$ is $F$.  The algebra $\mathcal{K}_F^{(m)}$ is easily seen to be *-isomorphic to the \emph{twisted group C*-algebra} of $F$ by the multiplier $(u,v)\in F\mapsto e^{-\frac{i}{2}\sigma(u,v)}$ \parencite[][\S III.1.7]{landsmanMathematicalTopicsClassical1998}.

Next, for each $F\in \mathcal{L}_0(E)$, define
\begin{align}
    \mathcal{R}^{(m)}_F = C^*\left(\bigcup_{F_0\in \mathcal{L}_0(F)}\mathcal{K}_{F_0}^{(m)}\right).
\end{align}
If $F_1\subseteq F_2$, then clearly $\mathcal{R}^{(m)}_{F_1}\subseteq \mathcal{R}^{(m)}_{F_2}$ so we can construct the inductive limit
\begin{align}
    \mathfrak{R}^{(m)} = \overline{\bigcup_{F\in \mathcal{L}_0(E)}\mathcal{R}_F^{(m)}},
\end{align}
where the closure is taken in the norm topology on the vacuum representation.
In other words, $\mathfrak{R}^{(m)}$ is the smallest C*-algebra containing all the operators $K^{(m)}(h)$, where the base of $h$ \emph{ranges over all finite-dimensional symplectically non-degenerate subspaces $F\in \mathcal{L}_0(E)$}.  Note that $\mathfrak{R}^{(m)}$ is a C*-algebra, but not a von Neumann algebra.  We will next construct a Poincar\'{e} net whose quasilocal algebra is $\mathfrak{R}^{(m)}$.

We define a representation of the identity component of the Poincar\'{e} group on $\mathfrak{R}^{(m)}$ by associating each Poincar\'e transformation $(\Lambda,x)$ with an automorphism $\alpha_{\Lambda,x}^{(m)}$ of $\mathfrak{R}^{(m)}$ defined by
\begin{align}
\label{eqn:Poincare transformations on smeared elements}
    \alpha_{\Lambda,x}^{(m)}(K^{(m)}(h)) = K^{(m)}(h\circ\tau_{(\Lambda,x)^{-1}}^{(m)})
\end{align}
for a tame function $h\in A^\infty_c(E)$.  

For this definition to make sense, one only needs to know the following.

\begin{prop}
\label{prop:symplectomorphisms lift to algebra maps}
    Whenever $\delta: E\to E$ is a symplectomorphism and $h\in A^\infty_c(E)$ with $E^{(m)}_h\in\mathcal{L}_0(E)$, then $h\circ \delta\in A^\infty_c(E)$ with $E^{(m)}_{h\circ \delta}\in \mathcal{L}_0(E)$.  Hence, if $K^{(m)}(h)\in\mathcal{R}^{(m)}$, then $K^{(m)}(h\circ\delta)\in\mathcal{R}^{(m)}$.
\end{prop}

\begin{proof}
    Clearly, if $h\in A^\infty_c(E)$ is a tame function with base $E_h^{(m)}\in \mathcal{L}_0(E)$ then since $\delta$ is a symplectomorphism of $E$, it follows that $h\circ\delta$ is a tame function with base $\delta\big(E_h^{(m)}\big)\subseteq E$.  The base $\delta\big(E_h^{(m)}\big)$ is symplectically non-degenerate since $E_h^{(m)}$ is symplectically non-degenerate, and hence $\delta\big(E_h^{(m)}\big)\in\mathcal{L}_0(E)$. Moreover, the representative of $h\circ\delta$ is still smooth and compactly supported.  
\end{proof}
\noindent It now follows from Propoposition \ref{prop:symplectomorphisms lift to algebra maps} that our definition in Eq.~\eqref{eqn:Poincare transformations on smeared elements} thus makes sense with the symplectomorphism $\delta = \tau_{(\Lambda,x)^{-1}}^{(m)}$.


To associate operators with local regions, for any double cone $\mathcal{O}$ with base $U$, we define the local algebra $\mathfrak{R}^{(m)}(\Lambda\mathcal{O} + x)$ as
\begin{align}
    \mathfrak{R}^{(m)}(\Lambda\mathcal{O} + x) = C^*\left(\bigcup_{F\in\mathcal{L}_0(E_U)} \alpha_{\Lambda,x}^{(m)}\left(\mathcal{K}^{(m)}_F\right) \right).
\end{align}
In particular, $\mathfrak{R}^{(m)}(\mathcal{O})$ is the smallest C*-algebra containing all the operators $K^{(m)}(h)$, where the base of $h$ ranges over all finite-dimensional symplectically non-degenerate subspaces of $E_U$.  One can check easily that $(\mathfrak{R}^{(m)},\alpha^{(m)})$ is a Poincar\'{e} covariant net.

The local C*-algebras $\mathcal{W}^{(m)}(\mathcal{O})$, $\mathcal{A}^{(m)}(\mathcal{O})$, and $\mathfrak{R}^{(m)}(\mathcal{O})$ are each distinct even though they all generate the same weak operator closure in the vacuum representation.  It is well known that $\mathcal{W}^{(m)}(\mathcal{O})$ and $\mathcal{A}^{(m)}(\mathcal{O})$ generate the same von Neumann algebra.  To see that $\mathfrak{R}^{(m)}(\mathcal{O})$ does too, consider a double cone $\mathcal{O}$ with base given by a bounded open subset $U\subseteq \Sigma$ and an arbitrary $u\in F\in \mathcal{L}_0(U)$.  Since $W(u)\in \mathcal{W}^{(m)}(\mathcal{O})$, it suffices to show that $W(u)$ can be obtained as a weak operator limit of elements of $\mathfrak{R}^{(m)}(\mathcal{O})$.  In fact, one can construct such a weak operator limit of operators $K^{(m)}(h_n)\in \mathcal{K}^{(m)}_F\subseteq \mathfrak{R}^{(m)}(\mathcal{O})$ by defining $h_n\in A^\infty_c(E)$, each with fixed base $E_{h_n} = F$, to have representatives $\tilde{h}_n$ that yield a sequence approaching a delta function at a point $u\in F$, e.g., normalized Gaussians with variance proportional to $\frac{1}{n}$.  It follows that $K^{(m)}(h_n)\to W(u)$ as $n\to\infty$ so that $\mathfrak{R}^{(m)}(\mathcal{O})$ is weak operator dense in $\mathcal{W}^{(m)}(\mathcal{O})$.  On the other hand, clearly $\mathcal{W}^{(m)}(\mathcal{O})$ is weak operator dense in $\mathfrak{R}^{(m)}(\mathcal{O})$ since the weak operator integral defining $K^{(m)}(h)\in\mathfrak{R}^{(m)}(\mathcal{O})$ can be constructed as a weak operator limit of Riemann sums involving  only terms $W(u)\in\mathcal{W}^{(m)}(\mathcal{O})$.

We have already said that $\mathcal{K}^{(m)}_F$ is *-isomorphic to the twisted group C*-algebra of $F$, and both are known to be *-isomorphic to the algebra $\mathcal{B}_0(L^2(\R^{\dim F/2}))$ of compact operators on $L^2(\R^{\dim F/2})$ via the standard Schr\"odinger representation $\pi^{S,m}_F: \mathcal{K}_F^{(m)}\to \mathcal{B}_0(L^2(\R^{\dim F/2}))$ \parencite[][\S II.2]{landsmanMathematicalTopicsClassical1998} defined by
\begin{align}
    \big(\pi^{S,m}_F(K^{(m)}(h))\psi\big)(x) = \int_{\R^{\dim F/2}} h(x-y,p)e^{ip(x+y)/2}\psi(y)\ dp dy
\end{align}
for $h\in A_c^\infty(E)$ with $E^{(m)}_h = F$ and $\psi\in L^2(\R^{\dim F/2})$.  Actually, the operator $\pi^{S,m}_F(K^{(m)}(h))$ is even Hilbert-Schmidt with integral kernel given by 
\begin{align}
    k_h(x,y) := \int h(x-y,p)e^{ip(x+y)/2}dp.
\end{align}
The Hilbert-Schmidt norm is given by $||\pi^{S,m}_F(K^{(m)}(h))||_{2} = ||k_h||_{L^2(F\times F)} = (2\pi)^{\dim F/4}||\tilde{h}||_{L^2(F)}$.  Moreover, since $\pi^{S,m}_F$ is a faithful representation of $\mathcal{K}^{(m)}_F$, it follows that
\begin{align}
    ||K^{(m)}(h)|| = ||\pi^{S,m}_F(K^{(m)}(h))||\leq ||\pi^{S,m}_F(K^{(m)}(h))||_{2} = (2\pi)^{\dim F/4}||\tilde{h}||_{L^2(F)},
\end{align}
which is itself bounded uniformly across values of $m$.

We pause to ward off a potential confusion here.  It is well known that the local von Neumann algebras generated by $\mathcal{W}^{(m)}(\mathcal{O})$, and hence also $\mathfrak{R}^{(m)}(\mathcal{O})$, in the vacuum representation are type $III$.  It follows that those algebras contain no finite projections, and hence no compact operators in the vacuum representation.  So of course $K^{(m)}(h)\in\mathfrak{R}^{(m)}(\mathcal{O})$ cannot be compact in the vacuum representation.  The representation $\pi^{S,m}_{E^{(m)}_h}$ in which the operator $K^{(m)}(h)$ is compact is only a representation of the \emph{subalgebra} $\mathcal{K}^{(m)}_{E^{(m)}_h}$.  Still, we can build representations of larger algebras to obtain a better understanding of the operator $K^{(m)}(h)$ in other representations.


We can extend the Schr\"odinger representation $\pi^{S,m}_F$ in a natural way to the larger C*-algebra $\mathcal{R}^{(m)}_F$.  Suppose $K^{(m)}(h)\in \mathcal{R}^{(m)}_{F}$ so that $h\in A_c^\infty(E)$ has $E^{(m)}_h \subseteq F$.  Letting $\big(E^{(m)}_h\big)' = \big(E^{(m)}_h\big)^{\sigma\perp}\cap F$, we can take the orthogonal decomposition $F = E^{(m)}_h\oplus \big(E^{(m)}_h\big)'$ with respect to the symplectic form.  Then since $\dim F = \dim E^{(m)}_h + \dim \big(E^{(m)}_h\big)'$, we have a natural tensor product structure $L^2(\R^{\dim F/2})= L^2(\R^{\dim E^{(m)}_h/2})\otimes L^2(\R^{\dim \big(E^{(m)}_h\big)'/2})$.  And thus, we can define $\tilde{\pi}^{S,m}_F: \mathcal{R}^{(m)}_F\to \mathcal{B}(L^2(\R^{\dim F/2}))$ by
\begin{align}
    \tilde{\pi}^{S,m}_F(K^{(m)}(h)) = \pi^{S,m}_{E^{(m)}_h}(K^{(m)}(h))\otimes I_{\big(E^{(m)}_h\big)'},
\end{align}
where $I_{\big(E^{(m)}_h\big)'}$ is the identity operator on $L^2(\R^{\dim \big(E^{(m)}_h\big)'/2})$.  The operator $\tilde{\pi}^{S,m}_F(K^{(m)}(h))$ is generally not compact because of the presence of the tensor factor of the identity operator on an infinite-dimensional space, and we can likewise think of the vacuum representation $\pi^{(m)}(K^{(m)}(h))$ of the same operator as containing a tensor factor of the identity on an infinite dimensional space.  This explains how $\mathfrak{R}^{(m)}(\mathcal{O})$ can generate a Type $III$ von Neumann algebra with no finite projections.  
The representation $\tilde{\pi}^{S,m}_F$ is indeed an extension of $\pi^{S,m}_F$ because whenever $E^{(m)}_h = F$, then $\big(E^{(m)}_h\big)' = \{0\}$ so that $\tilde{\pi}^{S,m}_F(K^{(m)}(h)) = \pi^{S,m}_F(K^{(m)}(h))$ recovers the compact operator from the original Schr\"odinger representation.

We now have three distinct nets for the real scalar field: $\mathcal{W}^{(m)}$, $\mathcal{A}^{(m)}$, and $\mathfrak{R}^{(m)}$.  In the next section, we shall show that each generates a lower semicontinuous scaling bundle.

\section{Lower Semicontinuous Scaling Bundles for the Scalar Field}
\label{sec:lower semicontinuous scalar field}

In this section, we demonstrate a strategy for constructing scaling bundles that we hope may generalize to other models.  The primary technical challenge to construct a scaling bundle is the condition that for a section $\underline{A}\in\underline{\mathfrak{A}}$, the map
\begin{align}
    \lambda\mapsto ||\underline{A}_\lambda||_\lambda
\end{align}
must be lower semicontinuous and bounded.  We will show that algebras of sections satisfying this condition can be constructed for each of our Poincar\'e nets for the massive free scalar field.  We show that each of these nets gives rise to a lower semicontinuous scaling bundle under an appropriate action of rescaling transformations.

We find that in each net, a scaling limit net of the massive free scalar field is net isomorphic to the net for the massless free scalar field, as expected.  We demonstrate this by providing an explicit net bundle homomorphism between the scaling bundle for the massive free scalar field and another net bundle, which we will call the \emph{variable mass bundle}.  This latter bundle describes a limit of decreasing mass.  More specifically, the fiber net above $\lambda\in [0,1]$ for the variable mass bundle will be the net for the free scalar field of mass $\lambda m$.

This section provides an example of a general strategy for constructing scaling bundles and determining their scaling limits, which we hope may apply to other models.  Given a Poincar\'e net, we first find a collection of dilation transformations that intertwine the dynamical automorphisms with a change in some dynamical parameters (mass, in our example).  This is the sense in which we recover the traditional notion of ``running coupling constants''.  We then show that the changing dynamics give rise to sections $\lambda\mapsto \underline{A}_\lambda$ satisfying the condition that $\lambda\mapsto ||\underline{A}_\lambda||_\lambda$ is lower semicontinuous and bounded, so that they generate a lower semicontinuous bundle of Poincar\'e nets.  Finally, we use the collection of dilation automorphisms to construct a bundle homomorphism between the bundle of varying dynamics and the scaling bundle of the original Poincar\'e net.  Theorem \ref{thm:extension of net bundle} then implies that the two bundles have isomorphic largest extensions and the largest scaling limit is net isomorphic to the Poincar\'e net for the appropriate limiting value of the dynamical parameters.

We begin in \S\ref{subsec:generating sections} to motivate and define sections that will generate scaling bundles for the massive free scalar field.  In \S\ref{subsec:variable mass bundle for the massive free scalar field}, we then define the variable mass bundle, which is a net bundle whose fibers correspond to free scalar fields of different masses.  That is, the variable mass bundle has fiber over $\lambda\in [0,1]$ that is a Poincar\'e net for the mass $\lambda m$ free scalar field.  Finally, in \S\ref{subsec:scaling bundle for the massive free scalar field}, we use the restriction of the variable mass bundle to $(0,1]$ to construct our lower semicontinuous scaling bundle for the massive free scalar field.  This implies that there is a net epimorphism from the largest scaling limit---i.e., the Poincar\'e net above $\lambda = 0$ extending the scaling bundle to $[0,1]$---to the $\lambda = 0$ fiber of the variable mass bundle, which is the $m=0$ free scalar field.

\subsection{Generating Sections}
\label{subsec:generating sections}

We are now in a position to define scaling bundles for the Poincar\'e nets for the massive free scalar field.  To do so, we use the dilation *-isomorphisms $\{\sigma_\lambda\}_{\lambda>0}$ defined by \textcite[][p. 8]{buchholzScalingAlgebrasRenormalization1998}.  For each $\lambda>0$, the map $\sigma_\lambda: \mathcal{W}\to \mathcal{W}$ is defined by the extension of
\begin{align}
\label{eqn:definition of sigma}
    \sigma_\lambda\left(W(u)\right) = W(\delta_\lambda u),
\end{align}
for $u\in E$, where
\begin{align}
    \left(\delta_\lambda(f,g)\right)(\mathbf{x}) = (\lambda^{-\frac{s+1}{2}} f(\lambda^{-1}\mathbf{x}),\lambda^{-\frac{s-1}{2}} g(\lambda^{-1}x)
\end{align}
for all $\mathbf{x}\in\R^s$.  This map satisfies
\begin{align}
\label{eqn:sigma intertwines alphas}
    \sigma_\lambda\circ\alpha^{(\lambda m)}_{\Lambda,x} = \alpha^{(m)}_{\Lambda,\lambda x}\circ\sigma_\lambda
\end{align}
for all $\lambda\in (0,1]$ and
\begin{align}
    \sigma_\lambda\left(\mathcal{W}^{(\lambda m)}(\mathcal{O})\right) = \mathcal{W}^{(m)}(\lambda\mathcal{O})
\end{align}
for local regions $\mathcal{O}$, which follows from the previous equation and the fact that
\begin{align}
    \supp(\delta_\lambda u)= \lambda\cdot \supp(u)
\end{align}
for all $u\in E$.

\textcite[Lemma 3.2]{buchholzScalingAlgebrasRenormalization1998} use these dilation transformations to define particular renormalization group flows, which serve as elements of their scaling algebra for the massive free scalar field.  We wish to use the dilation transformations to define flows that are at least formal analogs of the maps $\underline{A}: \lambda\mapsto \sigma_\lambda(A)$, where $A$ comes from each of our nets $\mathcal{W}^{(m)}$, $\mathcal{A}^{(m)}$, and $\mathfrak{R}^{(m)}$.  We clearly already have the resources to define rescaled sections of the Weyl algebra $\mathcal{W}$ by defining, for each $W(u)\in\mathcal{W}$, the section
\begin{align}
\label{eqn: weyl sections}
    \underline{W}(u): \lambda\mapsto \sigma_\lambda(W(u)) = W(\delta_\lambda u).
\end{align}
Following our earlier notation, we write $\underline{W}_\lambda(u) = W(\delta_\lambda u)$.
Next, for $S^{(m)}(q,u)\in \mathcal{A}^{(m)}$, define the rescaled section 
    \begin{align} 
    \label{eqn: s sections}
    \underline{S}^{(m)}(q,u): \lambda \mapsto S^{(m)}(q_\lambda,\delta_\lambda u)=\int q(x) \pi^{(m)}\big(\alpha^{(m)}_{1,\lambda x}  \circ\sigma_\lambda(W(u)) \big) d^{1+s}x,
    \end{align}
where $q_\lambda(x) = \lambda^{-(1+s)}q(\lambda^{-1}x)$.  We also write $\underline{S}_\lambda^{(m)}(q,u) = S^{(m)}(q_\lambda,\delta_\lambda u)$.
 Finally, consider $K^{(m)}(h)\in \mathfrak{R}^{(m)}$, where $h$ has canonical presentation $h(u) = \tilde{h}\big(P^{(m)}_{E^{(m)}_h}u\big)$.  Since the base $E^{(m)}_{h\circ\delta_\lambda^{-1}}$ generally varies with $\lambda$, we must consider $h_\lambda(u) := \tilde{h}\big(P^{(\lambda m)}_{E^{(m)}_h}u\big)$ so that $E^{(\lambda m)}_{h_\lambda} = E^{(m)}_h$ is constant for all $\lambda$ in order to preserve the localization of $K^{(m)}(h)$.  We define the rescaled section
\begin{align}
\label{eqn: k sections}
    \underline{K}^{(m)}(h): \lambda\mapsto K^{(m)}(h_\lambda\circ\delta_\lambda^{-1}) = \int_{E^{(m)}_{h_\lambda \circ\delta_\lambda^{-1}}} h_\lambda( u)\pi^{(m)}\big(\sigma_\lambda(W(u))\big) d\mu_{E^{(m)}_{h_\lambda\circ\delta_\lambda^{-1}}}.
\end{align}
We again write $\underline{K}^{(m)}_\lambda(h) = K^{(m)}(h_\lambda\circ\delta_\lambda^{-1})$.

We motivate the preceding definitions by showing they match the action of the *-isomorphism $\sigma_\lambda$ when it is suitably extended to $\mathcal{A}^{(m)}$ and $\mathcal{R}^{(m)}$.  To do so, we rely on the following fact.
\begin{prop}
    For every $\lambda>0$, there is a unitary transformation $V_\lambda: \mathcal{H}^{(\lambda m)}\to \mathcal{H}^{(m)}$ such that
    \begin{align}
        V_\lambda \pi^{(\lambda m)}(W(u)) V_\lambda^*= \pi^{(m)}(\sigma_\lambda(W(u))
    \end{align}
    for all $W(u)\in\mathcal{W}$.
\end{prop}

\begin{proof}
    Recall from the GNS construction for the state $\omega^{(m)}$ on the Weyl algebra $\mathcal{W}$ that $\mathcal{H}^{(m)}$ consists in the completion of $\mathcal{W}$ in the inner product
    \begin{align}
    \langle A,B\rangle_{(m)} = \omega^{(m)}\left(A^*B\right)
    \end{align}
    for $A,B\in\mathcal{W}$ with $\mathcal{W}$ acting by left multiplication.  Likewise, $\mathcal{H}^{(\lambda m)}$ follows the same construction with only the state in the definition of the inner product replaced by $\omega^{(\lambda m)}$.  Thus, we can define the map $V_\lambda: \mathcal{H}^{(\lambda m)}\to \mathcal{H}^{(m)}$ by
    \begin{align}
        V_\lambda A = \sigma_\lambda (A),
    \end{align}
    using the definition of $\sigma_\lambda$ on $\mathcal{W}$ in Eq.~\eqref{eqn:definition of sigma}.  It follows that $V_\lambda$ is unitary because for any $A,B\in \mathcal{W}$,
    \begin{align}
        \langle V_\lambda A, V_\lambda B\rangle_{(m)} = \omega^{(m)}(\sigma_\lambda(A^*B)) = \omega^{(\lambda m)}(A^*B).
    \end{align}
    Moreover, $V_\lambda$ intertwines the representations $\pi^{(\lambda m)}$ and $\pi^{(m)}\circ\sigma_\lambda$ of the Weyl algebra because for all $A\in\mathcal{W}$ and $u\in E$,
    \begin{align}
        V_\lambda \pi^{(\lambda m)} (W(u))V_\lambda^* A &= V_\lambda \pi^{(\lambda m)}(W(u))\sigma_\lambda^{-1}(A)\\
        &=\sigma_\lambda\left( W(u) \sigma_\lambda^{-1}(A)\right)\\
        &= \pi^{(m)}(\sigma_\lambda(W(u))) A.
    \end{align}
\end{proof}
\noindent Using the unitary operators $V_\lambda$, we may lift the action of $\sigma_\lambda$ to $\mathcal{A}^{(m)}$ and $\mathfrak{R}^{(m)}$.
\begin{prop}
    \begin{enumerate}
    \item The map $\sigma_\lambda: \mathcal{A}^{(\lambda m)}\to\mathcal{A}^{(m)}$ defined by
    \begin{align*}
        \sigma_\lambda(A) = V_\lambda A V_\lambda^*
    \end{align*}
    for all $A\in \mathcal{A}^{(\lambda m)}$ satisfies
    \begin{align}
        \sigma_\lambda(S^{(\lambda m)}(q,u)) = S^{(m)}(q_\lambda,\delta_\lambda u).
    \end{align}
   \item The map $\sigma_\lambda: \mathfrak{R}^{(\lambda m)}\to\mathfrak{R}^{(\lambda m)}$ defined by
    \begin{align*}
        \sigma_\lambda(A) = V_\lambda A V_\lambda^*
    \end{align*}
    for all $A\in \mathcal{R}^{(m)}$ satisfies
    \begin{align}
        \sigma_\lambda(K^{(\lambda m)}(h)) = K^{(m)}(h\circ\delta_\lambda^{-1}).
    \end{align}
    \end{enumerate}
\end{prop}

\begin{proof}
Both claims follow from the definition of the weak operator integrals.
\begin{enumerate}
    \item For all $\psi,\varphi\in\mathcal{H}^{(m)}$,
    \begin{align}
        \langle \psi,V_\lambda S^{(\lambda m)}(q,u) V_\lambda^* \varphi\rangle_{(m)} &= \int q(x) \langle \psi, V_\lambda\pi^{(\lambda m)}\big(\alpha^{(\lambda m)}_{1,x}(W(u))\big) V_\lambda^*\varphi\rangle_{(m)}d^{1+s}x\\
        &= \int q(x) \langle \psi, \pi^{(m)}\big(\sigma_\lambda\circ\alpha^{(\lambda m)}_{1,x}(W(u))\big)\varphi\rangle_{(m)}d^{1+s}x \\
        &=\int q(x) \langle \psi, \pi^{(m)}\big(\alpha^{(m)}_{1,\lambda x}\circ \sigma_\lambda(W(u))\big)\varphi\rangle_{(m)}d^{1+s}x \label{eqn:integral variable x}\\
        &=\int q_\lambda(y) \langle \psi, \pi^{(m)}\big(\alpha^{(m)}_{1, y}(W(\delta_\lambda u))\big)\varphi\rangle_{(m)}d^{1+s}y \label{eqn:integral variable lambda x}\\
        &=\langle \psi, S^{(m)}(q_\lambda,\delta_\lambda u) \varphi\rangle_{(m)}.
        \end{align}
    Here, we have used a simple change of variables $y= \lambda x$ to derive line \eqref{eqn:integral variable lambda x} from \eqref{eqn:integral variable x}.
    \item For all $\psi,\varphi\in\mathcal{H}^{(m)}$,
    \begin{align}
        \langle \psi,V_\lambda K^{(\lambda m)}(h) V_\lambda^* \varphi\rangle_{(m)} &= \int_{E^{(\lambda m)}_h} h(u) \langle  \psi,V_\lambda \pi^{(\lambda m)}(W(u))V_\lambda^* \varphi\rangle_{(m)}d\mu_{E^{(\lambda m)}_h}(u)\\
        &=\int_{E^{(\lambda m)}_h} h(u) \langle \psi, \pi^{(m)}(W(\delta_\lambda u)) \varphi\rangle_{(m)} d\mu_{E^{(\lambda m)}_h}(u)\label{eqn: integral variable u}\\
        &= \int_{E^{(m)}_{h\circ\delta_\lambda^{-1}}} h(\delta_\lambda^{-1} v) \langle \psi,\pi^{(m)}(W(v)) \varphi\rangle_{(m)} d\mu_{E^{(m)}_{h\circ\delta_\lambda^{-1}}}(v)\label{eqn: integral variable delta_lambda u}\\
        &= \langle \psi, K^{(m)}(h\circ\delta_\lambda^{-1}) \varphi\rangle_{(m)}.
    \end{align}
    Again, we use a simple change of variables $v = \delta_\lambda u$ to derive line \eqref{eqn: integral variable delta_lambda u} from \eqref{eqn: integral variable u}, noting that the symplectomorphism $\delta_\lambda$ preserves the volume form, and hence preserves the Lebesgue measure.
\end{enumerate}
\end{proof}

\noindent Note that we have $\sigma_\lambda(K^{(\lambda m)}(h_\lambda)) = K^{(m)}(h_\lambda\circ\delta_\lambda^{-1})$.  If $h$ has canonical presentation $h(u) = \tilde{h}\big(P^{(m)}_{F}u\big)$ with $F = E^{(m)}_h$, then $\delta_\lambda\circ P^{(\lambda m)}_{F} = P^{(m)}_{\delta_\lambda(F)}\circ \delta_\lambda$.  It follows that $h_\lambda\circ\delta_\lambda^{-1}$ has canonical presentation $h_\lambda\circ\delta_\lambda^{-1}(u) = \big(\tilde{h}\circ\delta_\lambda^{-1}\big)\circ P^{(m)}_{\delta_\lambda(F)}$.  Thus, if $K^{(m)}(h)\in \mathfrak{R}^{(m)}(\mathcal{O})$ for a double cone $\mathcal{O}$ with base $U$ so that $F\subseteq E_U$, then $\delta_\lambda(F)\subseteq E_{\lambda U}$ and hence $K^{(m)}(h_\lambda\circ\delta_\lambda^{-1})\in \mathfrak{R}^{(m)}(\lambda \mathcal{O})$. Therefore, our sections have the desired localization properties for a scaling bundle.

We will proceed to construct a lower semicontinuous scaling bundle for each net $\mathcal{W}^{(m)}$, $\mathcal{A}^{(m)}$, and $\mathfrak{R}^{(m)}$ using the dilation transformations.

\subsection{Variable mass bundles for the massive free scalar field}
\label{subsec:variable mass bundle for the massive free scalar field}

To begin, notice that the sections in Eq.~\eqref{eqn: weyl sections}-\eqref{eqn: k sections} are all of the form $\lambda\mapsto \sigma_\lambda(\underline{A}_\lambda)$, where we take $\underline{A}_\lambda$ to be one of the following:
\begin{align}
    \lambda&\mapsto W(u) &&  \text{ for fixed }u\in E\\
    \lambda&\mapsto S^{(\lambda m)}(q,u) && \text{ for fixed }q\in C^\infty_c(\R^{s+1}), u\in E\\
    \lambda&\mapsto K^{(\lambda m)}(h_\lambda)&& \text{ for fixed }h\in A_c^\infty(E).   
\end{align}
Recall that the algebra of sections of a scaling bundle must be closed under the Poincar\'e transformations.  So in order for a scaling bundle to contain the sections of the form $\lambda\mapsto \sigma_\lambda(\underline{A}_\lambda)$, it must also contain sections of the form
\begin{align}
\label{eqn:definition of scaling bundle generators}
    \lambda\mapsto \alpha^{(m)}_{\Lambda,\lambda x}\circ \sigma_\lambda\left(\underline{A}_\lambda\right),
\end{align}
where $(\Lambda,x)$ is an arbitrary Poincar\'e transformation.  We will take the *-algebra generated from sections of the form in Eq.~\eqref{eqn:definition of scaling bundle generators} under the operations of pointwise addition, multiplication, and involution.  When $A$ ranges over elements of $\mathcal{W}^{(m)}$, $\mathcal{A}^{(m)}$, and $\mathfrak{R}^{(m)}$, we call the resulting *-algebras of sections $\widetilde{\underline{\mathcal{W}}}_s^{(m)}$, $\widetilde{\underline{\mathcal{A}}}_s^{(m)}$, and $\widetilde{\underline{\mathfrak{R}}}_s^{(m)}$, respectively.  Here, the subscript `s' stands for `scaling', since we will eventually show that these sections each generate a scaling bundle.

Using Eq.~\eqref{eqn:sigma intertwines alphas}, we recognize that the sections in Eq.~\eqref{eqn:definition of scaling bundle generators} are the same as
\begin{align}
    \lambda\mapsto \sigma_\lambda\circ \alpha^{(\lambda m)}_{\Lambda, x}\left(\underline{A}_\lambda\right).
\end{align}
We wish to show that the norm of each of these sections is lower semicontinuous in $\lambda$, but since $\sigma_\lambda$ is a *-isomorphism and hence norm-preserving, it suffices to show that the norm of sections of the form
\begin{align}
\label{eqn: K generating sections}
    \lambda\mapsto \alpha_{\Lambda,x}^{(\lambda m)}\left(\underline{A}_\lambda\right)
\end{align}
is lower semicontinuous in $\lambda$ as it ranges over the base $(0,1]$.  In order to understand the limit $\lambda\to 0$, we notice that Eq.~\eqref{eqn: K generating sections} makes sense even for $\lambda = 0$ so that we can take as our goal showing that the norm of those sections is lower semicontinuous on the larger domain of all $\lambda\in[0,1]$.  We generate *-algebras from all sections of the form in Eq.~\eqref{eqn: K generating sections}, again under the operations of pointwise addition, multiplication, and involution.  When $A$ ranges over elements of $\mathcal{W}^{(\lambda m)}$, $\mathcal{A}^{(\lambda m)}$, and $\mathfrak{R}^{(\lambda m)}$, we call the resulting *-algebras of sections $\widetilde{\underline{\mathcal{W}}}_v^{(m)}$, $\widetilde{\underline{\mathcal{A}}}_v^{(m)}$, and $\widetilde{\underline{\mathfrak{R}}}_v^{(m)}$, respectively. Here, the subscript `$v$' stands for `variable mass', and these sections will be thought of as generating the variable mass bundle mentioned in the preamble to \S\ref{sec:lower semicontinuous scalar field}.  We wish to prove the following.

\begin{thm}
\label{thm:lower semicontinuity}
    For every $\underline{A}$ belonging to $\widetilde{\underline{\mathcal{W}}}_v^{(m)}$, $\widetilde{\underline{\mathcal{A}}}_v^{(m)}$, or $\widetilde{\underline{\mathfrak{R}}}_v^{(m)}$, the map $\lambda\mapsto ||\underline{A}_\lambda||_\lambda$ is lower semicontinuous on $[0,1]$.
\end{thm}


\noindent 
To prove lower semicontinuity of the map $\lambda\mapsto ||\underline{A}_\lambda||_\lambda$, we follow a strategy from \textcite[][p. 337-338]{binzFieldtheoreticWeylQuantization2004} of evaluating expectation values of a suitably large collection of states to find appropriate bounds.  One such collection of states, essentially matching those employed also by \textcite[][\S3]{buchholzScalingAlgebrasRenormalization1998}, is the collection of vector states in the fixed vacuum representation for $m_0>0$. (Buchholz and Verch actually fix $m_0=0$, but we find it useful to take $m_0>0$ for application to $s=1$ when the $m_0=0$ inner product is not well-defined; the argument is essentially the same.)   For each of our nets, we rely on the following lemma.

\begin{lem} \label{lem:taus are continuous}
For every $u \in E$, the map $\lambda \mapsto \tau^{(\lambda m)}_{\Lambda ,x} u \in E$ is continuous on $[0,1]$ in the norm defined by the inner product $\langle\cdot,\cdot\rangle_{m_0}$ in Eq.~\eqref{eqn:inner product} for any $m_0>0$.
\end{lem}

\begin{proof}
For $u = (f,g)\in E$, the norm defined by the inner product in Eq.~\eqref{eqn:inner product} can be written as
    \begin{align} 
    ||u||_{m_0}^2 = \int \hat{\mu}_{m_0}^{-1} | \hat{f} + i \hat{\mu}_{m_0} \hat{g} |^2 d^s p,
    \end{align}
where $\hat{\mu}_{m_0}(p) = (m_0^2+p^2)^{1/2}$, and $\hat{f} = \mathcal{F}[f], \hat{g}=\mathcal{F}[g]$ are the Fourier transforms of $f,g$ on $\mathbb R^s$.  We calculate for translations:
\begin{align}
    \big|\big|&( \tau^{(\lambda m)}_{1, x} - \tau^{(\lambda_0 m)}_{1, x})u\big|\big|_{m_0}^2 
    = \big|\big|( \tau^{(\lambda m)}_{t} \circ \tau_{\mathbf x} - \tau^{(\lambda_0 m)}_{t} \circ \tau_{\mathbf x})u\big|\big|_{m_0}^2 \\
    &= \big|\big|\tau_{\mathbf x} \big( \tau^{(\lambda m)}_{t}(u) - \tau^{(\lambda_0 m)}_{t} (u) \big)\big|\big|_{m_0}^2 \\
    &= \big|\big| \tau^{(\lambda m)}_{t}(u) - \tau^{(\lambda_0 m)}_{t}(u) \big|\big|_{m_0}^2 \\
    &= \int \hat{\mu}_{m_0}(p)^{-1} \bigg| \big(\cos(t \hat{\mu}_{ \lambda m}(p)) \hat{f}(p) - \hat{\mu}_{ \lambda m}(p) \sin(t \hat{\mu}_{ \lambda m}(p)) \hat{g}(p)\big) \nonumber \\
    & \hspace{.25in} - \big(\cos(t \hat{\mu}_{ \lambda_0 m}(p)) \hat{f}(p) - \hat{\mu}_{ \lambda_0 m}(p) \sin(t \hat{\mu}_{ \lambda_0 m}(p)) \hat{g}(p)\big) \nonumber \\
    & \hspace{.25in} +i \big(\hat{\mu}_{m_0}(p) \hat{\mu}_{\lambda m}(p)^{-1} \sin(t \hat{\mu}_{\lambda m}(p)) \hat{f}(p)  +  \hat{\mu}_{m_0}(p) \cos(t \hat{\mu}_{\lambda m}(p)) \hat{g}(p)\big)  \nonumber \\
    & \hspace{.25in} - i \big(\hat{\mu}_{m_0}(p) \mu_{\lambda_0 m}(p)^{-1} \sin(t \hat{\mu}_{\lambda_0 m}(p)) \hat{f}(p)  + \hat{\mu}_{m_0}(p) \cos(t \hat{\mu}_{\lambda_0 m}(p)) \hat{g}(p)\big) \bigg|^2 d^s p \\
    &= \int \hat{\mu}_{m_0}(p)^{-1} \bigg| \bigg( \cos(t \hat{\mu}_{ \lambda m}(p)) - \cos(t \hat{\mu}_{ \lambda_0 m}(p)) \bigg) \hat{f} (p)\nonumber \\
    & \hspace{.25in} - \bigg( \hat{\mu}_{ \lambda m}(p) \sin(t \hat{\mu}_{ \lambda m}(p)) - \hat{\mu}_{ \lambda_0 m}(p) \sin(t \hat{\mu}_{ \lambda_0 m}(p)) \bigg) \hat{g}(p) \nonumber \\
    & \hspace{.25in} + i \hat{\mu}_{m_0}(p) \bigg( \hat{\mu}_{\lambda m}(p)^{-1} \sin(t \hat{\mu}_{\lambda m}(p)) - \mu_{\lambda_0 m}(p)^{-1} \sin(t \hat{\mu}_{\lambda_0 m}(p)) \bigg) \hat{f}(p) \nonumber \\
    & \hspace{.25in}  + i \hat{\mu}_{m_0}(p) \bigg( \cos(t \hat{\mu}_{\lambda m}(p)) -  \cos(t \hat{\mu}_{\lambda_0 m}(p)) \bigg) \hat{g}(p) \Bigg|^2 d^s p.
\end{align}

\noindent Since $f,g\in C^\infty_c(\Sigma)$, one can easily find a function dominating the integrand for all $\lambda$.  By the dominated convergence theorem, it follows that the norm above vanishes in the limit $\lambda \to \lambda_0$ since $\hat{\mu}_{\lambda m}$ approaches $\hat{\mu}_{\lambda_0 m}$ pointwise. This shows that $\lambda \mapsto \tau^{(\lambda m)}_{1, x} u$ is continuous with respect to $||\cdot ||_{m_0}$.  One can make an analogous argument for $\lambda\mapsto \tau^{(\lambda m)}_{\Lambda,0}$, which we leave to the reader.
\end{proof}

\begin{proof}[Proof of Theorem \ref{thm:lower semicontinuity}]
    Fix $m_0>0$.  
    By \textcite[][Prop. 5.2.4]{bratteliOperatorAlgebrasQuantum1981}, for any state $\psi$ in the Hilbert space $\mathcal{H}^{(m_0)}$ of the mass $m_0$ vacuum representation,
 it follows from Lemma \ref{lem:taus are continuous} that
 \begin{align}
     \lambda\mapsto ||\pi^{(m_0)}\big(W(\tau_{\Lambda,x}^{(\lambda m)}u)\big)\psi||
 \end{align}
 is continuous, where the norm is calculated in $\mathcal{H}^{(m_0)}$.  It is easy to see that the same holds for any $\underline{A}\in \widetilde{\underline{\mathcal{W}}}_v^{(m)}$.  For adjoints, we note that $W(u)^* = W(-u)$, and for products note that $||W(\tau^{(\lambda m)}_{\Lambda,x}u)||\leq 1$ provides a uniform bound.  It then follows that for any $\underline{A}$ that is a finite sum of products of the generating sections in Eq.~\eqref{eqn: K generating sections}, we have that the map
 \begin{align}
  \label{eqn:strongcontinuity}
     \lambda\mapsto ||\pi^{(m_0)}(\underline{A}_\lambda)\psi||
 \end{align}
 is also continuous.

 Moreover, $\pi^{(m_0)}$ is faithful because $\mathcal{W}$ is simple, so it follows that
 \begin{align}
     ||\underline{A}_\lambda||_\lambda = \sup_{\psi\in\mathcal{H}^{(m_0)}; 
||\psi|| = 1} ||\pi^{(m_0)}(\underline{A}_\lambda)\psi||.
 \end{align}
 Thus, for fixed $\underline{A}\in \widetilde{\underline{\mathcal{W}}}_v^{(m)}$, the map
\begin{align}
    \lambda\mapsto ||\underline{A}_\lambda||_\lambda
\end{align}
is a pointwise supremum of continuous functions, and hence is lower semicontinuous.
 
Since the vacuum states $\omega^{(\lambda m)}$ and $\omega^{(m_0)}$ are locally normal \parencite{EcFr74}, we can also consider elements of $\mathcal{A}^{(\lambda m)}(\mathcal{O})$ in the mass $m_0$ vacuum representation by employing the quasiequivalence of local algebras $\pi^{(m_0),\lambda}_\mathcal{O}: \overline{\mathcal{W}^{(\lambda m)}}(\mathcal{O})\to \overline{\mathcal{W}^{(m_0)}(\mathcal{O})}$, where the bar denotes the weak operator closure.  Since $\mathcal{A}^{(\lambda m)}\subseteq \overline{\mathcal{W}^{(\lambda m)}}$ and $\mathfrak{R}^{(\lambda m)}\subseteq \overline{\mathcal{W}^{(\lambda m)}}$, for all $S^{(\lambda m)}(q,u)\in\mathcal{A}^{(\lambda m)}$ and $K^{(\lambda m)}(h_\lambda)\in \mathfrak{R}^{(\lambda m)}$, we can choose a large enough region $\mathcal{O}$ to find 
\begin{align}
\label{eqn:integral S sections}
    \pi^{(m_0),\lambda}_\mathcal{O}(S^{(\lambda m)}(q,u)) &= \int q(x) \pi^{(m_0),\lambda}_\mathcal{O}\big(\alpha^{(\lambda m)}_{1,x} W(u)\big)d^{1+s}x\\
    \pi^{(m_0),\lambda}_\mathcal{O}(K^{(\lambda m)}(h_\lambda)) &= \int_{E^{(\lambda m)}_{h_\lambda}} h_\lambda(u) \pi^{(m_0),\lambda}_\mathcal{O} (W(u))d\mu_{E^{(\lambda m)}_{h_\lambda}}.
\end{align}
A straightforward calculation using the Weyl relations and an application of the dominated convergence theorem with strong continuity in Eq.~\eqref{eqn:strongcontinuity} shows that for any $\underline{A}$ in $\widetilde{\mathcal{A}}_v^{(m)}$ or $\widetilde{\mathfrak{R}}_v^{(m)}$, the map
 \begin{align}
     \lambda\mapsto ||\pi^{(m_0),\lambda}_\mathcal{O}(\underline{A}_\lambda)\psi||
 \end{align}
 is continuous as well, where $\underline{A}_\lambda$ is taken to belong to $\mathcal{A}^{(\lambda m)}(\mathcal{O})$ or $\mathfrak{R}^{(\lambda m)}(\mathcal{O})$.

Likewise, since $\pi_\mathcal{O}^{(m_0)}$ is faithful, 
 \begin{align}
     ||\underline{A}_\lambda||_\lambda = \sup_{\psi\in\mathcal{H}^{(m_0)}; 
||\psi|| = 1} ||\pi^{(m_0),\lambda}_\mathcal{O}(\underline{A}_\lambda)\psi||.
 \end{align}
 Thus, for a fixed section $\underline{A}$ in $ \widetilde{\underline{\mathcal{A}}}_v^{(m)}$ or $\widetilde{\mathfrak{R}}_v^{(m)}$, we also find that the map
\begin{align}
    \lambda\mapsto ||\underline{A}_\lambda||_\lambda
\end{align}
is a pointwise supremum of continuous functions, and hence is lower semicontinuous.

\end{proof}

Now we may generate C*-algebras of sections of lower semicontinuous bundles.  Define \begin{align}
    \underline{\mathcal{W}}_v^{(m)} &= \overline{\text{span}\{ f\underline{A}\ |\ f\in C([0,1]), \underline{A}\in \underline{\mathcal{W}}_v^{(m)}\}}\\
     \underline{\mathcal{A}}_v^{(m)} &= \overline{\text{span}\{ f\underline{A}\ |\ f\in C([0,1]), \underline{A}\in \underline{\mathcal{A}}_v^{(m)}\}}\\
     \underline{\mathfrak{R}}_v^{(m)} &= \overline{\text{span}\{ f\underline{A}\ |\ f\in C([0,1]), \underline{A}\in \underline{\mathfrak{R}}_v^{(m)}\}},
\end{align}
where the bar denotes closure in the $\sup$ norm over $\lambda\in [0,1]$.  Note that norms are nonnegative and uniform limits of lower semicontinuous functions are lower semicontinuous.  Clearly, it follows that for each section $\underline{A}$ in $\underline{\mathcal{W}}_v^{(m)}$, $\underline{\mathcal{A}}_v^{(m)}$, or $\underline{\mathfrak{R}}_v^{(m)}$, the map $\lambda\mapsto ||\underline{A}_\lambda||_\lambda$ is also lower semicontinuous.  We summarize the result in the following statement.

\begin{cor}
\label{cor:existence and uniqueness of variable mass bundle}
    There are lower semicontinuous bundles of C*-algebras $(\underline{\mathcal{W}}_v^{(m)},\chi,[0,1])$,  $(\underline{\mathcal{A}}_v^{(m)},\chi,[0,1])$, and $(\underline{\mathfrak{R}}_v^{(m)},\chi,[0,1])$.
\end{cor}
\noindent We call these the \emph{variable mass bundles} because the generating sections involve Poincar\'e automorphisms with varying mass $\lambda m$ for $\lambda\in [0,1]$.  We will next proceed to use the variable mass bundles to construct scaling bundles.


Before we move on to the construction of scaling bundles, we note that the variable mass bundles $\underline{\mathcal{W}}^{(m)}_v$ and $\underline{\mathfrak{R}}^{(m)}_v$ are merely lower semicontinuous bundles, and \emph{not} uniformly continuous bundles; the same applies to the scaling bundles we will construct from them.  We take a moment to explicitly show that there are sections that do not have continuous norms.  The first result for $\underline{\mathcal{W}}_v^{(m)}$ follows directly from the fact that the map $u\mapsto W(u)$ is not continuous in the norm on the Weyl algebra because $||W(u) - W(v)|| = 2$ whenever $u\neq v$.  

\begin{prop}
    There is a section $\underline{A}\in \underline{\mathcal{W}}_v^{(m)}$ such that the map $\lambda\mapsto ||\underline{A}_\lambda||_\lambda$ is \emph{not} continuous.
\end{prop}

\begin{proof}
    Take a fixed $\lambda_0\in [0,1]$, nonzero $u\in E$, and Poincar\'e transformation $(\Lambda,x)$.  Define $v = \tau^{(\lambda_0 m)}_{(\Lambda,x)^{-1}}(u)\neq 0$.  Consider the section
    \begin{align}
        \underline{A}_\lambda = \alpha^{(\lambda m)}_{1,0}(W(v)) - \alpha^{(\lambda m)}_{\Lambda,x}(W(u))
    \end{align}
    for all $\lambda\in [0,1]$.  Clearly, at $\lambda=\lambda_0$, we have
    \begin{align}
        ||\underline{A}_{\lambda_0}||_{\lambda_0} = ||W(v) - W(\tau^{(\lambda_0 m)}_{(\Lambda,x)^{-1}}u)|| = ||W(v) - W(v)|| = 0.
    \end{align}
   But for any $\lambda\neq 0$, we have
    \begin{align}
        ||\underline{A}_\lambda||_\lambda = ||W(v) - W(\tau^{(\lambda m)}_{(\Lambda,x)^{-1}}u)|| = 2,
    \end{align}
    since $v\neq \tau^{(\lambda m)}_{(\Lambda,x)^{-1}}u$.
\end{proof}

With slightly more work, one can show that $\underline{\mathfrak{R}}^{(m)}_v$ also fails to be a uniformly continuous bundle.  To see this, we follow a similar argument, now using the fact that for a section defined by $\lambda \mapsto \alpha^{(\lambda m)}_t(K^{(\lambda m)}(h))$, the base of the associated function $h\circ \tau^{(\lambda m)}_{-t}$ varies in $E$ as $\lambda$ varies.  This essentially allows for an appropriate choice of vector spoiling the continuity of the norm of some algebraic combinations.

\begin{prop}
    There is a section $\underline{A}\in \underline{\mathfrak{R}}^{(m)}_v$ such that the map $\lambda\mapsto ||\underline{A}_\lambda||_\lambda$ is \emph{not} continuous.
\end{prop}

\begin{proof}
    Take a fixed $\lambda_0\in [0,1]$, $t>0$ and nonzero $f\in C_c^\infty(\Sigma)$.  Consider any $h\in A_c^\infty(E)$ whose base is $F:=\text{span}\{(f,0),(0,f)\}$.  Define $h' := h\circ \tau_t^{(\lambda_0 m)}$, and consider the sections
    \begin{align}
        \underline{B}_\lambda = \alpha^{(\lambda m)}_{1,0}(K^{(\lambda m)}(h_\lambda)) && \underline{C}_\lambda = \alpha^{(\lambda m)}_t(K^{(\lambda m)}(h'_\lambda)).
    \end{align}
    for all $\lambda\in [0,1]$.  Since $\underline{\mathfrak{R}}_v^{(m)}$ must be closed under sums, we can form the section
    \begin{align}
        \underline{A}_\lambda = \underline{B}_\lambda - \underline{C}_\lambda = \alpha^{(\lambda m)}_{1,0}(K^{(\lambda m)}(h_\lambda)) - \alpha^{(\lambda m)}_t(K^{(\lambda m)}(h'_\lambda)),
    \end{align}
    which must belong to $\underline{\mathfrak{R}}_v^{(m)}$.  At $\lambda =\lambda_0$, we have
    \begin{align}
        ||\underline{A}_{\lambda_0}||_{\lambda_0} = ||K^{(\lambda_0 m)}(h_{\lambda_0}) - K^{(\lambda_0 m)}(h'_{\lambda_0}\circ \tau^{(\lambda_0 m)}_{-t})||_{\lambda_0} = || K^{(\lambda_0 m)}(h_{\lambda_0}) - K^{(\lambda_0 m)}(h_{\lambda_0})||_{\lambda_0} = 0. 
    \end{align}
    On the other hand, we will establish a positive lower bound for $||\underline{A}_{\lambda}||_\lambda$ when $\lambda\neq 0$.

    Notice that the base of $h_\lambda$ appearing in $\underline{B}_\lambda$ is $E_{h_\lambda}^{(\lambda m)} = E^{(m)}_{h} = F$ for each $\lambda\in [0,1]$.  On the other hand, the base of $h'_\lambda\circ \tau_{-t}^{(\lambda m)}$ appearing in $\underline{C}_\lambda$ is
    \begin{align}
        E^{(\lambda m)}_{h'_\lambda} = \tau_t^{(\lambda m)}\circ\tau_{-t}^{(\lambda_0 m)}\big(F\big).
    \end{align}
    Whenever $\lambda\neq \lambda_0$, we claim that $\tau_{t}^{(\lambda m)}\circ\tau_{-t}^{(\lambda_0 m)}\big(F\big)\neq F.$  To see this, suppose, for contradiction, that $\tau_{t}^{(\lambda m)}\circ\tau_{-t}^{(\lambda_0 m)}\big(F\big)= F.$  Then considering the basis $\{(f,0),(0,f)\}$, we must have
    \begin{align}
        \tau_{t}^{(\lambda m)}\circ\tau_{-t}^{(\lambda_0 m)}(f,0) &= (a_1f,a_2 f)\\ \tau_{t}^{(\lambda m)}\circ\tau_{-t}^{(\lambda_0 m)}(0,f) &= (b_1f,b_2f)
    \end{align}
    for some constants $a_1,a_2,b_1,b_2\in\R$.  It would then follow that we can divide by $f$ on $\supp f$ to find
    \begin{align}
        \tau_{t}^{(\lambda m)}\circ\tau_{-t}^{(\lambda_0 m)}(1_{\supp f},0) &= (a_1\cdot 1_{\supp f},a_2 \cdot 1_{\supp f})\\ 
        \tau_{t}^{(\lambda m)}\circ\tau_{-t}^{(\lambda_0 m)}(0,1_{\supp f}) &= (b_1\cdot 1_{\supp f},b_2\cdot 1_{\supp f}),
    \end{align}
    where we use $1_{\supp f}$ to denote the constant function assigning the value $1$ and supported in the open region $\supp f$.  Taking Fourier transforms, one sees that $\tau_{t}^{(\lambda m)}\circ\tau_{-t}^{(\lambda_0 m)}$ would need to act like a constant function of $p$ on some open set, which is impossible because the components of the Fourier transform of $\tau_{t}^{(\lambda m)}\circ\tau_{-t}^{(\lambda_0 m)}$ are real-analytic functions of $|p|^2$.  If they were constant on any open set, they would need to be constant everywhere, which clearly yields a contradiction.  We conclude that $\tau_{t}^{(\lambda m)}\circ\tau_{-t}^{(\lambda_0 m)}(F)\neq F$ when $\lambda\neq \lambda_0$.

    Now, letting $F^{\sigma\perp}$ denote the symplectic orthogonal complement of $F$, we define a map $T_\lambda: F^{\sigma\perp}\to F^*$ by
    \begin{align}
    T_\lambda(v)(u) = \sigma(v,\tau_{t}^{(\lambda m)}\circ\tau_{-t}^{(\lambda_0 m)}(u))
    \end{align}
    for all $v\in F^{\sigma\perp}$ and $u\in F$.  Suppose, for contradiction, that $T_\lambda = 0$.  Then we would have $\tau_{t}^{(\lambda m)}\circ\tau_{-t}^{(\lambda_0 m)}(u)\in (F^{\sigma\perp})^{\sigma\perp} = F$ for all $u\in F$, and so the fact that $F$ is finite-dimensional and $\tau_{t}^{(\lambda m)}\circ\tau_{-t}^{(\lambda_0 m)}$ is invertible would imply $\tau_{t}^{(\lambda m)}\circ\tau_{-t}^{(\lambda_0 m)}(F) =F$, contradicting the previous paragraph.  Thus, $T_\lambda\neq 0$, and hence we can choose for each $\lambda\neq \lambda_0$ a vector, denoted $v_\lambda\in F^{\sigma\perp}$, such that $T_\lambda(v_\lambda)\neq 0$.  We normalize so that $||T_\lambda(v_\lambda)|| = 1$ in $F^*$.

    Now we consider the unitaries $U_\lambda := \pi^{(\lambda m)}(W(v_\lambda))$, and note that since $v_\lambda\in F^{\sigma\perp}$, it follows that $U_\lambda$ commutes with $\pi^{(\lambda m)}(W(u))$ for all $u\in F$, and hence $U_\lambda \underline{B}_\lambda U_\lambda^* = \underline{B}_\lambda$.  We use the elementary bound
    \begin{align}
        ||U_\lambda \underline{A}_\lambda U_\lambda^* - \underline{A}_\lambda||_\lambda\leq 2 ||\underline{A}_\lambda||_\lambda,
    \end{align}
    which implies
    \begin{align}
    \label{eqn:section norm lower bound}
        ||U_\lambda \underline{C}_\lambda U_\lambda^* - \underline{C}_\lambda||_\lambda = ||U_\lambda \underline{B}_\lambda U_\lambda^*-U_\lambda \underline{A}_\lambda U_\lambda^* - (\underline{B}_\lambda-\underline{A}_\lambda)||_\lambda \leq 2 ||\underline{A}_\lambda||_\lambda.
    \end{align}
    Using the fact that $\underline{C}_\lambda = \alpha_t^{(\lambda m)}\circ \alpha_{-t}^{(\lambda_0 m)}(\underline{B}_\lambda)$, we find that the left hand side of Eq.~\eqref{eqn:section norm lower bound} is
    \begin{align}
        ||U_\lambda &\underline{C}_\lambda U_\lambda^* - \underline{C}_\lambda||_\lambda = ||\alpha_{t}^{(\lambda_0 m)}\circ \alpha_{-t}^{(\lambda m)}(U_\lambda \underline{C}_\lambda U_\lambda^* - \underline{C}_\lambda)||_\lambda\nonumber \\
        &=|| \pi^{(\lambda m)}\big(W\big(\tau_{t}^{(\lambda_0 m)}\circ \tau_{-t}^{(\lambda m)}(v_\lambda)\big)\big)\underline{B}_\lambda \pi^{(\lambda m)}\big(W\big(\tau_{t}^{(\lambda_0 m)}\circ \tau_{-t}^{(\lambda m)}(v_\lambda)\big)^*\big) - \underline{B}_\lambda||_\lambda, 
    \end{align}
    and the Weyl relations imply
    \begin{align}
        W\big(\tau_{t}^{(\lambda_0 m)}\circ \tau_{-t}^{(\lambda m)}(v_\lambda)\big)W(u)W\big(\tau_{t}^{(\lambda_0 m)}\circ \tau_{-t}^{(\lambda m)}(v_\lambda)\big)^*&= e^{-i\sigma\big(\tau_{t}^{(\lambda_0 m)}\circ \tau_{-t}^{(\lambda m)}(v_\lambda),u\big)}W(u)\nonumber\\& = e^{-iT_\lambda(v_\lambda)(u)}W(u)
    \end{align}
    so that
    \begin{align}
        \pi^{(\lambda m)}\big(W\big(\tau_{t}^{(\lambda_0 m)}\circ \tau_{-t}^{(\lambda m)}(v_\lambda)\big)\big)\underline{B}_\lambda \pi^{(\lambda m)}\big(W\big(\tau_{t}^{(\lambda_0 m)}\circ \tau_{-t}^{(\lambda m)}&(v_\lambda)\big)^*\big) - \underline{B}_\lambda\nonumber\\
        &= K^{(\lambda m)}\big((e^{-iT_\lambda(v_\lambda)} - 1) h_\lambda\big).
    \end{align}
    Thus, Eq.~\eqref{eqn:section norm lower bound} becomes
    \begin{align}
    \label{eqn: refined section norm lower bound}
        2||\underline{A}_\lambda||_\lambda \geq ||K^{(\lambda m)}\big((e^{-iT_\lambda(v_\lambda)} - 1)\big) h_\lambda||_\lambda.
    \end{align}
    
    Clearly, for any $\ell\in F^*$, the function $(e^{-i\ell}-1)h_\lambda$ on $F$ is not identically zero whenever $\ell\neq 0$.  So one can see from the finite-dimensional Schr\"odinger representation $\pi^{S,\lambda m}_F$ that \begin{align}
        G(\ell):=||K^{(\lambda m)}\big((e^{-i\ell} - 1)\big) h_\lambda||_\lambda > 0
    \end{align}
    for all $\ell \in F^*$ with $||\ell|| = 1$.  Moreover, $G$ is continuous because
    \begin{align}
        |G(\ell)-G(\ell')| &\leq ||K^{(\lambda m)}\big((e^{-i\ell}-e^{-i\ell'})h_\lambda\big)||_\lambda\nonumber\\
        &\leq (2\pi)^{\dim F/4}||(e^{-i\ell}-e^{-i\ell'})h_\lambda||_{L^2(F)}
    \end{align}
    approaches zero whenever $\ell'\to \ell$ by the dominated convergence theorem on $F$.  Since the Banach-Alaoglu theorem implies that the unit ball in $F^*$ is compact, it follows that
    \begin{align}
        c := \min_{\ell\in F^*; ||\ell|| = 1} G(\ell)>0,
    \end{align}
    and one can easily see from the definition of $h_\lambda$ that $c$ is independent of $\lambda$.  Thus, Eq.~\eqref{eqn: refined section norm lower bound} implies that
    \begin{align}
        ||\underline{A}_\lambda||_\lambda\geq c/2
    \end{align}
    for all $\lambda\neq \lambda_0$.  Therefore, $\lambda\mapsto ||\underline{A}_\lambda||_\lambda$ is discontinuous at $\lambda_0$ where $||\underline{A}_{\lambda_0}||_{\lambda_0} = 0$.
\end{proof}

\subsection{Scaling bundles for the massive free scalar field}
\label{subsec:scaling bundle for the massive free scalar field}

We now construct lower semicontinuous scaling bundles for the scalar field via the variable mass bundles.  More precisely, we define the C*-algebras of sections
\begin{align}
    \underline{\mathcal{W}}_s^{(m)} &= \left\{[\lambda\mapsto \sigma_\lambda(\underline{A}_\lambda)]\ |\ \underline{A}\in \underline{\mathcal{W}}_v^{(m)}\right\}\\
    \underline{\mathcal{A}}_s^{(m)} &= \left\{[\lambda\mapsto \sigma_\lambda(\underline{A}_\lambda)]\ |\ \underline{A}\in \underline{\mathcal{A}}_v^{(m)}\right\}\\
       \underline{\mathfrak{R}}_s^{(m)} &= \left\{[\lambda\mapsto \sigma_\lambda(\underline{A}_\lambda)]\ |\ \underline{A}\in \underline{\mathfrak{R}}_v^{(m)}\right\}.
\end{align}
Clearly, we have $\widetilde{\underline{\mathcal{W}}}_s^{(m)}\subseteq \underline{\mathcal{W}}_s^{(m)}$, $\widetilde{\underline{\mathcal{A}}}_s^{(m)}\subseteq \underline{\mathcal{A}}_s^{(m)}$, and $\widetilde{\underline{\mathfrak{R}}}_s^{(m)}\subseteq \underline{\mathfrak{R}}_s^{(m)}$.  For every section $\underline{A}$ in one of these algebras, we denote the evaluation map $\underline{A}\mapsto \underline{A}_\lambda$ by $\phi_\lambda$ for $\lambda\in (0,1]$.  Since $\sigma_\lambda$ is a *-algebra, it is isometric in the C*-norms on the fibers.  It follows that $\lambda\mapsto ||\underline{A}_\lambda||_\lambda$ is lower semicontinuous.  In other words, by Cor. \ref{cor:existence and uniqueness of variable mass bundle}, we have the following result.
\begin{thm}
        There are lower semicontinuous bundles of C*-algebras $(\underline{\mathcal{W}}_s^{(m)},\phi,(0,1])$,  $(\underline{\mathcal{A}}_s^{(m)},\phi,(0,1])$, and $(\underline{\mathfrak{R}}_s^{(m)},\phi,(0,1])$.
\end{thm}

Further, each bundle $(\underline{\mathcal{W}}_s^{(m)},\phi,(0,1])$,  $(\underline{\mathcal{A}}_s^{(m)},\phi,(0,1])$, and $(\underline{\mathfrak{R}}_s^{(m)},\phi,(0,1])$ carries the structure of a bundle of Poincar\'e nets by the following definitions.  First, we define a net structure by the conditions
\begin{align}
    \underline{A}\in \underline{\mathcal{W}}_s^{(m)}(\mathcal{O})&\text{ iff }\phi_\lambda(\underline{A})\in\underline{\mathcal{W}}_s^{(m)}(\lambda \mathcal{O})\text{ for every }\lambda\in (0,1]\\
    \underline{A}\in \underline{\mathcal{A}}_s^{(m)}(\mathcal{O})&\text{ iff }\phi_\lambda(\underline{A})\in\underline{\mathcal{A}}_s^{(m)}(\lambda \mathcal{O})\text{ for every }\lambda\in (0,1] \\
    \underline{A}\in \underline{\mathfrak{R}}_s^{(m)}(\mathcal{O})&\text{ iff }\phi_\lambda(\underline{A})\in\underline{\mathfrak{R}}_s^{(m)}(\lambda \mathcal{O})\text{ for every }\lambda\in (0,1].
\end{align}
Note that the generating sections for each bundle are local because $\sigma_\lambda$ rescales the support of each test function appropriately.

For every Poincar\'e transformation $(\Lambda,x)$, we define automorphisms $\underline{\alpha}_{\Lambda,x}^{(m),s}$ on $(\underline{\mathcal{W}}_s^{(m)},\phi,(0,1])$,  $(\underline{\mathcal{A}}_s^{(m)},\phi,(0,1])$, and $(\underline{\mathfrak{R}}_s^{(m)},\phi,(0,1])$, which we denote by the same symbol.  In each case, the automorphism is defined by
\begin{align}
   \phi_\lambda\big( \underline{\alpha}_{\Lambda,x}^{(m),s}(\underline{A})\big) = \alpha_{\Lambda,\lambda x}^{(m)}\big(\phi_\lambda(\underline{A})\big).
\end{align}
for each section $\underline{A}$ coming from the appropriate algebra.  It is easy to see that the maps $\underline{\alpha}_{\Lambda,x}^{(m),s}$ are well defined *-automorphisms.

\begin{prop}
    The structures $(\underline{\mathcal{W}}_s^{(m)},\underline{\alpha}^{(m),s},\phi,(0,1])$,  $(\underline{\mathcal{A}}_s^{(m)},\underline{\alpha}^{(m),s},\phi,(0,1])$, and \\$(\underline{\mathfrak{R}}_s^{(m)},\underline{\alpha}^{(m),s},\phi,(0,1])$ are lower semicontinuous scaling bundles.
\end{prop}

\begin{proof}
    The argument is identical for the bundles $\underline{\mathfrak{A}}_s^{(m)} := \underline{\mathcal{W}}_s^{(m)}, \underline{\mathcal{A}}_s^{(m)}$, or $\underline{\mathfrak{R}}_s^{(m)}$, so we treat all three cases at once.  We further denote the fibers by $\mathfrak{A}^{(m)} = \mathcal{W}^{(m)}, \mathcal{A}^{(m)}$, or $\mathfrak{R}^{(m)}$.  The assignment $\mathcal{O}\mapsto \underline{\mathfrak{A}}_s^{(m)}(\mathcal{O})$ is clearly a Poincar\'e net since each fiber is a Poincar\'e net with the assignment $\mathcal{O}\mapsto \mathfrak{A}^{(m)}(\lambda \mathcal{O})$ in the fiber over $\lambda\in (0,1]$.  So it only remains for us to establish that, for each $\lambda\in (0,1]$, $\phi_\lambda: \underline{\mathfrak{A}}_s^{(m)}\to \mathfrak{A}^{(m)}$ is a net epimorphism.

    Note that the property
    \begin{align}
        \phi_\lambda\circ \underline{\alpha}_{\Lambda,x}^{(m),s} = \alpha_{\Lambda,\lambda x}\circ \phi_\lambda
    \end{align}
    is satisfied by construction.  Moreover, for any local region $\mathcal{O}$,
    \begin{align}
        \phi_\lambda\left(\underline{\mathfrak{A}}^{(m)}_s(\mathcal{O})\right) = \phi_\lambda\left(\bigcap_{\mu\in (0,1]} \phi_\mu^{-1}\left(\mathfrak{A}^{(m)}(\mu\mathcal{O})\right)\right)\subseteq \phi_\lambda\left(\phi_\lambda^{-1}(\mathfrak{A}^{(m)}(\lambda\mathcal{O}))\right)= \mathfrak{A}^{(m)}(\lambda\mathcal{O}).
    \end{align}
    On the other hand, $\mathfrak{A}^{(m)}(\lambda\mathcal{O})$ has a dense subset given by linear combinations of the generators $\alpha_{\Lambda,\lambda x}^{(m)}(A)$, where $A = W(u), S^{(m)}(q,u)$, or $K^{(m)}(h_\lambda)$. Since $\sigma_\lambda$ is a *-isomorphism on each fiber, we must have that $A = \sigma_\lambda(B)$ for some $B\in\mathfrak{A}^{(\lambda m)}$ in the corresponding fiber of the variable mass bundle.  Moreover, since the evaluation map $\chi_\lambda$ in the variable mass bundle is surjective, there is a section $\underline{B}$ in the variable mass bundle with $\underline{B}_\lambda = B$.  Hence the section defined by $\underline{A}_\lambda = \alpha_{\Lambda,\lambda x}^{(m)}\circ\sigma_\lambda(\underline{B}_\lambda) = \sigma_\lambda\circ \alpha^{(\lambda m)}_{\Lambda,x}(\underline{B}_\lambda)$ yields $\underline{A}\in\underline{\mathfrak{A}}_s^{(m)}$.  It follows from  the localization properties of $\sigma_\lambda$ that $\phi_\lambda(\underline{A}) \in \mathfrak{A}^{(m)}(\lambda\mathcal{O})$.  In other words, $\mathfrak{A}^{(m)}(\lambda\mathcal{O})\subseteq \phi_\lambda(\underline{\mathfrak{A}}_s^{(m)}(\mathcal{O}))$, which establishes that $\phi_\lambda(\underline{\mathfrak{A}}_s^{(m)}(\mathcal{O})) = \mathfrak{A}^{(m)}(\lambda\mathcal{O})$.  Therefore, $\phi_\lambda$ is a net epimorphism, and we conclude that $\underline{\mathfrak{A}}^{(m)}_s$ is a scaling bundle.
\end{proof}

Now we are in a position to characterize the scaling limit of our lower semicontinuous scaling bundles.  We again denote the algebras of sections of the scaling bundles by $\underline{\mathfrak{A}}^{(m)}_s := \underline{\mathcal{W}}_s^{(m)}, \underline{\mathcal{A}}_s^{(m)}$, or $\underline{\mathfrak{R}}_s^{(m)}$, and we now denote the corresponding algebra of sections of the variable mass bundles by $\underline{\mathfrak{A}}^{(m)}_v := \underline{\mathcal{W}}_v^{(m)}, \underline{\mathcal{A}}_v^{(m)}$, or $\underline{\mathfrak{R}}_v^{(m)}$.  We can always restrict the variable mass bundle $\underline{\mathfrak{A}}^{(m)}_v$ to the base space $(0,1]\subseteq [0,1]$ with the identical algebra of sections yet with the fiber $\chi_0(\underline{\mathfrak{A}}^{(m)}_v) = \mathfrak{A}^{(0)}$ removed.  We denote this restriction by $\underline{\mathfrak{A}}^{(m)}_v|_{(0,1]}$.  Then we can define a map $\underline{\sigma}: \underline{\mathfrak{A}}^{(m)}_v|_{(0,1]}\to \underline{\mathfrak{A}}^{(m)}_s$ between the algebras of sections induced by the fiberwise maps $\sigma_\lambda$ according to
\begin{align}
    \phi_\lambda(\underline{\sigma}(\underline{A})) = \sigma_\lambda(\chi_\lambda(\underline{A}))
\end{align}
for all $\underline{A}\in \underline{\mathfrak{A}}^{(m)}_v|_{(0,1]}$ and $\lambda\in (0,1]$.  By construction $(\operatorname{id},\underline{\sigma})$ is clearly a bundle isomorphism $(\underline{\mathfrak{A}}^{(m)}_v,\chi,(0,1])\to (\underline{\mathfrak{A}}^{(m)}_s,\phi,(0,1])$, where $\operatorname{id}: (0,1]\to (0,1]$ is the identity function on the restricted base space.

Moreover, $\underline{\mathfrak{A}}_v^{(m)}$ can be given a natural net bundle structure that makes $(\operatorname{id},\underline{\sigma})$ a net bundle isomorphism.  Specifically, define the net assignment for each local region $\mathcal{O}$ by
\begin{align}
    \underline{\mathfrak{A}}_v^{(m)}(\mathcal{O}) = \bigcap_{\lambda\in [0,1]}\chi_\lambda^{-1}\big(\mathfrak{A}^{(\lambda m)}(\mathcal{O})\big).
\end{align}
Generating sections for each variable mass bundle are again local according to these definitions.

Further, we define the Poincar\'e automorphisms $\underline{\alpha}_{\Lambda,x}^{(m),v}$ of $\underline{\mathfrak{A}}_v^{(m)}$ given by
\begin{align} \chi_\lambda\big(\underline{\alpha}_{\Lambda,x}^{(m),v}(\underline{A})\big) = \alpha_{\Lambda,x}^{(\lambda m)}\big(\chi_\lambda(\underline{A})\big)
\end{align}
for all $\underline{A}\in \underline{\mathfrak{A}}_v^{(m)}$.  The fact that for each $\lambda\in [0,1]$, the fiber $\mathfrak{A}^{(\lambda m)}$ is a Poincar\'e net implies that $\underline{\mathfrak{A}}_v^{(m)}$ is a net bundle.

It follows that $\underline{\sigma}$ is a net isomorphism from $\underline{\mathfrak{A}}_v^{(m)}|_{(0,1]}$ to $\underline{\mathfrak{A}}_s^{(m)}$.  This follows by construction since
\begin{align}
    \underline{\sigma}\big(\underline{\mathfrak{A}}^{(m)}_v|_{(0,1]}(\mathcal{O})\big) = \underline{\mathfrak{A}}_s^{(m)}(\mathcal{O})
\end{align}
for every local region $\mathcal{O}$, and
\begin{align}
    \underline{\sigma}\circ \underline{\alpha}_{\Lambda,x}^{(m),v} = \underline{\alpha}_{\Lambda,x}^{(m),s}\circ\underline{\sigma}
\end{align}
for all Poincar\'e transformations $(\Lambda,x)$.  We summarize the result as follows.

\begin{prop}
For each of the following pairs:
\begin{itemize}
    \item $\underline{\mathcal{W}}_v^{(m)}|_{(0,1]}$ and $\underline{\mathcal{W}}_s^{(m)}$,
    \item $\underline{\mathcal{A}}_v^{(m)}|_{(0,1]}$ and $\underline{\mathcal{A}}_s^{(m)}$,
    \item $\underline{\mathfrak{R}}_v^{(m)}|_{(0,1]}$ and $\underline{\mathfrak{R}}_s^{(m)}$,
\end{itemize}
the map $(\operatorname{id},\underline{\sigma})$ is a net bundle isomorphism.
\end{prop}

Recall that in each case for the variable mass bundle, the algebras of sections for the restricted and unrestricted bundles are identical in the sense that $\underline{\mathfrak{A}}_v^{(m)}|_{(0,1]} = \underline{\mathfrak{A}}_v^{(m)}$.  It follows that the map $(\operatorname{id},\underline{\sigma}^{-1})$ is a minimal extension of the lower semicontinuous bundle of C*-algebras $\underline{\mathfrak{A}}_s^{(m)}$ that also preserves the structure of the net bundle.  Let $\mathfrak{A}^{(m)}_0$ denote the $\lambda = 0$ fiber of the unique minimal extension of $\underline{\mathfrak{A}}_s^{(m)}$ satisfying (UP), obtained from Thm. \ref{thm:extension of net bundle}.  As $\mathfrak{A}^{(m)}_s$ ranges over $\underline{\mathcal{W}}_s^{(m)}, \underline{\mathcal{A}}_s^{(m)}$, and $\underline{\mathfrak{R}}_s^{(m)}$, we denote the $\lambda=0$ fibers $\mathfrak{A}^{(m)}_0$ by $\mathcal{W}^{(m)}_0$, $\mathcal{A}^{(m)}_0$, and $\mathfrak{R}^{(m)}_0$, respectively.  Now, it follows from (UP) that we have the following sense in which the $m=0$ net $\mathfrak{A}^{(0)}$ provides a scaling limit of the scaling bundle $\underline{\mathfrak{A}}_s^{(m)}$.
\begin{thm}
    For each of our scaling bundles $\underline{\mathcal{W}}_s^{(m)}, \underline{\mathcal{A}}_s^{(m)}$, and $\underline{\mathfrak{R}}_s^{(m)}$, there is a net epimorphism from the $\lambda = 0$ fiber of the unique minimal extension satisfying (UP) to the corresponding massless net.  Specifically,
    \begin{itemize}
        \item There is a net epimorphism from $\mathcal{W}^{(m)}_0$ to $\mathcal{W}^{(0)}$.
        \item There is a net epimorphism from $\mathcal{A}^{(m)}_0$ to $\mathcal{A}^{(0)}$.
        \item There is a net epimorphism from $\mathfrak{R}^{(m)}_0$ to $\mathfrak{R}^{(0)}$.
    \end{itemize}
\end{thm}

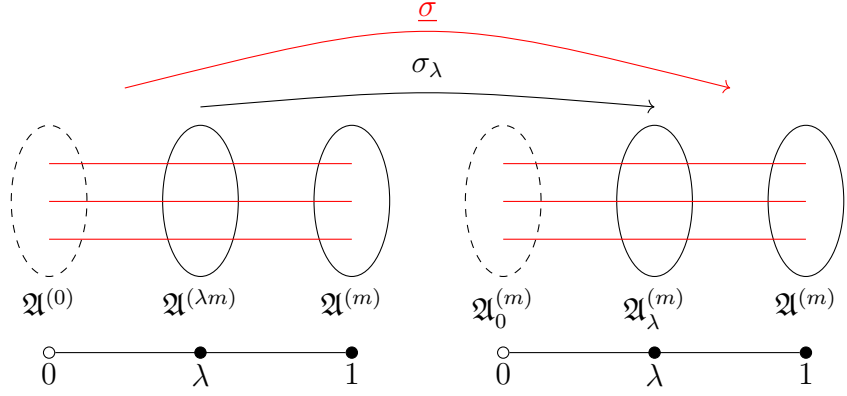
\begin{figure}
\centering
\begin{tikzpicture}
\node[circle,draw,inner sep=1.5pt] (b10) at (-5,-0.5) {};
\node[circle,draw, fill,inner sep=1.5pt] (b1l) at (-3,-0.5) {};
\node[circle,draw, fill,inner sep=1.5pt] (b11) at (-1,-0.5) {};
\node[anchor=north] at (b10) {$0$};
\node[anchor=north] at (b1l) {$\lambda$};
\node[anchor=north] at (b11) {$1$};
\node[circle,draw,inner sep=1.5pt] (b20) at (1,-0.5) {};
\node[circle,draw, fill,inner sep=1.5pt] (b2l) at (3,-0.5) {};
\node[circle,draw, fill,inner sep=1.5pt] (b21) at (5,-0.5) {};
\node[anchor=north] at (b20) {$0$};
\node[anchor=north] at (b2l) {$\lambda$};
\node[anchor=north] at (b21) {$1$};
\draw (b10) -- (b11);
\draw (b20) -- (b21);
\node[above=1cm, ellipse, draw, dashed, minimum height = 2cm, minimum width =1cm] at (b10) {};
\node[above=1cm, ellipse, draw, minimum height = 2cm, minimum width =1cm] at (b1l) {};
\node[above=1cm, ellipse, draw, minimum height = 2cm, minimum width =1cm] at (b11) {};
\node[above=1cm, ellipse, draw, dashed, minimum height = 2cm, minimum width =1cm] at (b20) {};
\node[above=1cm, ellipse, draw, minimum height = 2cm, minimum width =1cm] at (b2l) {};
\node[above=1cm, ellipse, draw, minimum height = 2cm, minimum width =1cm] at (b21) {};
\node[anchor=north] at (-5,0.5) {$ \mathfrak{A}^{(0)}$};
\node[anchor=north] at (-3,0.5) {$ \mathfrak{A}^{(\lambda m)}$};
\node[anchor=north] at (-1,0.5) {$ \mathfrak{A}^{(m)}$};
\node[anchor=north] at (1,0.5) {$\mathfrak{A}^{(m)}_0$};
\node[anchor=north] at (3,0.5) {$\mathfrak{A}^{(m)}_\lambda$};
\node[anchor=north] at (5,0.5) {$\mathfrak{A}^{(m)}$};
\draw[red] (-5,2) -- (-1,2);
\draw[red] (-5,1.5) -- (-1,1.5);
\draw[red] (-5,1) -- (-1,1);
\draw[red] (5,2) -- (1,2);
\draw[red] (5,1.5) -- (1,1.5);
\draw[red] (5,1) -- (1,1);
\node[color=red] at (0,4) {$\underline \sigma$};
\node[anchor=south] at (0,3) {$\sigma_\lambda$};
\draw[->,red] (-4,3) ..controls(0,4).. (4,3);
\draw[->] (-3,2.75) ..controls(0,3).. (3,2.75);
\end{tikzpicture}
\caption{The bundle isomorphism $\underline \sigma$ between the restriction of the variable mass bundle to $(0,1]$ and the scaling bundle.}
\end{figure}

\section{A Uniformly Continuous Bundle of Nonlocal Observables for the Scalar Field}
\label{sec:uniformly continuous scalar field}

In this section, we show that it is possible to construct families of rescaled observables for the real scalar field for which the map
\begin{align}
    \lambda\mapsto ||\underline{A}_\lambda||_\lambda
\end{align}
is actually uniformly continuous, thus leading to the construction of a uniformly continuous bundle of C*-algebras.  The cost of the construction we provide here is that we consider \emph{nonlocal} observables, so that the corresponding uniformly continuous bundle does not have a natural net bundle structure, and so the bundle does \emph{not} satisfy our definition of a scaling bundle.  On the other hand, the sections of the bundle have a natural interpretation in terms of rescalings arising from the very same dilation maps $\sigma_\lambda$ that we used to construct scaling bundles in \S\ref{sec:lower semicontinuous scalar field}, and so we believe they are appropriate for the analysis of the small-distance limit scaling limit.

\subsection{Anti-Wick Ordered Observables from Phase Space Smearing}

We begin by constructing a new algebra of nonlocal observables, whose properties we will analyze under rescalings.  To motivate the construction of this section, recall that the generating elements of $\mathfrak{R}^{(m)}$ consist in phase space smearings of Weyl operators that mimic Weyl quantization on finite-dimensional subspaces of $E$.  Generally, on a finite-dimensional linear symplectic phase space $\R^{2n}$, the Weyl quantization of a phase space function $h\in C_c^\infty(\R^{2n})$ is given by a weak operator integral in the Schr\"odinger representation $\pi^S$ of the Weyl algebra over $\R^{2n}$ by
\begin{align}
    K(h) = \int_{\R^{2n}} h(u) \pi^S(W(u)) d\mu(u),
\end{align}
where $\mu$ is the Lebesgue measure on $\R^{2n}$.  This integral form implements the Weyl ordering of the operator product, but this is not the only possible operator ordering.  It is well known that one can transform the integral formula to obtain the Berezin quantization, or Anti-Wick ordering, given by
\begin{align}
\label{eqn:berezin quantization finite dimensional}
    Q(h) = \int_{\R^{2n}} h(u)\pi^S(W(u)) e^{-||u||^2/4} d\mu(u),
\end{align}
where $||u||$ is the standard norm on $\R^{2n}$.  We would like to generalize the integral formula for Berezin quantization now to the infinite-dimensional phase space $E$ for the real scalar field.

As in \S\ref{sec:scalar field background}, one immediately encounters the challenge that there is no Lebesgue measure on the infinite-dimensional space $E$.  For Weyl quantization, we circumvented this issue by restricting attention to tame functions that depend on only a finite-dimensional subspace of $E$. This implicitly leveraged the fact that the Weyl quantization of a tame function $h$ acts like the identity operator on tensor factors corresponding to directions in $E$ orthogonal to the base of $h$.  The same does \emph{not} hold for Berezin quantization because the function $e^{-||u||^2/4}$ varies in all directions in the phase space, even outside the base of $h$.  So we pursue an alternative strategy of trying to make the weak operator integrals in Eq.~\eqref{eqn:berezin quantization finite dimensional} well-defined when the domain is generalized to $E$.  To do so, we note that while there is no Lebesgue measure on $E$, it \emph{is} possible to generalize the notion of a \emph{Gaussian measure} to $E$.  This is helpful because we can reexpress Berezin quantization with an integral formula relative to a Gaussian measure
\begin{align}
\label{eqn:berezin quantization finite dimensions}
    Q(h) = \int_{\R^{2n}} h(u) \pi^S(W(u)) e^{||u||^2/4}d\gamma_{2n}(u)
\end{align}
with $d\gamma_{2n}(u) = e^{-||u||^2/2} d\mu(u)$ the Gaussian measure with covariance determined by the inner product.  It is this formula that we will generalize to obtain well-defined Anti-Wick ordered observables by phase space smearings of Weyl unitaries for the real scalar field.

With this motivation in mind from finite-dimensional phase spaces, we return to the real scalar field for the remainder of this section with phase space $E$, inner product $\langle\cdot,\cdot\rangle_m$, and Weyl algebra $\mathcal{W}^{(m)}$. We define $\gamma^{(m)}$ as the isonormal distribution on $\mathfrak{h}^{(m)}$ whose covariance matches the real inner product $\Re(\langle\cdot,\cdot\rangle_m)$, as constructed by \textcite[][]{baezIntroductionAlgebraicConstructive1992}.  The isonormal distribution is a \emph{generalized Gaussian distribution} on an infinite-dimensional space.  It is important to note that $\gamma^{(m)}$ is \emph{not} a genuine measure because its covariance operator (the identity) is not trace-class.  Rather, $\gamma^{(m)}$ gives us an abstract notion of integration by defining a linear functional assigning bounded tame functions $h:E\to\mathbb{C}$ to their expectations taken relative to the Gaussian distribution, where we continue to denote expectations by
\begin{align}
     \int_E h(u) d\gamma^{(m)}(u)
\end{align}
While $\gamma^{(m)}$ is \emph{not} a genuine measure, there is a so-called Wiener space, a (non-unique) measure space $(X,\mathbb{P})$ with an injection $\iota: \mathfrak{h}^{(m)}\to X$ allowing for the development of a traditional theory of integration and $L^p$ spaces \parencite{baezIntroductionAlgebraicConstructive1992,AmJaNo15}.  For $p>0$, the collection of bounded, tame functions on $E$ is $L^p$ dense in $L^p(E)$.  The isonormal distribution $\gamma^{(m)}$ is uniquely picked out by the condition that the projection of $\gamma^{(m)}$ onto any finite-dimensional subspace $F\subseteq \mathfrak{h}^{(m)}$ has Gaussian density
\begin{align}
    G_F(u) = (2\pi)^{-\dim F/2} e^{-\langle u,u\rangle_m/2}.
\end{align}
with respect to the Lebesgue measure on $F$.  The isonormal distribution will play a crucial role in what follows, and we refer the reader to the literature for further background \parencite{baezIntroductionAlgebraicConstructive1992,daleckijMeasuresDifferentialEquations1991}.

We consider the nonlocal observables defined formally by
\begin{align}
\label{eqn:definition Berezin quantization scalar field}
    Q^{(m)}(h) = \int_E h(u)\pi^{(m)}(W(u))e^{\langle u,u\rangle_m/4}d\gamma^{(m)}(u)
\end{align}
for $h\in A^0(E)$.  Note that this is a direct infinite-dimensional generalization of the formula in Eq.~\eqref{eqn:berezin quantization finite dimensions} for Berezin quantization in finite-dimensional phase spaces.  Even when $h$ is tame, the function $e^{\langle u,u\rangle_m/4}$ varies in all directions in phase space, and so we must say more to ensure Eq.~\eqref{eqn:definition Berezin quantization scalar field} is well-defined.  We understand the integral as a weak operator limit over the net of increasing finite-dimensional subspaces $F\subseteq E$.  We take the weak operator integrals
\begin{align}
\label{eqn:definition of berezing quantization finite}
    Q^{(m)}_F(h) &= \int_F h(u)\pi^{(m)}(W(u))e^{\langle u,u\rangle_m/4}d\gamma^{(m)}_{|F}(u) \\
    &= (2\pi)^{-\dim F/2}\int_F h(u)\pi^{(m)}(W(u))e^{-\langle u,u\rangle_m/4}d\mu_F(u),
\end{align}
where $\gamma_{|F}$ is the projection of $\gamma$ onto $F$ and $d\mu_F(u)$ is the $\dim F$-dimensional Lebesgue measure.  We will sometimes denote the Lebesgue measure on $F$ below by $du:= d\mu_F(u)$ for notational simplicity.  One can follow precisely the same argument we gave previously for the weak operator integral defining $K^{(m)}(h)$ to establish that $Q_F^{(m)}(h)$ exists.
\begin{prop}
    For $h\in A^0(E)$, the weak operator integral in Eq.~\eqref{eqn:definition of berezing quantization finite} exists.
\end{prop}
\begin{proof}
    Suppose $h\in A^0(E)$.  Again, the map $u\mapsto \langle \varphi, \pi^{(m)}\left(W(u)\right)\psi\rangle$ is measurable for any $\varphi,\psi\in\mathcal{H}^{(m)}$, and so
\begin{align}
    \langle \varphi,Q_F^{(m)}(h)\psi\rangle &= (2\pi)^{-\dim F/2} \int_{F} h(u)e^{-\langle u,u\rangle_m/4} \langle \varphi, \pi^{(m)}\left(W(u)\right)\psi\rangle d\mu_{F}(u)\\
    &\leq (2\pi)^{-\dim F/2} ||\varphi||\cdot||\psi||\int_{F} h(u)e^{-\langle u,u\rangle_m/4} ||\pi^{(m)}\left(W(u)\right)||d\mu_{E^{(m)}_h}(u)\\
    &\leq (2\pi)^{-\dim F/2} ||\varphi||\cdot||\psi||\cdot||h'||_{L^1(F,\mu_{F})}
\end{align}
since $||W(u)|| = 1$, where $h'(u) := h(u)e^{-\langle u,u\rangle_m/4}$ is integrable because $h$ is bounded.  Thus $Q_F^{(m)}(h)$ determines a bounded sesquilinear form, and hence a bounded operator on $\mathcal{H}^{(m)}$.
\end{proof}
\noindent We shall prove below, moreover, that the weak operator limit $Q^{(m)}(h) := \lim_{F\to E} Q_F^{(m)}(h)$ exists so that Eq.~\eqref{eqn:definition Berezin quantization scalar field} indeed defines an operator on $\mathcal{H}^{(m)}$.

We restrict attention to $h\in A^0(E)$ such that $E^{(m)}_h\subseteq J_mE$ using $J_m$ from Eq.~\eqref{eqn:inner product}.  It is easy to see that this requirement implies there is a bounded, continuous function $\tilde{h}:\R^{n}\to \C$ and vectors $v_1,...,v_n\in E_h^{(m)}$  such that 
\begin{align}
    h(u) = \tilde{h}(\sigma(v_1,u),...,\sigma(v_n,u))
\end{align}
for all $u\in E$.

We denote the *-algebra generated by operators of the form $Q^{(m)}(h)$ by
\begin{align}
    \mathfrak{B}^{(m)} = \{ Q^{(m)}(h)\ |\ h\in A^0(E), E^{(m)}_h\subseteq J_m E\}.
\end{align}
Notice the condition that $E^{(m)}_h\subseteq J_m E$ guarantees that the base of $h$ is associated with a subspace $J_mE_h^{(m)}$ of the original space $E$ rather than its completion $\mathfrak{h}^{(m)}$ relative to the inner product of a given mass.

We can generate a useful representation of the Weyl algebra $\mathcal{W}^{(m)}$, and hence of $\mathfrak{B}^{(m)}$ by taking what \textcite[][\S1.11]{baezIntroductionAlgebraicConstructive1992} call the \emph{real wave representation}.  To define the representation, recall first the one-particle structure $\kappa_m: E\to L^2(\R^s;\C)$ from Eq.~\eqref{eqn: one particle structure}.  Consider the real part of the Hilbert space $G = L^2(\R^s;\R)$ with the $L^2$ inner product $\langle \cdot,\cdot \rangle$.  Form the isonormal distribution $\gamma$ associated with the real inner product $\langle\cdot,\cdot\rangle$.  The Hilbert space of the real wave representation is the space $\mathcal{H} = L^2(G,\gamma)$ with the representation:
\begin{align}
\big(\pi_G^{(m)}(W(f,g))\Psi\big)(\phi) = e^{i\langle f,g\rangle/2}e^{-i\langle \phi,\mu_m^{-1/2}f\rangle/\sqrt{2}}e^{\langle \phi,\mu_m^{1/2}g\rangle/\sqrt{2}}e^{-\langle g,\mu_m g\rangle/2} \Psi(\phi-\sqrt{2}\mu_m^{1/2}g)
\end{align}
for all $f,g\in E$, $\psi\in L^2(G,\gamma)$, and each $m\geq 0$.  It is well known that the real wave representation $\pi_G^{(m)}$ on $L^2(G,\gamma)$ is unitarily equivalent to the mass $m$ vacuum representation on $\mathcal{H}^{(m)}$, hence justifying our use of the notation $\pi_G^{(m)}$.  This means that for each $m$, there is a unitary $U_m: \mathcal{H}^{(m)}\to L^2(G,\gamma)$ such that $\pi^{( m)}_G = U_m \pi^{( m)}U_m^* $.  In what follows, we will also need to use the transported base of each $h\in A^0(E)$: we denote $G^{(m)}_h := \kappa_m\big (E^{(m)}_h\big)\oplus i\kappa_m\big (E^{(m)}_h\big)$.  Since $G^{(m)}_h$ is a complex-linear subspace, we also have a decomposition into real and imaginary parts $G^{(m)}_h = H_h^{(m)}\oplus i H_h^{(m)}$ for a real-linear subspace $H_h^{(m)}$.

We now establish that in the real wave representation, each operator $Q^{(\lambda m)}(h)$ is Hilbert-Schmidt.

\begin{prop}
    For $h\in A^0(E)$ with $E^{(m)}_h\subseteq J_m E$, the operator $U_m Q^{( m)}(h)U_m^*$ is an integral operator of the form
    \begin{align}
    \label{eqn:full Hilbert-Schmidt}
    (U_m Q^{(m)}(h)U_m^*\Psi)(\phi) = \int_G q^{(m)}_h(\phi',\phi) \Psi(\phi') d\gamma(\phi')
    \end{align}
    for all $\psi\in L^2(G,\gamma)$ with integral kernel
    \begin{align}
    \label{eqn:full kernel}
        q_h^{(m)}(\phi,\phi') = e^{||P_{h}(\phi + \phi')||^2/8}\int_{H_h^{(m)}} h\circ\kappa_m^{-1}\big(2^{1/2}f+i2^{-1/2}(\phi-\phi')\big)e^{-i\langle f,\phi+\phi'\rangle/2} d\gamma_{|H^{(m)}_h}(f)
    \end{align}
    satisfying $q_h^{(m)}\in L^2(G\times G,\gamma\times \gamma)$.  Here, $P_{h}: G\to H^{(m)}_h$ is the orthogonal projection onto $H^{(m)}_h$, $\gamma_{|H^{(m)}_h}$ is the projection of $\gamma$ onto $H^{(m)}_h$, and $||\cdot||^2$ is the norm associated with the inner product $\langle\cdot,\cdot\rangle$ on $G = L^2(\R^s;\R)$.
\end{prop}

\begin{proof}
    We consider first the net of finite dimensional subspaces $F\subseteq E$ in the weak operator integral in Eq.~\eqref{eqn:berezin quantization finite dimensions} defining $Q^{(m)}(h)$.  We restrict to subspaces $F$ with $J_m(F) = F$ so that we have $F = F_1\oplus J_m(F_1)$ for a subspace $F_1$ with $\gamma^{(m)}_{|F}(f,g) = \gamma^{(m)}_{|F_1}(f)\times \gamma^{(m)}_{|J_M(F_1)}(g)$. We will show
    \begin{align}
    \label{eqn:finite integral operator}
    (U_m Q^{(m)}_F(h)U_m^*\Psi)(\phi) = \int_{\kappa_m(F_1)} q^{(m)}_{h,F}(\phi',\phi) \Psi(\phi') d\gamma(\phi')
    \end{align}
    with 
    \begin{align}
    \label{eqn:finite kernel}
        q_{h,F}^{(m)}(\phi,\phi') = e^{||P_F(\phi+\phi')||^2/8}\int_{\kappa_m(F_1)} h\circ\kappa_m^{-1}\big(2^{1/2}f+i2^{-1/2}(\phi-\phi')\big)e^{-i\langle f,\phi+\phi'\rangle/2} d\gamma_{|\kappa_m(F_1)}(f).
    \end{align}
    Here we use $P_F:G\to \kappa_m(F_1)$ to denote the orthogonal projection onto $\kappa_m(F_1)$.
    Eq.~\eqref{eqn:finite kernel} follows by direct calculation using the fact that the pushforward under $\kappa_m$ of $\gamma^{(m)}$ is the fixed mass-independent distribution $\gamma^2:=\gamma\times \gamma$ on $L^2(\R^s;\C) = G\oplus i G$.  For any $\Psi,\Psi'\in L^2(G,\gamma)$, we can apply the Fubini-Tonelli theorem as well as a few changes of variables to find
    \begin{align}
        \langle  \Psi'&, U_m Q^{(m)}_F(h)U_m^*\Psi\rangle_\gamma = \int_{\kappa_m(F)} h(u)\left\langle \Psi', \pi^{(m)}_G(W(u))\Psi\right\rangle_\gamma e^{\langle u,u\rangle_m/4}d\gamma^2_{|\kappa_m(F)}(\kappa_m(u))\\
        &= \int_{\kappa_m(F)}\int_G h(f,g) \overline{\Psi'(\phi)}e^{i\langle f,g\rangle/2}e^{-i\langle \phi,\mu_m^{-1/2}f\rangle/\sqrt{2}}e^{\langle \phi,\mu_m^{1/2}g\rangle/\sqrt{2}}e^{-\langle g,\mu_m g\rangle/2}\nonumber\\
        &\hspace{1in} \times e^{\langle (f,g),(f,g)\rangle_m/4}\Psi(\phi-\sqrt{2}\mu_m^{1/2}g)\ d\gamma(\phi)\ d\gamma^2_{|\kappa_m(F)}(\mu_m^{-1/2}f + i\mu_m^{1/2}g)\\
    &=\int_{G}\int_{\kappa_m(F)} h(f,g) \overline{\Psi'(\phi)}e^{i\langle f,g\rangle/2}e^{-i\langle \phi,\mu_m^{-1/2}f\rangle/\sqrt{2}}e^{\langle \phi,\mu_m^{1/2}g\rangle/\sqrt{2}}e^{-\langle g,\mu_m g\rangle/2}\nonumber\\
        &\hspace{1in} \times e^{\frac{\langle f,\mu_m^{-1}f\rangle + \langle g,\mu_m g\rangle}{4}}\Psi(\phi-\sqrt{2}\mu_m^{1/2}g)\ d\gamma^2_{|\kappa_m(F)}(\mu_m^{-1/2}f + i\mu_m^{1/2}g)\ d\gamma(\phi)\\
        &=\int_{G} \overline{\Psi'(\phi)} \int_{\kappa_m(F)} h\circ\kappa_m^{-1}(f'+ig') e^{i\langle f',g'\rangle/2}e^{-i\langle \phi,f'\rangle/\sqrt{2}}e^{\langle \phi,g'\rangle/\sqrt{2}}e^{-\langle g',g'\rangle/2}\nonumber\\
        &\hspace{1.5in} \times e^{\frac{\langle f',f'\rangle + \langle g', g'\rangle}{4}}\Psi(\phi-\sqrt{2}g') e^{-\langle f',f'\rangle/2} e^{-\langle g',g'\rangle/2}\ df'\ dg' \ d\gamma(\phi)\\
        &= \int_{G} \overline{\Psi'(\phi)}\int_{\kappa_m(F)} h\circ\kappa_m^{-1}\big(2^{1/2}f'+i2^{-1/2}(\phi-\phi')\big)\nonumber\\
        &\hspace{1.5in}\times e^{i\langle f',\phi-\phi'\rangle/2}e^{-i\langle \phi,f'\rangle}e^{\langle \phi,\phi-\phi'\rangle/2}e^{-||\phi-\phi'||^2/4}e^{||f'||^2/2}e^{||\phi-\phi'||^2/8}\nonumber\\
        &\hspace{1.75in} \times \Psi(\phi') e^{-||f'||^2}e^{-||\phi-\phi'||^2/4}\ df'\ d\phi' \ d\gamma(\phi)\\
        &= \int_{G} \overline{\Psi'(\phi)}\int_{\kappa_m(F)} h\circ\kappa_m^{-1}\big(2^{1/2}f'+i2^{-1/2}(\phi-\phi')\big)e^{-i\langle f', \phi+\phi'\rangle/2}\nonumber\\
        &\hspace{1.75in} \times \Psi(\phi') e^{||\phi+\phi'||^2/8}e^{-||f'||^2/2}e^{-||\phi'||^2/2}\ df'\ d\phi' \ d\gamma(\phi).
    \end{align}
This suffices to establish Eq.~\eqref{eqn:finite integral operator}-\eqref{eqn:finite kernel}.

Now, in a tail of the net of increasing subspaces $F\subseteq E$, we will have $G^{(m)}_h\subseteq \kappa_m(F)$.  As before we can take a decomposition into real and imaginary parts $\kappa_m(F) = H_1\oplus iH_1$.  Since in the tail of the net $G^{(m)}_h\subseteq \kappa_m(F)$, we must have $H_h^{(m)}\subseteq H_1$ so that we can take the orthogonal decomposition $H_1 = H_h^{(m)}\oplus (H_h^{(m)})'$, where $(H^{(m)}_h)' := (H^{(m)}_h)^\perp\cap H_1$ is the orthogonal complement in $H_1$ with respect to the real part of the inner product.  In other words, we can write each $f'+i\phi'\in \kappa_m(F)$ with $f,\phi\in H_1$ as $f'_1 + f'_2+ i\phi'$ for $f'_1$ the orthogonal projection onto $H^{(m)}_h$ and $f'_2$ the orthogonal projection onto $(H^{(m)}_h)'$.  Then since $h$ is tame, it is constant along the directions in $(H^{(m)}_h)'$, and so we have
\begin{align}
    &\int_{\kappa_m(F)}  h\circ\kappa_m^{-1}\big(2^{1/2}f'+i2^{-1/2}(\phi-\phi')\big)\Psi(\phi')e^{-i\langle f', \phi+\phi'\rangle/2}e^{||\phi+\phi'||^2/8}e^{-||f'||^2/2}e^{-||\phi'||^2/2}\ df'\ d\phi'\\
    &= \int_{H_1}\int_{H^{(m)}_h}  h\circ\kappa_m^{-1}\big(2^{1/2}f_1'+i2^{-1/2}(\phi-\phi')\big)\Psi(\phi')e^{-i\langle f'_1,\phi + \phi'\rangle/2} e^{||\phi+\phi'||^2/8}\nonumber\\
    &\hspace{.5in}\label{eqn:kernel expression}\times  \int_{(H^{(m)}_h)'} e^{-i\langle f'_2,\phi + \phi'\rangle/2} d\gamma_{|(H^{(m)}_h)'}(f'_2)\ d\gamma_{|H^{(m)}_h}(f'_1)\ d\gamma_{|H_1}(\phi').
\end{align}
The inner integral yields
\begin{align}
    \int_{(H^{(m)}_h)'} e^{-i\langle f'_2,\phi + \phi'\rangle/2} d\gamma_{|(H^{(m)}_h)'}(f'_2)&=\int_{(H^{(m)}_h)'} e^{-i\langle f'_2,\phi + \phi'\rangle/2} e^{-||f'_2||^2/2}\ df'_2\\
    &=e^{-||P_h'(\phi+\phi')||^2/8},
\end{align}
where $P_h': G\to (H^{(m)}_h)'$ is the orthogonal projection onto $(H^{(m)}_h)'$.  Plugging this into the outer integral in Eq.~\eqref{eqn:kernel expression}, we find
\begin{align}
    &\int_{H_1}\int_{H^{(m)}_h}  h\circ\kappa_m^{-1}\big(2^{1/2}f'+i2^{-1/2}(\phi-\phi')\big)\Psi(\phi')e^{-i\langle f',\phi + \phi'\rangle/2}\\
    &\hspace{.75in}\times e^{||\phi+\phi'||^2/8}e^{-||P_h'(\phi+\phi')||^2/8}\ d\gamma_{H^{(m)}_h}(\phi')\ d\gamma_{|H_1}(f')\\
    &\hspace{.25in}=\int_{H_1}\int_{H^{(m)}_h}h\circ\kappa_m^{-1}\big(2^{1/2}f +i2^{-1/2}(\phi-\phi')\big)\Psi(\phi')e^{-i\langle f',\phi + \phi'\rangle/2}\\
    &\hspace{1in} \times e^{||P_h(\phi+\phi')||^2/8}\ d\gamma_{H^{(m)}_h}(\phi')\ d\gamma_{|H_1}(f')\\
    & \hspace{.25in}=\int_{H_1} q^{(m)}_h(\phi,\phi')\Psi(\phi')d\gamma(\phi').
\end{align}

Thus, in the tail of the net of increasing subspaces $F\subseteq E$ in which $G^{(m)}_h\subseteq \kappa_m(F)$, we can write
\begin{align}
U_mQ^{(m)}_F(h)U_m^* =\hat{Q}^{(m)}(h)\otimes R_{F}
\end{align}
acting on $L^2(G,\gamma)\cong L^2\big(H^{(m)}_h,\gamma_{|H^{(m)}_h}\big)\otimes L^2\big((H^{(m)}_h)^\perp,\gamma_{|(H^{(m)}_h)^\perp}\big)$ since $G = H^{(m)}_h\oplus (H^{(m)}_h)^\perp$.
Here, $\hat{Q}^{(m)}(h)$ is the operator on $L^2\big(H^{(m)}_h,\gamma_{|H^{(m)}_h}\big)$ given by
\begin{align}
    (\hat{Q}^{(m)}(h)\Psi)(\phi) = \int_{H^{(m)}_h} q_h^{(m)}(\phi',\phi)\Psi(\phi')\ d\gamma_{|H^{(m)}_h}(\phi')
\end{align}
for all $\Psi\in L^2\big(H^{(m)}_h,\gamma_{|H^{(m)}_h}\big)$.  And $R_F$ is the operator on $L^2\big((H^{(m)}_h)^\perp,\gamma_{|(H^{(m)}_h)^\perp}\big)$ given by
\begin{align}
    R_F = P_{1}^{(H^{(m)}_h)'}\otimes I_{L^2(H_1^\perp,\gamma_{|H_1^\perp})},
\end{align}
where we take the decomposition $L^2\big((H^{(m)}_h)^\perp,\gamma_{|(H^{(m)}_h)^\perp}\big)\cong L^2\big((H^{(m)}_h)',\gamma_{|(H^{(m)}_h)'}\big)\otimes L^2(H_1^\perp,\gamma_{H_1^\perp})$ since $(H^{(m)}_h)^\perp = (H^{(m)}_h)'\oplus H_1^\perp$.  The operator $P_{1}^{(H^{(m)}_h)'}$ is the projection onto the constant function $1$ (the vacuum) in $(H^{(m)}_h)'$, or in other words the Hilbert-Schmidt operator with kernel $k^{(H^{(m)}_h)'}_1(\phi,\phi') = 1$ as follows from the fact that $h$ is constant in the directions along $(H^{(m)}_h)'$.  The operator $I_{L^2(H_1^\perp,\gamma_{|H_1^\perp})}$ is the identity on the remaining tensor factor.

One can further show that the operators $R_F$ converge strongly to $P_1^{(H_h^{(m)})^\perp}$ as $F\to E$.  To see this, recall that bounded, tame functions are dense in $L^2\big((H^{(m)}_h)^\perp,\gamma_{(H^{(m)}_h)^\perp}\big)$ and consider the action of $R_F$ on a tame function $\Psi$ on $(H^{(m)}_h)^\perp$.  We can choose $F$ large enough that the base of $\Psi$ is contained in $(H^{(m)}_h)'$ so that for any subspace $F'\subseteq G$ with $F\subseteq F'$, $R_{F'}\Psi = P_1^{(H_h^{(m)})^\perp}\Psi$, since $\Psi$ was already assumed to be constant in the directions on which the identity operator acts.  Thus, for bounded, tame $\Psi$, $R_{F'}\Psi$ is constant in a tail of the net and so converges to $P_1^{(H_h^{(m)})^\perp}\Psi$ as $F\to E$.  The result extends to all $\Psi\in L^2\big((H^{(m)}_h)^\perp,\gamma_{(H^{(m)}_h)^\perp}\big)$ by density of the bounded, tame functions.  Moreover, since $R_F$ converges strongly to $P_1^{(H_h^{(m)})^\perp}$, it follows that $U_m Q_F^{(m)}(h)U_m^*$ converges strongly to the operator $U_mQ^{(m)}(h)U_m^*$ in Eq.~\eqref{eqn:full Hilbert-Schmidt}.
\end{proof}

In what follows, we will analyze the scaling limits of the operators $Q^{(m)}(h)\in\mathfrak{B}^{(m)}$ by defining analogs of the scaling bundles and variable mass bundles that we considered in \S\ref{sec:lower semicontinuous scalar field}.

\subsection{Generating Sections}

Begin by recalling our rescaling transformations $\sigma_\lambda$ on the Weyl algebra, defined from dilation transformations $\delta_\lambda$ on test functions in $E$.  We extended the transformation $\sigma_\lambda$ to act on weak operator integrals of Weyl unitaries by establishing the existence of a unitary $V_\lambda: \mathcal{H}^{(\lambda m)}\to \mathcal{H}^{(m)}$ such that $\sigma_\lambda(\cdot) = V_\lambda (\cdot) V_\lambda^*$.  We now proceed to use the same method to define rescaled families of operators from $\mathfrak{B}^{(m)}$ using $V_\lambda$.
\begin{prop}
    The map $\sigma_\lambda: \mathfrak{B}^{(\lambda m)}\to \mathfrak{B}^{(m)}$ defined by $\sigma_\lambda(A) = V_\lambda A V_\lambda^*$ satisfies
    \begin{align}
        \sigma_\lambda(Q^{(\lambda m)}(h)) = Q^{(m)}(h\circ\delta_\lambda^{-1}).
    \end{align}
\end{prop}

\begin{proof}
    For all $\psi,\varphi\in\mathcal{H}^{(m)}$,
    \begin{align}
        \langle \psi,V_\lambda Q^{(\lambda m)}(h)V_\lambda^*\varphi\rangle_{(m)} &= \int_E h(u) \langle \psi,V_\lambda \pi^{(\lambda m)}(W(u))V_\lambda^*\varphi\rangle_{(m)}e^{\langle u,u\rangle_{\lambda m}/4}d\gamma^{(\lambda m)}(u)\\
        &= \lim_{F\to E}\int_F h(u) \langle \psi,V_\lambda \pi^{(\lambda m)}(W(u))V_\lambda^*\varphi\rangle_{(m)}e^{-\langle u,u\rangle_{\lambda m}/4}d\mu_F(u)\label{eqn:change of variables 1}\\
        &= \lim_{F\to E}\int_{\delta_\lambda(F)} h(\delta_\lambda^{-1} v) \langle \psi,\pi^{(m)}(W( v))\varphi\rangle_{(m)}e^{-\langle v,v\rangle_m/4}d\mu_{\delta_\lambda(F)}(v)\label{eqn:change of variables 2}\\
        &= \int_E h(\delta_\lambda^{-1} v) \langle \psi,\pi^{(m)}(W( v))\varphi\rangle_{(m)}e^{\langle v,v\rangle_m/4}d\gamma^{(m)}(v)\\
        &= \langle \psi, Q^{(m)}(h\circ\delta_\lambda^{-1})\varphi\rangle_{(m)}.
    \end{align}
    Again, we use the change of variables $v = \delta_\lambda u$ to derive line \eqref{eqn:change of variables 2} from \eqref{eqn:change of variables 1}, noting that $\langle \delta_\lambda^{-1} u,\delta_\lambda^{-1} u\rangle_{m} = \langle u,u\rangle_{\lambda m}$ and again $\delta_\lambda$ preserves the Lebesgue measure.
\end{proof}

Just as in \S\ref{sec:lower semicontinuous scalar field}, we define the *-algebra of sections of a bundle $\underline{\widetilde{\mathfrak{B}}}_s^{(m)}$ as generated from sections of the form
\begin{align}
    \lambda\mapsto \alpha^{(m)}_{\Lambda,\lambda x}\circ \sigma_\lambda(Q^{(m)}(h)),
\end{align}
where $h\in A^0(E)$ ranges over all smearing functions with $E_h^{(m)}\subseteq J_mE$, and $(\Lambda,x)$ ranges over all Poincar\'e transformations.
We ultimately wish to prove that
\begin{align}
    \lambda\mapsto ||\underline{A}_\lambda||_\lambda
\end{align}
is uniformly continuous for every $\underline{A}\in\widetilde{\underline{\mathfrak{B}}}_s^{(m)}$.  As before, we proceed by moving from spacetime rescalings to variable mass sections.

\subsection{Variable mass bundle of nonlocal observables}
To analyze the continuity of norms of sections, we again use the rescaling map $\sigma_\lambda$ to note that for every section $\underline{A}\in\widetilde{\underline{\mathfrak{B}}}_s^{(m)}$,
\begin{align}
    ||\alpha_{\Lambda,\lambda x}^{(m)}\circ\sigma_\lambda(\underline{A}_\lambda)||_\lambda = ||\sigma_\lambda\circ \alpha_{\Lambda, x}^{(\lambda m)}(\underline{A}_\lambda)||_\lambda =||\alpha_{\Lambda, x}^{(\lambda m)}(\underline{A}_\lambda)||_\lambda.
\end{align}
So it suffices to generate the *-algebra of sections of a variable mass bundle $\underline{\widetilde{\mathfrak{B}}}^{(m)}_v$ from sections of the form
\begin{align}
\label{eqn:generating Q sections}
    \lambda\mapsto \alpha^{(\lambda m)}_{\Lambda,x}(Q^{(m)}(h)),
\end{align}
where $h\in A^0(E)$ ranges over all smearing functions with $E_h^{(m)}\subseteq J_mE$, and $(\Lambda,x)$ ranges over all Poincar\'e transformations.
We will show that these sections generate a uniformly continuous bundle of C*-algebras by establishing the following:
\begin{thm}\label{thm:uniform continuity}
    For any $\underline{A}\in \underline{\widetilde{\mathfrak{B}}}_v^{(m)}$, the map $\lambda\mapsto ||\underline{A}_\lambda||_\lambda$ is continuous on $[0,1]$. 
\end{thm}

\begin{proof}
We will actually prove the stronger statement that after transporting each mass $m$ representation to the real wave representation on $L^2(G,\gamma)$, the map $\lambda\mapsto U_{\lambda m}\underline{A}_\lambda U^*_{\lambda m}$ is continuous in Hilbert-Schmidt norm in the sense that $\lim_{\lambda\to \lambda_0}||U_{\lambda m} \underline{A}_\lambda U^*_{\lambda m}-U_{\lambda_0 m} \underline{A}_{\lambda_0}U_{\lambda_0 m}^*||_{HS} = 0.$  

Consider first a generating section $\underline{A}\in \widetilde{\underline{\mathfrak{B}}}^{(m)}_v$ of the form
\begin{align}
    \underline{A}_\lambda = \alpha^{(\lambda m)}_{\Lambda,x}(Q^{(\lambda m)}(h)).
\end{align}
We claim that $\lim_{\lambda\to \lambda_0} ||U_{\lambda m}\underline{A}_\lambda U_{\lambda m}^*- U_{\lambda_0 m}\underline{A}_{\lambda_0} U_{\lambda_0 m}^*||_{HS} = 0$.  We establish this from direct calculation with the integral kernels.  Since it is well known that the Hilbert-Schmidt norm of a Hilbert-Schmidt operator on $L^2(G,\gamma)$ is the $L^2$ norm of its integral kernel in $L^2(G\times G,\gamma\times \gamma)$, it suffices to show that
\begin{align}
    \label{eqn:L^2 convergence of kernels}
    \lim_{\lambda\to \lambda_0}\Big|\Big|q^{(\lambda m)}_{h\circ\tau^{\lambda m}_{\Lambda,x}}-q^{(\lambda_0 m)}_{h\circ\tau^{\lambda_0 m}_{\Lambda,x}}\Big|\Big|_{L^2(G\times G,\gamma\times \gamma)}= 0.
\end{align} We have the integral formula for the kernel in Eq.~\eqref{eqn:full kernel}, and we wish to apply two dominated convergence arguments. First, note that for any $F\subseteq L^2(G,\gamma)$ with $H_{h\circ\tau^{(\lambda m)}_{\Lambda,x}}\subseteq F$, we can take $F' = (H_{h\circ\tau^{(\lambda m)}_{\Lambda,x}})^\perp\cap F$.  We let $P_F$ and $P_{F'}$ denote the orthogonal projections onto $F$ and $F'$, respectively.  We have the decomposition $L^2(F,\gamma_{|F}) \cong L^2(H_{h\circ\tau^{(\lambda m)}_{\Lambda,x}},\gamma_{|H_{h\circ\tau^{(\lambda m)}_{\Lambda,x}}})\otimes L^2(F',\gamma_{|F'})$.  Thus,
\begin{align}
    e&^{||P_{F}(\phi+\phi')||^2/8}\int_{F} h\circ \tau^{(\lambda m)}_{\Lambda,x}\circ\kappa_m^{-1}\big(2^{1/2}f+i2^{-1/2}(\phi-\phi')\big)e^{-i\langle f,\phi+\phi'\rangle/2} d\gamma_{|F}(f)\\
    &= e^{||P_{F}(\phi + \phi')||^2/8}\int_{H_{h\circ\tau^{(\lambda m)}_{\Lambda,x}}^{(m)}} h\circ \tau^{(\lambda m)}_{\Lambda,x}\circ\kappa_m^{-1}\big(2^{1/2}f_0+i2^{-1/2}(\phi-\phi')\big)\\
    &\hspace{1.5in} \times e^{-i\langle f_0,\phi+\phi'\rangle/2} d\gamma_{|H_{h\circ\tau^{(\lambda m)}_{\Lambda,x}}^{(m)}}(f_0) \int_{F'} e^{-i\langle f_1,\phi+\phi'\rangle/2} d\gamma_{|F'}(f_1)\\
    &= e^{\big|\big|P_{h\circ\tau^{(\lambda m)}_{\Lambda,x}}(\phi+\phi')\big|\big|^2/8}\int_{H_{h\circ\tau^{(\lambda m)}_{\Lambda,x}}^{(m)}} h\circ \tau^{(\lambda m)}_{\Lambda,x}\circ\kappa_m^{-1}\big(2^{1/2}f+i2^{-1/2}(\phi-\phi')\big)\\
    &\hspace{3in} \times e^{-i\langle f,\phi+\phi'\rangle/2} d\gamma_{|H^{(m)}_{h\circ\tau^{(\lambda m)}_{\Lambda,x}}}(f)\\
    &=q_{h\circ\tau^{(\lambda m)}_{\Lambda,x}}^{(\lambda m)}(\phi,\phi') 
\end{align}
since
\begin{align}
    \int_{F'} e^{-i\langle f_1,\phi+\phi'\rangle/2} d\gamma_{|F'}(f_1) = e^{-||P_{F'}(\phi + \phi')||^2/8}.
\end{align}
Hence, taking the limit $F\to E$, it makes sense to write
\begin{align}
\label{eqn:common domain integral formula for kernel}
    q_{h\circ\tau^{(\lambda m)}_{\Lambda,x}}^{(\lambda m)}(\phi,\phi') = e^{||\phi+\phi'||^2/8}\int_{G} h\circ \tau^{(\lambda m)}_{\Lambda,x}\circ\kappa_m^{-1}\big(2^{1/2}f+i2^{-1/2}(\phi-\phi')\big)e^{-i\langle f,\phi+\phi'\rangle/2} d\gamma(f).
\end{align}
This allows us to treat the kernel $q_{h\circ\tau^{(\lambda m)}_{\Lambda,x}}^{(\lambda m)}$ for each $\lambda\in [0,1]$ as calculated from an integral over the common domain $G$.

Recall also the result of Lemma~\ref{lem:taus are continuous}, and note that an analogous argument can be made for the strong continuity of $\lambda\mapsto \kappa_{\lambda m}$.  Using these two facts, the continuity of $h$ implies that the integrand in Eq.~\eqref{eqn:common domain integral formula for kernel} is pointwise continuous.  Moreover, the integrand is clearly bounded by $\sup_{u\in E}|h(u)|$.  Hence, it follows from the dominated convergence theorem that $\lambda\mapsto q^{(\lambda m)}_{h\circ\tau^{\lambda m}_{\Lambda,x}}$ is pointwise continuous on $G\times G$.  Moreover, since for all $\phi,\phi'\in G$,
\begin{align}
    \left|q^{(\lambda m)}_{h\circ\tau^{\lambda m}_{\Lambda,x}}(\phi,\phi')-q^{(\lambda_0 m)}_{h\circ\tau^{\lambda_0 m}_{\Lambda,x}}(\phi,\phi')\right|\leq 2\sup_{u\in E} |h(u)|,
\end{align}
the dominated convergence theorem again applies to yield Eq.~\eqref{eqn:L^2 convergence of kernels}.

Thus, we have found
\begin{align}
    \Big|||\underline{A}_\lambda||_\lambda - ||\underline{A}_{\lambda_0}||_{\lambda_0}\Big| &\leq \Big|||U_{\lambda m}\underline{A}_\lambda U^*_{\lambda m}|| - ||U_{\lambda_0 m}\underline{A}_{\lambda_0} U^*_{\lambda_0 m}||\Big|\\
    &\leq ||U_{\lambda m}\underline{A}_\lambda U^*_{\lambda m}-U_{\lambda_0 m}\underline{A}_{\lambda_0} U^*_{\lambda_0 m}||\\
    &\leq ||U_{\lambda m}\underline{A}_\lambda U^*_{\lambda m}-U_{\lambda_0 m}\underline{A}_{\lambda_0} U^*_{\lambda_0 m}||_{HS}\to 0.
\end{align}
Thus, $\lambda\mapsto ||\underline{A}_\lambda||_\lambda$ is continuous for $\lambda\in [0,1]$

Suppose now $\underline{B} = \underline{A} + \underline{C}$ is a sum of generating sections $\underline{A}$ and $\underline{C}$ of the form in Eq.~\eqref{eqn:generating Q sections}.  Then a simple application of the triangle inequality implies
\begin{align}
    ||U_{\lambda m}\underline{B}_\lambda U^*_{\lambda m}-U_{\lambda_0 m}\underline{B}_{\lambda_0} U^*_{\lambda_0 m}||_{HS} &\leq ||U_{\lambda m}\underline{A}_\lambda U^*_{\lambda m}-U_{\lambda_0 m}\underline{A}_{\lambda_0} U^*_{\lambda_0 m}||_{HS}\nonumber\\
    &\hspace{.5in}+ ||U_{\lambda m}\underline{C}_\lambda U^*_{\lambda m}-U_{\lambda_0 m}\underline{C}_{\lambda_0} U^*_{\lambda_0 m}||_{HS}\to 0,
\end{align}
By the reasoning above, we have found $\lambda\mapsto ||\underline{B}_\lambda||_\lambda$ is continuous for $\lambda\in [0,1]$

Suppose next that $\underline{B} = \underline{A}\cdot\underline{C}$ is a product of generating sections $\underline{A}$ and $\underline{C}$ of the form in Eq.~\eqref{eqn:generating Q sections}.  Then
\begin{align}
    ||U_{\lambda m}\underline{B}_\lambda U^*_{\lambda m}-&U_{\lambda_0 m}\underline{B}_{\lambda_0} U^*_{\lambda_0 m}||_{HS} = ||U_{\lambda m}\underline{A}_\lambda\underline{C}_\lambda U^*_{\lambda m}-U_{\lambda_0 m}\underline{A}_{\lambda_0} \underline{C}_{\lambda_0} U^*_{\lambda_0 m}||_{HS}\\
    &=||U_{\lambda m}\underline{A}_\lambda U_{\lambda m}^* U_{\lambda_m}\underline{C}_\lambda U^*_{\lambda m}-U_{\lambda_0 m}\underline{A}_{\lambda_0}U_{\lambda_0 m}^*U_{\lambda m}\underline{C}_\lambda U^*_{\lambda m}\nonumber\\
    &\hspace{.5in}+U_{\lambda_0 m}\underline{A}_{\lambda_0}U_{\lambda_0 m}^*U_{\lambda m}\underline{C}_\lambda U^*_{\lambda m}-U_{\lambda_0 m}\underline{A}_{\lambda_0}U_{\lambda_0}^*U_{\lambda_0} \underline{C}_{\lambda_0} U^*_{\lambda_0 m}||_{HS}\\
    &\leq\Big(\sup_{\lambda\in[0,1]}||\underline{C}_\lambda||_{HS}\Big)\cdot||U_{\lambda m}\underline{A}_\lambda U^*_{\lambda m}-U_{\lambda_0 m}\underline{A}_{\lambda_0}U^*_{\lambda_0 m}||_{HS}\nonumber\\
    &\hspace{.5in}+ ||\underline{A}_{\lambda_0}||_{HS}\cdot||U_{\lambda m}\underline{C}_\lambda U^*_{\lambda m}-U_{\lambda_0 m} \underline{C}_{\lambda_0} U^*_{\lambda_0 m}||_{HS}\\
    &\to 0.
\end{align}
Here, we use the fact that $\lambda\mapsto ||\underline{C}_\lambda||_{HS}$ is bounded since it is continuous for $\lambda\in [0,1]$.  Hence, we find that $\lambda\mapsto ||\underline{B}_\lambda||_\lambda$ is continuous for $\lambda\in [0,1]$ for any finite sum of products of generating sections.
\end{proof}

Just as in the previous sections, we can now generate a variable mass bundle with C*-algebra of sections
\begin{align}
    \underline{\mathfrak{B}}^{(m)}_v = \overline{\text{span}\left\{f\underline{A}\ |\ f\in C([0,1]), \underline{A}\in \widetilde{\underline{\mathfrak{B}}}_v
    \right\}}.
\end{align}
We summarize the result in the following corollary.

\begin{cor}
    The structure $(\underline{\mathfrak{B}}^{(m)}_v,\chi,[0,1])$ is a uniformly continuous bundle of C*-algebras.
\end{cor}

\subsection{Uniform continuity of rescaled sections of nonlocal observables}

Mutatis mutandis, since $\sigma_\lambda$ is a *-isomorphism, we can generate a further uniformly continuous bundle over $(0,1]$ with C*-algebra of sections
\begin{align}
\underline{\mathfrak{B}}^{(m)}_s = \overline{\text{span}\{f\underline{A}\ |\ f\in UC_b((0,1]), \underline{A}\in \widetilde{\underline{\mathfrak{B}}}_s\}}
\end{align}
and with fibers $\mathfrak{B}^{(m)}$ for fixed $m$.  For each $\underline{A}\in \underline{\mathfrak{B}}^{(m)}_v$, we know the function $\lambda\mapsto ||\underline{A}_\lambda||_\lambda$ is continuous for $\lambda\in [0,1]$, and so removing the point at $\lambda=0$ yields a section $\underline{B}_\lambda=\sigma_\lambda(\underline{A}_\lambda)$ such that $\underline{B}\in \underline{\mathfrak{B}}^{(m)}_s$ and the map $\lambda\mapsto ||\underline{B}_\lambda||_\lambda = ||\underline{A}_\lambda||$ is therefore uniformly continuous for $\lambda\in(0,1]$.  By the definition, every section $\underline{B}\in \underline{\mathfrak{B}}^{(m)}_s$ is of the form $\underline{B}_\lambda = \sigma_\lambda(\underline{A}_\lambda)$ for some $\underline{A}\in \underline{\mathfrak{B}}^{(m)}_v$, so we have established the following.

\begin{thm}
    The structure $(\underline{\mathfrak{B}}^{(m)}_s,\phi,(0,1])$ is a uniformly continuous bundle of C*-algebras.
\end{thm}

Since the value of a section at any given $\lambda\in (0,1]$ is a non-local observable, $\underline{\mathfrak{B}}^{(m)}_s$ does not have the structure of a Poincar\'e net, and so does not form a scaling bundle by our definition.  Nevertheless, the uniform continuity of the bundle implies that $\underline{\mathfrak{B}}^{(m)}_s$ has a canonical extension to the base $[0,1]$, where we denote the fiber at $\lambda = 0$ by $\mathfrak{B}^{(m)}_0$.  It follows that the fiber $\mathfrak{B}_0^{(m)}$ is *-isomorphic to the algebra of non-local observables $\mathfrak{B}^{(0)}$ for the $m=0$ theory, carrying the Poincar\'e automorphisms $\alpha^{(0)}_{\Lambda,x}$.

To see this, we follow the same strategy as in \S\ref{sec:lower semicontinuous scalar field}.  First, define a *-isomorphism $\underline{\sigma}: \underline{\mathfrak{B}}^{(m)}_v\to \underline{\mathfrak{B}}^{(m)}_s$ between the C*-algebras of sections by
\begin{align}
    \underline{\sigma}(\underline{A})_\lambda = \sigma_\lambda(\underline{A}_\lambda)
\end{align}
for all $\underline{A}\in \underline{\mathfrak{B}}^{(m)}_v$ and $\lambda\in (0,1]$.  Denote again the restriction of the bundle $(\underline{\mathfrak{B}}^{(m)}_v,\chi,[0,1])$ to $(0,1]$ by $\underline{\mathfrak{B}}^{(m)}_v|_{(0,1]}$, which has the same C*-algebra of sections $\underline{\mathfrak{B}}^{(m)}_v$.  We then have the following result.

\begin{prop}
    The map $(\operatorname{id},\underline{\sigma})$ is a bundle isomorphism from $\underline{\mathfrak{B}}^{(m)}_v|_{(0,1]}$ to $\underline{\mathfrak{B}}^{(m)}_s$.
\end{prop}

It follows that $(\operatorname{id},\underline{\sigma}^{-1})$ is a minimal extension of the bundle $\underline{\mathfrak{B}}^{(m)}_s$, and by Thm.~\ref{thm:steeger and feintzeig extension theorem}, the minimal extension, as well as the fiber at $\lambda = 0$, are unique up to isomorphism.  In other words, we have the following sense in which the $m=0$ algebra $\mathfrak{B}^{(0)}$ provides a scaling limit of the bundle $\underline{\mathfrak{B}}^{(m)}_s$.

\begin{thm}
    There is a *-isomorphism $\sigma_0$ from $\mathfrak{B}^{(m)}_0$ to $\mathfrak{B}^{(0)}$.
\end{thm}

Moreover, even though we do not have a canonical net structure on the sections $\underline{\mathfrak{B}}_s^{(m)}$ because there is no canonical net structure on the C*-algebra of nonlocal observables $\mathfrak{B}^{(m)}$, we can still have well-defined Poincar\'e automorphisms.  In particular, we can still define automorphisms $\underline{\alpha}_{\Lambda,x}^{(m),s}$ of $\underline{\mathfrak{B}}_s^{(m)}$ by
\begin{align}
\underline{\alpha}_{\Lambda,x}^{(m),s}(\underline{A})_\lambda = \alpha_{\Lambda,\lambda x}^{(m)}(\underline{A}_\lambda)
\end{align}
for all $\underline{A}\in \underline{\mathfrak{B}}_s^{(m)}$.  And we can define automorphisms $\underline{\alpha}_{\Lambda,x}^{(m),v}$ of $\underline{\mathfrak{B}}_v^{(m)}$ by
\begin{align}
\underline{\alpha}_{\Lambda,x}^{(m),v}(\underline{A})_\lambda = \alpha_{\Lambda, x}^{(\lambda m)}(\underline{A}_\lambda)
\end{align}
for all $\underline{A}\in \underline{\mathfrak{B}}_v^{(m)}$.  As before, we have the relation
\begin{align}
    \underline{\sigma}\circ \underline{\alpha}^{(m),v}_{\Lambda,x} = \underline{\alpha}_{\Lambda,x}^{(m),s}\circ \underline{\sigma}
\end{align}
for all Poincar\'e transformations $(\Lambda,x)$.  It follows now from Prop.~\ref{prop:steeger and feintzeig functoriality of extension} that there are unique limit morphisms $\alpha^{(m),0}_{\Lambda,x}$ of the limiting fiber $\mathfrak{B}^{(m)}_0$ such that
\begin{align}
    \sigma_0\circ \alpha^{(0)}_{\Lambda,x} = \alpha^{(m),0}_{\Lambda,x}\circ \sigma_0
\end{align}
for all Poincar\'e tranfsormations $(\Lambda,x)$.  In this sense, the functoriality of the extension of uniformly continuous bundles of C*-algebras implies that the nonlocal observables in $\mathfrak{B}^{(m)}_0$ carry the $m=0$ dynamics.


\printbibliography

@article{Bu17,
	author = {Buchholz, D.},
	journal = {Communications in Mathematical Physics},
	number = {2},
	pages = {691-716},
	title = {{The Resolvent Algebra for Oscillating Lattice Systems: Dynamics, Ground and Equilibrium States}},
	volume = {353},
	year = {2017}}

@article{Bu18,
	author = {Buchholz, D.},
	journal = {Communications in Mathematical Physics},
	number = {3},
	pages = {949-981},
	title = {{The Resolvent Algebra of Non-relativistic Bose Fields: Observables, Dynamics and States}},
	volume = {362},
	year = {2018}}

@incollection{BuGr15,
	author = {Buchholz, D. and Grundling, H.},
	booktitle = {The Message of Quantum Science: Attempts Towards a Synthesis},
	editor = {Blanchard, B. and Fr\"{o}hlich, J.},
	pages = {33-45},
	publisher = {Springer},
	title = {{Quantum systems and Resolvent Algebras}},
	year = {2015}}

@Article{BuVe98,
  author        = {Buchholz, D. and Verch, R.},
  journal       = {Reviews in Mathematical Physics},
  title         = {{Scaling Algebras and Renormalization Group in Algebraic Quantum Field Theory. II. Instructive Examples}},
  year          = {1998},
  number        = {6},
  pages         = {775-800},
  volume        = {10},
}

@article{BuGr08,
	author = {Buchholz, D. and Grundling, H.},
	journal = {Journal of Functional Analysis},
	pages = {2725-2779},
	title = {{The resolvent algebra: A new approach to canonical quantum systems}},
	volume = {254},
	year = {2008}}

@article{vN19,
	author = {T. D. H. van Nuland},
	doi = {10.1016/j.jfa.2019.02.022},
	journal = {Journal of Functional Analysis},
	month = {oct},
	number = {8},
	pages = {2815--2838},
	publisher = {Elsevier {BV}},
	title = {Quantization and the resolvent algebra},
	volume = {277},
	year = {2019}}

@article{BuFr20,
	author = {Detlev Buchholz and Klaus Fredenhagen},
	doi = {10.1007/s00220-020-03700-9},
	journal = {Communications in Mathematical Physics},
	month = {mar},
	number = {2},
	pages = {947--969},
	publisher = {Springer Science and Business Media {LLC}},
	title = {{A C*-algebraic Approach to Interacting Quantum Field Theories}},
	volume = {377},
	year = {2020}}

@article{BuFr20a,
	author = {Detlev Buchholz and Klaus Fredenhagen},
	doi = {10.1016/j.exmath.2020.06.002},
	journal = {Expositiones Mathematicae},
	month = {jun},
	number = {2},
	pages = {150--167},
	publisher = {Elsevier {BV}},
	title = {{Classical dynamics, arrow of time, and genesis of the Heisenberg commutation relations}},
	volume = {38},
	year = {2020}}

@article{Kay1979,
	author = {Kay, B.},
	journal = {Journal of Mathematical Physics},
	pages = {1712-3},
	title = {{A uniqueness result in the Segal-Weinless approach to linear Bose fields}},
	volume = {20},
	year = {1979}}

@book{Rieffel1993,
	address = {Providence, RI},
	author = {Rieffel, M.},
	publisher = {American Mathematical Society},
	series = {Memoirs of the American Mathematical Society},
	title = {{Deformation quantization for actions of $\mathbb{R}^d$}},
	year = {1993}}

@article{EcFr74,
	author = {Eckmann, J.-P. and Fr\"{o}hlich, J.},
	journal = {Annales de l'institut Henri Poincar\'{e}. Section A, Physique Th\'{e}orique},
	number = {2},
	pages = {201--209},
	title = {{Unitary equivalence of local algebras in the quasifree representation}},
	volume = {20},
	year = {1974}}

@article{vN22,
	author = {Teun D. H. van Nuland},
	doi = {10.1007/s11005-022-01525-2},
	journal = {Letters in Mathematical Physics},
	month = {apr},
	number = {2},
	publisher = {Springer Science and Business Media {LLC}},
	title = {{Strict deformation quantization of abelian lattice gauge fields}},
	volume = {112},
	year = {2022}}

@Article{vNSt24,
  author    = {van Nuland, T. D. H. and Stienstra, R.},
  journal   = {Annales Henri Poincaré},
  title     = {{Classical and Quantised Resolvent Algebras for the Cylinder}},
  year      = {2024},
  issn      = {1424-0661},
  month     = apr,
  number    = {2},
  pages     = {479--526},
  volume    = {26},
  doi       = {10.1007/s00023-024-01434-1},
  publisher = {Springer Science and Business Media LLC},
}

@Article{AmNi08,
  author    = {Ammari, Zied and Nier, Francis},
  journal   = {Annales Henri Poincaré},
  title     = {{Mean Field Limit for Bosons and Infinite Dimensional Phase-Space Analysis}},
  year      = {2008},
  issn      = {1424-0661},
  month     = Nov,
  number    = {8},
  pages     = {1503--1574},
  volume    = {9},
  doi       = {10.1007/s00023-008-0393-5},
  publisher = {Springer Science and Business Media LLC},
}

@Article{KrRa78,
  author  = {Kr\'ee, P. and R\k{a}czka, R.},
  journal = {Annales de l'I.H.P. Section A-Physique th\'eorique},
  title   = {{Kernels and symbols of operators in quantum field theory}},
  year    = {1978},
  number  = {1},
  pages   = {41--66},
  volume  = {28},
}

@Article{AmJaNo15,
  author    = {Amour, Laurent and Jager, Lisette and Nourrigat, Jean},
  journal   = {Journal of Functional Analysis},
  title     = {{On bounded pseudodifferential operators in Wiener spaces}},
  year      = {2015},
  issn      = {0022-1236},
  month     = Nov,
  number    = {9},
  pages     = {2747--2812},
  volume    = {269},
  doi       = {10.1016/j.jfa.2015.08.004},
  publisher = {Elsevier BV},
}

@Article{BuFr20b,
  author    = {Buchholz, Detlev and Fredenhagen, Klaus},
  journal   = {Annales Henri Poincaré},
  title     = {{Dynamical C*-algebras and Kinetic Perturbations}},
  year      = {2020},
  issn      = {1424-0661},
  month     = Dec,
  number    = {3},
  pages     = {1001--1033},
  volume    = {22},
  doi       = {10.1007/s00023-020-01002-3},
  publisher = {Springer Science and Business Media LLC},
}

@Article{BrDuFrRe22,
  author    = {Brunetti, Romeo and Dütsch, Michael and Fredenhagen, Klaus and Rejzner, Kasia},
  journal   = {Letters in Mathematical Physics},
  title     = {{C*-algebraic approach to interacting quantum field theory: inclusion of Fermi fields}},
  year      = {2022},
  issn      = {1573-0530},
  month     = Oct,
  number    = {5},
  volume    = {112},
  doi       = {10.1007/s11005-022-01590-7},
  publisher = {Springer Science and Business Media LLC},
}

@article{kirchberg_operations_1995,
    title = {Operations on continuous bundles of {C}*-algebras},
    volume = {303},
    issn = {0025-5831, 1432-1807},
    url = {http://link.springer.com/10.1007/BF01461011},
    doi = {10.1007/BF01461011},
    language = {en},
    number = {1},
    urldate = {2024-02-29},
    journal = {Mathematische Annalen},
    author = {Kirchberg, Eberhard and Wassermann, Simon},
    month = sep,
    year = {1995},
    pages = {677--697},
}

@article{buchholzScalingAlgebrasRenormalization1998,
	title = {Scaling Algebras and Renormalization Group in Algebraic Quantum Field Theory. {II}. Instructive Examples},
	volume = {10},
	issn = {0129-055X, 1793-6659},
	url = {http://arxiv.org/abs/hep-th/9708095},
	doi = {10.1142/S0129055X98000252},
	pages = {775--800},
	number = {6},
	journaltitle = {Reviews in Mathematical Physics},
	shortjournal = {Rev. Math. Phys.},
	author = {Buchholz, Detlev and Verch, Rainer},
	urldate = {2020-03-23},
	date = {1998-08},
	langid = {english},
	eprinttype = {arxiv},
	eprint = {hep-th/9708095},
}

@article{buchholzScalingAlgebrasRenormalization1995,
	title = {Scaling Algebras and Renormalization Group in Algebraic Quantum Field Theory},
	volume = {07},
	issn = {0129-055X, 1793-6659},
	url = {http://arxiv.org/abs/hep-th/9501063},
	doi = {10.1142/S0129055X9500044X},
	pages = {1195--1239},
	number = {8},
	journaltitle = {Reviews in Mathematical Physics},
	shortjournal = {Rev. Math. Phys.},
	author = {Buchholz, Detlev and Verch, Rainer},
	urldate = {2020-03-23},
	date = {1995-11},
	langid = {english},
	eprinttype = {arxiv},
	eprint = {hep-th/9501063},
}

@article{buchholzQuarksGluonsColour1996,
	title = {Quarks, gluons, colour: Facts or fiction?},
	volume = {469},
	issn = {05503213},
	url = {http://arxiv.org/abs/hep-th/9511002},
	doi = {10.1016/0550-3213(96)00143-5},
	shorttitle = {Quarks, gluons, colour},
	pages = {333--353},
	number = {1},
	journaltitle = {Nuclear Physics B},
	shortjournal = {Nuclear Physics B},
	author = {Buchholz, Detlev},
	urldate = {2020-08-04},
	date = {1996-06},
	eprinttype = {arxiv},
	eprint = {hep-th/9511002},
}

@article{binzFieldtheoreticWeylQuantization2004,
	title = {Field-theoretic Weyl Quantization as a Strict and Continuous Deformation Quantization},
	volume = {5},
	issn = {1424-0637, 1424-0661},
	url = {http://link.springer.com/10.1007/s00023-004-0171-y},
	doi = {10.1007/s00023-004-0171-y},
	pages = {327--346},
	number = {2},
	journaltitle = {Annales Henri Poincaré},
	shortjournal = {Ann. Henri Poincaré},
	author = {Binz, Ernst and Honegger, Reinhard and Rieckers, Alfred},
	urldate = {2020-10-22},
	date = {2004-04},
	langid = {english},
}

@article{manuceauSmallestAlgebraCanonical1973,
	title = {The smallest C*-algebra for canonical commutations relations},
	volume = {32},
	issn = {0010-3616, 1432-0916},
	url = {http://link.springer.com/10.1007/BF01645594},
	doi = {10.1007/BF01645594},
	pages = {231--243},
	number = {3},
	journaltitle = {Communications in Mathematical Physics},
	shortjournal = {Commun.Math. Phys.},
	author = {Manuceau, J. and Sirugue, M. and Testard, D. and Verbeure, A.},
	urldate = {2021-01-15},
	date = {1973-09},
	langid = {english},
}

@article{bayenDeformationTheoryQuantization1978,
	title = {Deformation theory and quantization. I. Deformations of symplectic structures},
	volume = {111},
	issn = {0003-4916},
	url = {https://www.sciencedirect.com/science/article/pii/0003491678902245},
	doi = {10.1016/0003-4916(78)90224-5},
	pages = {61--110},
	number = {1},
	journaltitle = {Annals of Physics},
	shortjournal = {Annals of Physics},
	author = {Bayen, F and Flato, M and Fronsdal, C and Lichnerowicz, A and Sternheimer, D},
	urldate = {2021-04-13},
	date = {1978-03-01},
	langid = {english},
}

@article{rieffelQuestionsQuantization1998,
	title = {Questions on quantization},
	url = {http://arxiv.org/abs/quant-ph/9712009},
	journaltitle = {{arXiv}:quant-ph/9712009},
	author = {Rieffel, Marc A.},
	urldate = {2021-04-13},
	date = {1998-04-23},
	langid = {english},
	eprinttype = {arxiv},
	eprint = {quant-ph/9712009},
}

@article{steegerExtensionsBundlesAlgebras2021,
	title = {Extensions of Bundles of C*-algebras},
	issn = {0129-055X, 1793-6659},
	url = {http://arxiv.org/abs/2102.11993},
	doi = {10.1142/S0129055X21500252},
	pages = {2150025},
	journaltitle = {Reviews in Mathematical Physics},
	shortjournal = {Rev. Math. Phys.},
	author = {Steeger, Jeremy and Feintzeig, Benjamin H.},
	urldate = {2021-06-02},
	date = {2021-05-26},
	eprinttype = {arxiv},
	eprint = {2102.11993},
}

@article{fewsterAlgebraicQuantumField2019,
	title = {Algebraic Quantum Field Theory -- an introduction},
	url = {https://arxiv.org/abs/1904.04051v2},
	author = {Fewster, Christopher J. and Rejzner, Kasia},
	urldate = {2022-02-11},
	date = {2019-04-08},
	langid = {english},
}

@article{binzConstructionUniquenessWeyl2004,
	title = {Construction and uniqueness of the C $^{\textrm{*}}$ -Weyl algebra over a general pre-symplectic space},
	volume = {45},
	issn = {0022-2488, 1089-7658},
	url = {http://aip.scitation.org/doi/10.1063/1.1757036},
	doi = {10.1063/1.1757036},
	pages = {2885--2907},
	number = {7},
	journaltitle = {Journal of Mathematical Physics},
	shortjournal = {Journal of Mathematical Physics},
	author = {Binz, Ernst and Honegger, Reinhard and Rieckers, Alfred},
	urldate = {2022-05-03},
	date = {2004-07},
	langid = {english},
}

@article{dantoniScalingAlgebrasSuperselection2006,
	title = {Scaling Algebras and Superselection Sectors: Study of a Class of Models},
	volume = {18},
	issn = {0129-055X, 1793-6659},
	url = {https://www.worldscientific.com/doi/abs/10.1142/S0129055X06002723},
	doi = {10.1142/S0129055X06002723},
	shorttitle = {{SCALING} {ALGEBRAS} {AND} {SUPERSELECTION} {SECTORS}},
	pages = {565--594},
	number = {5},
	journaltitle = {Reviews in Mathematical Physics},
	shortjournal = {Rev. Math. Phys.},
	author = {D'Antoni, Claudio and Morsella, Gerardo},
	urldate = {2022-10-19},
	date = {2006-06},
	langid = {english},
}

@article{rieffelContinuousFieldsAlgebras1989,
	title = {Continuous fields of C*-algebras coming from group cocycles and actions},
	volume = {283},
	issn = {0025-5831, 1432-1807},
	url = {http://link.springer.com/10.1007/BF01442857},
	doi = {10.1007/BF01442857},
	pages = {631--643},
	number = {4},
	journaltitle = {Mathematische Annalen},
	shortjournal = {Math. Ann.},
	author = {Rieffel, Marc A.},
	urldate = {2022-11-23},
	date = {1989-04},
	langid = {english},
}

@article{bostelmannDilationSymmetriesArising2010,
	title = {On Dilation Symmetries Arising from Scaling Limits},
	volume = {294},
	issn = {0010-3616, 1432-0916},
	url = {http://link.springer.com/10.1007/s00220-009-0899-9},
	doi = {10.1007/s00220-009-0899-9},
	pages = {21--60},
	number = {1},
	journaltitle = {Communications in Mathematical Physics},
	shortjournal = {Commun. Math. Phys.},
	author = {Bostelmann, Henning and D’Antoni, Claudio and Morsella, Gerardo},
	urldate = {2023-01-19},
	date = {2010-02},
	langid = {english},
}

@article{bostelmannScalingAlgebrasPointlike2009,
	title = {Scaling Algebras and Pointlike Fields: A Nonperturbative Approach to Renormalization},
	volume = {285},
	issn = {0010-3616, 1432-0916},
	url = {http://link.springer.com/10.1007/s00220-008-0613-3},
	doi = {10.1007/s00220-008-0613-3},
	shorttitle = {Scaling Algebras and Pointlike Fields},
	pages = {763--798},
	number = {2},
	journaltitle = {Communications in Mathematical Physics},
	shortjournal = {Commun. Math. Phys.},
	author = {Bostelmann, Henning and D’Antoni, Claudio and Morsella, Gerardo},
	urldate = {2023-01-19},
	date = {2009-01},
	langid = {english},
}

@article{contiAsymptoticMorphismsSuperselection2014,
	title = {Asymptotic morphisms and superselection theory in the scaling limit},
	volume = {8},
	issn = {1661-6952},
	url = {https://ems.press/doi/10.4171/jncg/169},
	doi = {10.4171/JNCG/169},
	pages = {735--770},
	number = {3},
	journaltitle = {Journal of Noncommutative Geometry},
	shortjournal = {J. Noncommut. Geom.},
	author = {Conti, Roberto and Morsella, Gerardo},
	urldate = {2023-01-19},
	date = {2014},
	langid = {english},
}

@article{contiScalingLimitSubsystems2009,
	title = {Scaling Limit for Subsystems and Doplicher–Roberts Reconstruction},
	volume = {10},
	issn = {1424-0637, 1424-0661},
	url = {http://link.springer.com/10.1007/s00023-009-0418-8},
	doi = {10.1007/s00023-009-0418-8},
	pages = {485--511},
	number = {3},
	journaltitle = {Annales Henri Poincaré},
	shortjournal = {Ann. Henri Poincaré},
	author = {Conti, Roberto and Morsella, Gerardo},
	urldate = {2023-01-19},
	date = {2009-06},
	langid = {english},
}

@misc{morsellaStructureChargesUltraviolet2002,
	title = {The Structure of Charges in the Ultraviolet and an Intrinsic Notion of Confinement},
	url = {http://arxiv.org/abs/math-ph/0201042},
	number = {{arXiv}:math-ph/0201042},
	publisher = {{arXiv}},
	author = {Morsella, Gerardo},
	urldate = {2023-01-19},
	date = {2002-05-27},
	langid = {english},
	eprinttype = {arxiv},
	eprint = {math-ph/0201042},
}

@article{haagAlgebraicApproachQuantum1964,
	title = {An Algebraic Approach to Quantum Field Theory},
	volume = {5},
	issn = {0022-2488, 1089-7658},
	url = {http://aip.scitation.org/doi/10.1063/1.1704187},
	doi = {10.1063/1.1704187},
	pages = {848--861},
	number = {7},
	journaltitle = {Journal of Mathematical Physics},
	shortjournal = {Journal of Mathematical Physics},
	author = {Haag, Rudolf and Kastler, Daniel},
	urldate = {2023-02-07},
	date = {1964-07},
	langid = {english},
}

@book{landsmanFoundationsQuantumTheory2017,
	location = {New York, {NY}},
	title = {Foundations of quantum theory},
	isbn = {978-3-319-51776-6},
	publisher = {Springer Berlin Heidelberg},
	author = {Landsman, N. P.},
	date = {2017},
}

@book{haagLocalQuantumPhysics1996,
	location = {Berlin ; New York},
	edition = {2nd rev. and enl. ed},
	title = {Local quantum physics: fields, particles, algebras},
	isbn = {978-3-540-61451-7},
	series = {Texts and monographs in physics},
	shorttitle = {Local quantum physics},
	pagetotal = {390},
	publisher = {Springer},
	author = {Haag, Rudolf},
	date = {1996},
}

@book{landsmanMathematicalTopicsClassical1998,
	location = {New York},
	title = {Mathematical topics between classical and quantum mechanics},
	isbn = {978-0-387-98318-9},
	series = {Springer monographs in mathematics},
	pagetotal = {529},
	publisher = {Springer},
	author = {Landsman, N. P.},
	date = {1998},
}

@article{steegerClassicalLimitSingular2021,
	title = {Is the classical limit “singular”?},
	volume = {88},
	issn = {0039-3681},
	url = {https://www.sciencedirect.com/science/article/pii/S0039368121000716},
	doi = {10.1016/j.shpsa.2021.05.007},
	pages = {263--279},
	journaltitle = {Studies in History and Philosophy of Science},
	shortjournal = {Studies in History and Philosophy of Science},
	author = {Steeger, Jeremy and Feintzeig, Benjamin H.},
	urldate = {2023-02-25},
	date = {2021-08-01},
	langid = {english},
}

@book{baezIntroductionAlgebraicConstructive1992,
	location = {Princeton, N.J},
	title = {Introduction to algebraic and constructive quantum field theory},
	isbn = {978-0-691-08546-3},
	series = {Princeton series in physics},
	pagetotal = {291},
	publisher = {Princeton University Press},
	author = {Baez, John C. and Segal, Irving Ezra and Zhou, Zhengfang},
	date = {1992},
}

@book{reedMethodsModernMathematical1972,
	location = {New York},
	title = {Methods of modern mathematical physics},
	isbn = {978-0-12-585001-8},
	publisher = {Academic Press},
	author = {Reed, Michael and Simon, Barry},
	date = {1972},
}

@article{contiAsymptoticMorphismsSuperselection2020,
	title = {Asymptotic morphisms and superselection theory in the scaling limit {II}: analysis of some models},
	volume = {376},
	issn = {0010-3616, 1432-0916},
	url = {http://arxiv.org/abs/1901.07292},
	doi = {10.1007/s00220-019-03564-8},
	shorttitle = {Asymptotic morphisms and superselection theory in the scaling limit {II}},
	pages = {1767--1801},
	number = {3},
	journaltitle = {Communications in Mathematical Physics},
	shortjournal = {Commun. Math. Phys.},
	author = {Conti, Roberto and Morsella, Gerardo},
	urldate = {2023-09-20},
	date = {2020-06},
	eprinttype = {arxiv},
	eprint = {1901.07292 [hep-th, physics:math-ph]},
}

@article{wernerQuantumHarmonicAnalysis1984,
	title = {Quantum harmonic analysis on phase space},
	volume = {25},
	issn = {0022-2488, 1089-7658},
	url = {https://pubs.aip.org/jmp/article/25/5/1404/227463/Quantum-harmonic-analysis-on-phase-space},
	doi = {10.1063/1.526310},
	pages = {1404--1411},
	number = {5},
	journaltitle = {Journal of Mathematical Physics},
	author = {Werner, R.},
	urldate = {2024-02-12},
	date = {1984-05-01},
	langid = {english},
}

@article{kirchbergOperationsContinuousBundles1995,
	title = {Operations on continuous bundles of C*-algebras},
	volume = {303},
	issn = {0025-5831, 1432-1807},
	url = {http://link.springer.com/10.1007/BF01461011},
	doi = {10.1007/BF01461011},
	pages = {677--697},
	number = {1},
	journaltitle = {Mathematische Annalen},
	shortjournal = {Math. Ann.},
	author = {Kirchberg, Eberhard and Wassermann, Simon},
	urldate = {2024-02-29},
	date = {1995-09},
	langid = {english},
}

@article{fellStructureAlgebrasOperator1961,
	title = {The structure of algebras of operator fields},
	volume = {106},
	issn = {0001-5962},
	url = {http://projecteuclid.org/euclid.acta/1485889234},
	doi = {10.1007/BF02545788},
	pages = {233--280},
	number = {3},
	journaltitle = {Acta Mathematica},
	shortjournal = {Acta Math.},
	author = {Fell, J. M. G.},
	urldate = {2024-02-29},
	date = {1961},
	langid = {english},
}

@book{bratteliOperatorAlgebrasQuantum1981,
	location = {New York Berlin Paris [etc.]},
	edition = {2nd ed},
	title = {Operator Algebras and Quantum Statistical Mechanics: Equilibrium States, Models in Quantum Statistical Mechanics},
	volume = {2},
	isbn = {978-0-387-10381-5 978-3-540-10381-3},
	publisher = {Springer},
	author = {Bratteli, Ola and Robinson, Derek William},
	date = {1981},
}

@book{dixmierAlgebras1977,
	location = {Amsterdam ; New York : New york},
	title = {C*-algebras},
	isbn = {978-0-7204-0762-4},
	series = {North-Holland mathematical library},
	pagetotal = {492},
	number = {v. 15},
	publisher = {North-Holland ; sole distributors for the U.S.A. and Canada, Elsevier North Holland},
	author = {Dixmier, Jacques},
	date = {1977},
}

@article{dantoniScalingAlgebrasCharged2004,
	title = {Scaling Algebras for Charged Fields and Short-Distance Analysis for Localizable and Topological Charges},
	volume = {5},
	rights = {http://www.springer.com/tdm},
	issn = {1424-0637, 1424-0661},
	url = {http://link.springer.com/10.1007/s00023-004-0183-7},
	doi = {10.1007/s00023-004-0183-7},
	pages = {809--870},
	number = {5},
	journaltitle = {Annales Henri Poincaré},
	shortjournal = {Ann. Henri Poincaré},
	author = {D’Antoni, Claudio and Morsella, Gerardo and Verch, Rainer},
	urldate = {2024-05-17},
	date = {2004-09},
	langid = {english},
}

@incollection{halvorsonAlgebraicQuantumField2007,
	title = {Algebraic Quantum Field Theory},
	isbn = {978-0-444-51560-5},
	url = {https://linkinghub.elsevier.com/retrieve/pii/B9780444515605500117},
	pages = {731--864},
	booktitle = {Philosophy of Physics},
	publisher = {Elsevier},
	author = {Halvorson, Hans},
	urldate = {2025-02-18},
	date = {2007},
	langid = {english},
	doi = {10.1016/B978-044451560-5/50011-7},
}

@book{fellRepresentationsAlgebrasLocally1988,
	location = {Boston},
	title = {Representations of *-algebras, locally compact groups, and Banach *-algebraic bundles},
	isbn = {978-0-12-252721-0 978-0-12-252722-7},
	series = {Pure and applied mathematics},
	pagetotal = {2},
	number = {125-126},
	publisher = {Academic Press},
	author = {Fell, J. M. G. and Doran, Robert S.},
	date = {1988},
}

@book{arakiMathematicalTheoryQuantum1999,
	location = {Oxford ; New York},
	title = {Mathematical theory of quantum fields},
	isbn = {978-0-19-851773-3},
	series = {The international series of monographs on physics},
	pagetotal = {236},
	number = {101},
	publisher = {Clarendon Press},
	author = {Araki, Huzihiro},
	date = {1999},
}

@article{bayenDeformationTheoryQuantization1978b,
	title = {Deformation theory and quantization. {II}. Physical applications},
	volume = {111},
	rights = {https://www.elsevier.com/tdm/userlicense/1.0/},
	issn = {00034916},
	url = {https://linkinghub.elsevier.com/retrieve/pii/0003491678902257},
	doi = {10.1016/0003-4916(78)90225-7},
	pages = {111--151},
	number = {1},
	journaltitle = {Annals of Physics},
	shortjournal = {Annals of Physics},
	author = {Bayen, F and Flato, M and Fronsdal, C and Lichnerowicz, A and Sternheimer, D},
	urldate = {2025-02-18},
	date = {1978-03},
	langid = {english},
}

@article{kontsevichDeformationQuantizationPoisson2003,
	title = {Deformation Quantization of Poisson Manifolds},
	volume = {66},
	issn = {0377-9017},
	url = {http://link.springer.com/10.1023/B:MATH.0000027508.00421.bf},
	doi = {10.1023/B:MATH.0000027508.00421.bf},
	pages = {157--216},
	number = {3},
	journaltitle = {Letters in Mathematical Physics},
	shortjournal = {Letters in Mathematical Physics},
	author = {Kontsevich, Maxim},
	urldate = {2025-02-18},
	date = {2003-12},
	langid = {english},
}

@article{fedosovSimpleGeometricalConstruction1994,
	title = {A simple geometrical construction of deformation quantization},
	volume = {40},
	issn = {0022-040X},
	url = {https://projecteuclid.org/journals/journal-of-differential-geometry/volume-40/issue-2/A-simple-geometrical-construction-of-deformation-quantization/10.4310/jdg/1214455536.full},
	doi = {10.4310/jdg/1214455536},
	number = {2},
	journaltitle = {Journal of Differential Geometry},
	shortjournal = {J. Differential Geom.},
	author = {Fedosov, Boris V.},
	urldate = {2025-02-18},
	date = {1994-01-01},
	langid = {english},
}

@book{daleckijMeasuresDifferentialEquations1991,
	location = {Dordrecht Boston London},
	title = {Measures and differential equations in infinite dimensional spaces},
	isbn = {978-0-7923-1517-9},
	series = {Mathematics and its applications},
	number = {76},
	publisher = {Kluwer academic publ},
	author = {Dalecky, Y. and Fomin, S.V.},
	date = {1991},
}

\end{document}